\documentclass[aps,prb,twocolumn,superscriptaddress,floatfix,nofootinbib,longbibliography,letter]{revtex4-1}
\usepackage{epsfig,lipsum,amsmath,amssymb,color,dsfont,physics}
\definecolor{jadclr}{rgb}{0,0.5,0.5}
\definecolor{jadcolor}{rgb}{0.85,0.33,0.1}
\definecolor{philipp}{rgb}{1,.4,.3}

\usepackage[dvipsnames]{xcolor}
\usepackage[bookmarks=true,colorlinks,linkcolor=OrangeRed,urlcolor=NavyBlue,citecolor=RoyalBlue]{hyperref}
\usepackage{bm}
\usepackage{booktabs}
\usepackage{enumerate}
\usepackage{array}
\usepackage{dsfont}
\usepackage{microtype}
\usepackage[normalem]{ulem}
\newcommand{\mbeq}{\overset{!}{=}}
\def\d{\mathrm d}
\definecolor{mypurple}{rgb}{0.49,0.18,0.56}
\definecolor{mygold}{rgb}{0.93,0.69,0.13}
\definecolor{mygreen}{rgb}{0,0.5,0}
\definecolor{myblue}{rgb}{0,0,0.75}
\definecolor{mymagenta}{cmyk}{0,1,0,0.12}
\definecolor{mygray}{rgb}{0.5,0.5,0.5}

\newcommand{\canc}[1]{}

\usepackage{mathtools}

\usepackage{mathrsfs}
\makeatletter
\newcommand{\thickhline}{
	\noalign {\ifnum 0=`}\fi \hrule height 1pt
	\futurelet \reserved@a \@xhline
}
\newcolumntype{"}{@{\hskip\tabcolsep\vrule width 1pt\hskip\tabcolsep}}
\makeatother
\begin{document}
\title{Origin of staircase prethermalization in lattice gauge theories}
\author{Jad C.~Halimeh} 
\affiliation{INO-CNR BEC Center and Department of Physics, University of Trento, Via Sommarive 14, I-38123 Trento, Italy}
\affiliation{Kirchhoff Institute for Physics, Ruprecht-Karls-Universit\"{a}t Heidelberg, Im Neuenheimer Feld 227, 69120 Heidelberg, Germany}
\affiliation{Institute for Theoretical Physics, Ruprecht-Karls-Universit\"{a}t Heidelberg, Philosophenweg 16, 69120 Heidelberg, Germany}

\author{Philipp Hauke}
\affiliation{INO-CNR BEC Center and Department of Physics, University of Trento, Via Sommarive 14, I-38123 Trento, Italy}
\affiliation{Kirchhoff Institute for Physics, Ruprecht-Karls-Universit\"{a}t Heidelberg, Im Neuenheimer Feld 227, 69120 Heidelberg, Germany}
\affiliation{Institute for Theoretical Physics, Ruprecht-Karls-Universit\"{a}t Heidelberg, Philosophenweg 16, 69120 Heidelberg, Germany}

\begin{abstract}
	Quantum many-body systems with exact local gauge symmetries exhibit rich out-of-equilibrium physics such as constrained dynamics and disorder-free localization. In a joint submission [J.~C.~Halimeh and P.~Hauke, \href{https://arxiv.org/abs/2004.07248}{arXiv:2004.07248}], we present evidence of \textit{staircase prethermalization} in a $\mathrm{Z}_2$ lattice gauge theory subjected to a small breaking of gauge invariance. Here, we consolidate this finding and the associated emergent nonperturbative timescales analytically and numerically. By means of a Magnus expansion, we demonstrate how exact resonances between different gauge-invariant supersectors are the main reason behind the emergence of staircase prethermalization. Furthermore, we showcase the robustness of our conclusions against various initial conditions including different system sizes, matter fillings, and gauge-invariance sectors, in addition to various boundary conditions, such as different maximal on-site matter occupations. We also elaborate on how our conclusions are unique to local-symmetry models and why they break down in the case of global-symmetry breaking. We moreover extend our results to $\mathrm{U}(1)$ lattice gauge theories, illustrating the generality of our findings. Our work offers an analytic footing into the constrained dynamics of lattice gauge theories and provides proof of a certain intrinsic robustness of gauge-theory dynamics to errors in experimental settings.
\end{abstract}

\date{\today}
\maketitle

\tableofcontents

\section{Introduction}\label{sec:Intro}
Quantum simulators for lattice gauge theories provide a fundamentally different viewpoint on subatomic physics from a low-energy perspective.\cite{Wiese_review,Dalmonte_review,Zohar_review,Banuls_review} 
Not only does this approach generate new cross-fertilizations between different physics communities, but it also promises to enable cheaper and faster experimental investigations \cite{Martinez2016,Bernien2017,Klco2018,Goerg2019,Kokail2019,Klco2019,Schweizer2019,Mil2019,Yang2020} into open questions from high-energy physics. 
One major challenge for the low-energy experimental realization of gauge theories is the need to ensure gauge invariance, the local conservation law that ties charged matter and gauge fields to each other. 
The requirement to engineer the experimental system such that it respects the gauge symmetry\cite{Halimeh2020a,Halimeh2020e} is in stark contrast to subatomic physics, where---with the exception of few examples\cite{Foerster1980,Poppitz2008,Wetterich2017}---gauge invariance is mostly accepted as a law given by nature\cite{Weinberg_book,Rothe_book} (see Refs.~\onlinecite{Narayanan1995,Golterman2001} for examples of discussions of gauge-invariance violations in lattice numerics).
However, in quantum-simulator setups where both matter and gauge fields are realized as active degrees of freedom,\cite{Schweizer2019,Mil2019,Yang2020} enforcing perfect gauge invariance is impossible, as it would require unrealistic fine tuning of experimental parameters. 
Despite several remedies already available to mitigate the effect of inherent gauge invariance-breaking errors,\cite{Banerjee2012,Hauke2013,Stannigel2014,Kuehn2014,Kuno2015,Yang2016,Kuno2017,Negretti2017,Barros2019} understanding the time evolution of gauge violation in lattice gauge theories is an open challenge.  

\begin{figure}[htp]
	\centering
	\hspace{-.25 cm}
	\includegraphics[width=.48\textwidth]{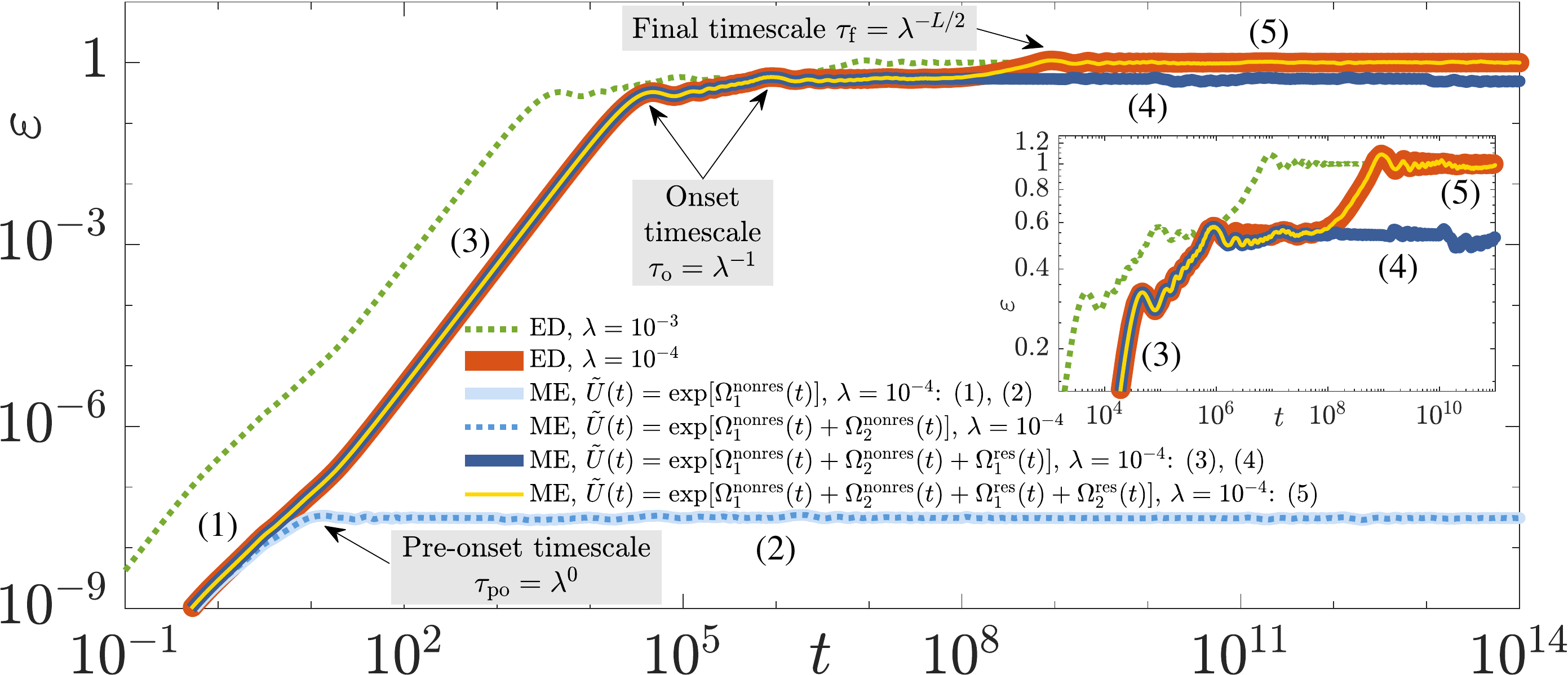}\quad
	\hspace{-.15 cm}
	\caption{(Color online). Paradigm example of staircase prethermalization. The system is a lattice gauge theory whose gauge invariance is slightly broken by a term of strength $\lambda$. The key observable is the gauge-symmetry violation $\varepsilon$. 
	We identify the following typical time regimes:   
	{\bf{(1)}} An initial increase $\varepsilon\propto (\lambda t)^2$ follows from time-dependent perturbation theory (see also Ref.~\onlinecite{Halimeh2020a}). 
	{\bf{(2)}} First features appear at a `pre-onset' timescale $\lambda^0$ when fast oscillating terms---due to large energy gaps $\Delta E$ in the unperturbed lattice gauge theory---start being averaged away, $t\Delta E \gg 1$. 
	{\bf{(3)}} Due to resonant processes ($\Delta E=0$), gauge invariance continues to rise as $\varepsilon\propto (\lambda t)^2$, leading to a breakdown of time-dependent perturbation theory. The subsequent time regimes can be described through effective Hamiltonians $H_\mathrm{eff}^{(s)}\propto \lambda^s$, derivable from corresponding resonant terms $\Omega_s^\mathrm{res}$ in a Magnus expansion (see Sec.~\ref{sec:ME}).
	{\bf{(4)}} 
	At a time $t\gg \tau_\mathrm{o}=\lambda^{-1}$, the dynamics generated by $H_\mathrm{eff}^{(1)}$	saturates to a steady state. The onset plateau is reached. 
	{\bf{(5)}} At a timescale $\tau_\mathrm{i}=\lambda^{-2}$, $H_\mathrm{eff}^{(2)}$ becomes dominant and drives the system away from the onset plateau towards a new steady state. 
	The phenomenon repeats for higher orders in the Magnus expansion until the final plateau is reached at a timescale $\tau_\mathrm{f}=\lambda^{-L/2}$. 
	For this plot, we have used numerical data for the $\mathrm{Z}_2$ LGT with $L=4$ matter sites, but the buildup of staircase prethermalization at timescales $\lambda^0,\lambda^{-1},\ldots,\lambda^{-L/2}$ is general for any $L$ (see Sec.~\ref{sec:Z2LGT}) and is also valid in the $\mathrm{U}(1)$ LGT (see Sec.~\ref{sec:U1LGT}). ME: Magnus expansion. ED: exact diagonalization. 
	} 
	\label{fig:Schematics} 
\end{figure}

In a joint submission,\cite{Halimeh2020b} we demonstrate how a small breaking of gauge invariance with strength $\lambda$ drives a $\mathrm{Z}_2$ lattice gauge theory into the nonperturbative behavior of \textit{staircase prethermalization} (see also Fig.~\ref{fig:Schematics}). In this phenomenon, the gauge violation and local observables enter long-lived prethermal plateaus occurring at timescales $\lambda^{-s}$, with $s$ an integer in the range $0\leq s \leq L/2$, where $L$ is the (even) number of matter sites. The plateau reached at timescale $\lambda^{-L/2}$ is the final nonthermal steady state, where the gauge violation is maximal. The phenomenon of staircase prethermalization is of relevance not only for near-future quantum simulators, but also in the quest for understanding the real-time dynamics of gauge theories. In particular, it complements recent works indicating slow dynamics and many-body localization in systems with an abundance of local constraints---but without any terms that break gauge invariance.\cite{Smith2017a,Smith2017b,Brenes2018,Turner2018,Karpov2020} 

In this paper, we provide a substantially increased understanding of our original findings\cite{Halimeh2020b} through several ways. 
First, we provide in Sec.~\ref{sec:ME} a detailed analytic derivation based on a Magnus expansion to corroborate and further explain the aforementioned timescales arising in the prethermal staircase. This permits a full explanation of the timescales observed in staircase prethermalization, and even brings about an excellent quantitative agreement with exact results. Sections~\ref{sec:Z2LGT} to~\ref{sec:U1LGT} illuminate this phenomenon through numerical results for various parameters and initial conditions, as well as discrete and continuous gauge symmetries, demonstrating that it is a generic phenomenon for models with local symmetries. In Secs.~\ref{sec:GlobSym} and~\ref{sec:MBL}, we put our results into context of systems that break a global symmetry and of recent results on constrained dynamics in lattice gauge theories, respectively. We conclude with a short discussion in Sec.~\ref{sec:conclusion}. Our paper also includes four Appendices. Appendix~\ref{sec:glossary} constitutes a glossary of terminology we repeatedly use throughout the text that is either novel or not yet well established in the community. Appendix~\ref{sec:furtherME} provides further numerical results on the Magnus expansion, while Appendix~\ref{sec:Numerics} offers more details on the exact diagonalization (ED) numerics. Appendix~\ref{sec:TDPT} shows a derivation in time-dependent perturbation theory (TDPT) to describe the short-time dynamics.

\section{Staircase prethermalization: A Magnus expansion analysis}\label{sec:ME}
In this Section, we detail our analytic treatment in Magnus expansion that fully elucidates the phenomenon of staircase prethermalization first observed in our joint submission Ref.~\onlinecite{Halimeh2020b}. 
The Magnus expansion explains the observed timescales through the appearance of exact resonances between different gauge-invariance supersectors (see Appendix~\ref{sec:glossary} for precise definition) in the unperturbed gauge theory. Even more, it provides striking quantitative agreement with the corresponding ED results over all stages of the time evolution; cf.~Fig.~\ref{fig:Schematics}.

\subsection{Quench protocol and results}
The scenario that leads to staircase prethermalization as observed in Ref.~\onlinecite{Halimeh2020b} is as follows. 
The system of interest is a gauge theory, living in a one-dimensional spatial lattice with sites $j=1,\ldots,L$. The theory is described by Hamiltonian $H_0$ and generators of the gauge transformation $G_j$ at each matter site $j$, which fulfil $\left[H_0,G_j\right]=0$, $\forall j$. In other words, the $G_j$ represent a set of local symmetries that are conserved under the dynamics of $H_0$. 
For the example of the $\mathrm{Z}_2$ gauge theory considered in Sec.~\ref{sec:Z2LGT}, $G_j$ assumes the two eigenvalues $g_j=0,2$; see Eq.~\eqref{eq:Gauss}.
We consider a situation where this ideal local gauge invariance is perturbed by a term $\lambda H_1$, with $\left[H_1,G_j\right]\neq 0$ and where $\lambda$ controls the strength of the perturbation. 
The initial state is chosen as a pure state from one gauge-invariant sector (see Glossary in Appendix~\ref{sec:glossary}) with fixed $g_j$ values. The time evolution under the full Hamiltonian $H=H_0+\lambda H_1$ will generate deviations from the initial $g_j$ eigenvalues beyond the gauge-invariant supersector containing the initial sector. We denote this as a gauge violation. 

As discussed in Ref.~\onlinecite{Halimeh2020b} and seen in Fig.~\ref{fig:Schematics}, this scenario leads to a sequence of plateaus in local observables such as the gauge violation. While TDPT can cleanly explain the short-time behavior (see Appendix~\ref{sec:TDPT}), the long evolution times where these plateaus appear require more powerful analytic methods. 
A Magnus expansion of the time-evolution operator fulfils this purpose, as shown by our detailed derivations presented in the following. 

\begin{table*}\footnotesize
	\caption{Energy gaps of states connected by $H_1$ in first- and second-order processes after starting in the $P_0$ supersector of the $\mathrm{Z}_2$ LGT with $L=4$ matter sites. The notation is as follows: $E_{\alpha\beta}=E_{\beta,l}-E_{\alpha,q}$ denotes the energy gap due to a first-order process in $H_1$ from a gauge-invariant sector within supersector $\{\alpha\}$ to a gauge-invariant sector within supersector $\{\beta\}$, such that the corresponding eigenstates $\ket{\alpha,q}$ and $\ket{\beta,l}$ lead to a nonvanishing amplitude $\bra{\alpha,q}H_1\ket{\beta,l}$; cf.~Eq.~\eqref{eq:Omega1}. Similarly, $E_{\alpha\beta\gamma}=E_{\gamma,p}-E_{\alpha,q}$ stands for the gap due to a second-order process in $H_1$ going first from a gauge-invariant sector within supersector $\{\alpha\}$ to a gauge-invariant sector within supersector $\{\beta\}$, and then from the latter to a gauge-invariant sector within supersector $\{\gamma\}$, such that the amplitudes $\bra{\alpha,q}H_1\ket{\beta,l}$ and $\bra{\beta,l}H_1\ket{\gamma,p}$ are nonvanishing; cf.~Eq.~\eqref{eq:Omega_2_LehmannBasis}. While there are many exact resonances (last column), the system accesses no processes with small but nonvanishing gap (second-to-last column). Here, we have chosen $\lambda_\text{thresh}=0.01$ because when the gauge-violation strength is above this threshold some of the plateaus are compromised in the prethermal staircase (see, e.g., Fig.~\ref{fig:FigL4} in Sec.~\ref{sec:Z2LGT}).}
	\label{tab:energyGapsZ2L4}
	\begin{tabular}{| l | l | l | l | l |}\hline
		Gap & Nonzero minimum & Total number of accessible states & Number of states with $0<E<\lambda_\text{thresh}$ & Number of states with $E=0$ \\ \hline\hline
		$E_{02}$ & $0.039825$ & $768$ & $0$ & $64$ ($8.33\%$ of total) \\ \hline
		$E_{020}$ & $0.479219$ & $9216$ & $0$ & $1152$ ($12.50\%$ of total)\\ \hline
		$E_{022}$ & $0.039825$ & $48384$ & $0$ & $4608$ ($9.52\%$ of total) \\ \hline
		$E_{024}$ & $0.479219$ & $9204$ & $0$& $1151$ ($23.76\%$ of total) \\ \hline
	\end{tabular}
\end{table*}

\subsection{Magnus expansion}

We start by rewriting the full time-evolution operator $U(t)=\mathrm{e}^{-\mathrm{i}(H_0+\lambda H_1)t}$ in an interaction picture with respect to the gauge-invariant part $H_0$, 
\begin{align}
&U(t)=\mathrm{e}^{-\mathrm{i}H_0t}\tilde{U}(t),\\\label{eq:Utilde}
&\tilde{U}(t)=\mathcal{T}\big\{\mathrm{e}^{-\mathrm{i}\lambda\int_0^t \d\tau H_1(\tau)}\big\}.
\end{align}
In this exact rewriting, all processes that induce transitions between various gauge-invariant supersectors are contained in $\tilde{U}(t)$. 
This time-evolution operator is in general a complicated object, being governed by the time-dependent Hamiltonian $H_1(t)=\mathrm{e}^{\mathrm{i}H_0t}H_1\mathrm{e}^{-\mathrm{i}H_0t}$, which rotates with frequencies generated by $H_0$. 
The aim of this Section is to derive a description of $\tilde{U}(t)=\mathrm{e}^{\Omega(t)}$ through an effective skew-Hermitian operator $\Omega(t)$, from which the emergence of the plateaus becomes transparent. 

To this end, we employ the Magnus expansion,\cite{Blanes2009} which entails finding the effective skew-Hermitian operator as a perturbative expansion in orders of $\lambda H_1$, 
$\Omega(t)=\sum_{n=1}^\infty\Omega_n(t)$ 
with 
\begin{align}\label{eq:OmegaOne}
&\Omega_1(t)=-\mathrm{i}\lambda\int_0^t \d t_1H_1(t_1),\\\label{eq:Omega2}
&\Omega_2(t)=-\frac{\lambda^2}{2}\int_0^t \d t_1\int_0^{t_1}\d t_2 \big[H_1(t_1),H_1(t_2)\big],
\end{align}
and so on. Each order in the Magnus expansion contains an infinite resummation of terms from TDPT. This property enables accurate descriptions of time evolutions far beyond the abilities of usual TDPT, although the convergence of the Magnus expansion in many-body systems typically needs to be checked by numerical means.\cite{Blanes2009}
As we will see in the following, the Magnus expansion reveals a separation of timescales that gives a clear physical picture for the occurrence of stable gauge-violation plateaus. 


\begin{table*}\footnotesize
	\caption{Same as Table~\ref{tab:energyGapsZ2L4} but for $L=6$ matter sites. Only a few small but nonvanishing gaps play a role (second-to-last column), while there are many more exact resonances (last column). Note that there are many additional resonances in third-order processes (such as $E_{0246}$), which give rise to the third and final prethermal plateau, but these are not shown due to the large computational overhead required to compute them.}
	\label{tab:energyGapsZ2L6}
	\begin{tabular}{| l | l | l | l | l |}\hline
		Gap & Nonzero minimum & Total number of accessible states & Number of states with $0<E<\lambda_\text{thresh}$ & Number of states with $E=0$ \\ \hline\hline
		$E_{02}$ & $0.003545$ & $11502$ & $24$ (0.21$\%$ of total) & $480$ ($4.17\%$ of total) \\ \hline
		$E_{020}$ & $0.009376$ & $459382$ & $2304$ ($0.5\%$ of total) & $33386$ ($7.27\%$ of total)\\ \hline
		$E_{022}$ & $0.000275$ & $5171556$ & $4104$ ($0.08\%$ of total) & $138216$ ($2.67\%$ of total) \\ \hline
		$E_{024}$ & $0.000275$ & $5035596$ & $12094$ ($0.24\%$ of total) & $138176$ ($2.74\%$ of total) \\ \hline
	\end{tabular}
\end{table*}

\subsubsection{First-order Magnus term and onset of first plateau}\label{sec:MEA}

From TDPT (see Appendix~\ref{sec:TDPT}), we know that at short times the leading order in $\lambda$ dominates the time evolution of gauge violation. Thus, we first focus on $\Omega_1(t)$, which can be written as
\begin{align}
\label{eq:Omega1}
\Omega_1(t)=&\,-\mathrm{i}\lambda\int_0^t \d t_1\sum_{\alpha,\beta}\sum_{q,l}\mathrm{e}^{\mathrm{i}(E_{\alpha,q}-E_{\beta,l})t_1}\nonumber\\
&\,\qquad\times\bra{\alpha,q}H_1\ket{\beta,l}\ket{\alpha,q}\bra{\beta,l},
\end{align}
in a common energy eigenbasis $\{\ket{\alpha,q}\}$ of $H_0$ and $G_j$, $\forall j$, where $\alpha$ denotes the fixed gauge-invariant sector of the eigenstate $\ket{\alpha,q}$, and $q$ represents all remaining good quantum numbers such as the associated eigenenergy $E_{\alpha,q}$. The time evolution generated by $\Omega_1(t)$ distinguishes two crucially different cases, depending on whether $H_1$ accesses an exact resonance ($E_{\alpha,q}= E_{\beta,l}$) or not ($E_{\alpha,q}\neq E_{\beta,l}$). 

In the case of $E_{\alpha,q}\neq E_{\beta,l}$, we get
\begin{align}
\Omega_1^\text{nonres}(t)=&\,-\lambda\sum_{\alpha,\beta}\sum_{q,l}\frac{\mathrm{e}^{\mathrm{i}(E_{\alpha,q}-E_{\beta,l})t}-1}{E_{\alpha,q}-E_{\beta,l}}\nonumber\\
&\,\qquad\times\bra{\alpha,q}H_1\ket{\beta,l}\ket{\alpha,q}\bra{\beta,l}\,.
\end{align}
At short times, the oscillating phases $\mathrm{e}^{\mathrm{i}(E_{\alpha,q}-E_{\beta,l})t}$ generate a time evolution out of the initial gauge-invariant supersector when $\tilde{U}(t)$ acts on a quantum state, leading to the increase {\bf{(1)}} in Fig.~\ref{fig:Schematics}. 
For the considered system sizes, the gaps $E_{\alpha,q}-E_{\beta,l}$ are, however, almost all either 0 (to a numerical precision of $10^{-10}$) or on the order of $0.01 J$ or larger, with $J>0$ the energy unit; see Tables \ref{tab:energyGapsZ2L4} and \ref{tab:energyGapsZ2L6} for the $\mathrm{Z}_2$ LGT, as an example. Thus, at times $t\gg 1/J$, the oscillating terms $\mathrm{e}^{\mathrm{i}(E_{\alpha,q}-E_{\beta,l})t}$ average out, and the nonresonant first-order contribution in the Magnus expansion becomes a completely time-independent operator. Similar arguments hold for nonresonant terms in the higher orders of the Magnus expansion. Thus, at a \textit{pre-onset} timescale $\tau_\mathrm{po}=\lambda^0$ the dynamics of gauge violation can produce a first plateau, marked as {\bf{(2)}} in Fig.~\ref{fig:Schematics}. Since the norm of $\Omega_1^\text{nonres}(t)$ remains bounded, its physics can be well captured in TDPT (see Appendix~\ref{sec:TDPT}). 
Whether this plateau is realized depends on whether the concrete microscopic parameters realize a separation of scales from the following terms that go beyond TDPT (see Secs.~\ref{sec:Z2LGT} and~\ref{sec:U1LGT} for examples). 

To describe the physics beyond the first feature at $\tau_\mathrm{po}$, we need to treat separately the \emph{resonant} contributions where $E_{\alpha,q}=E_{\beta,l}$. For these, we get from Eq.~\eqref{eq:Omega1} 
\begin{align}
\Omega_1^\text{res}(t)=&\,-\mathrm{i}t\lambda\sum_{\alpha,\beta}\sum_{q,l}\bra{\alpha,q}H_1\ket{\beta,l}\ket{\alpha,q}\bra{\beta,l}.
\end{align}
In their contribution to $\tilde{U}(t)$, these terms act as an effective time-independent Hamiltonian $H_\mathrm{eff}^{(1)}=\mathrm{i}\Omega_1^\text{res}(t)/t$ with strength $\lambda$. The dynamics generated by $H_\mathrm{eff}^{(1)}$ induces further gauge violations [see {\bf{(3)}} in Fig.~\ref{fig:Schematics}] until, at times $t\gg \lambda^{-1}$, it settles into a steady state. This leads to the onset plateau at a timescale $\tau_\mathrm{o}=\lambda^{-1}$, see {\bf{(4)}} in Fig.~\ref{fig:Schematics}.   
Since the norm of $\Omega_1^\text{res}(t)$ increases as $\lambda t$, it is at the same scale that simple TDPT breaks down.


\begin{widetext}
\subsubsection{Second-order Magnus term and duration of first plateau}\label{sec:MEB}

As we show in this Section, the second-order terms in the Magnus expansion start playing a crucial role once the first-order terms have settled into a steady state. In the eigenbasis of $H_0$, Eq.~\eqref{eq:Omega2} can be rewritten as
\begin{align}\nonumber
\Omega_2(t)=&\,-\frac{\lambda^2}{2}\int_0^t \d t_1\int_0^{t_1}\d t_2\sum_{\alpha,\beta,\gamma}\sum_{q,l,p}\bra{\alpha,q}H_1\ket{\beta,l}\bra{\beta,l}H_1\ket{\gamma,p}\ket{\alpha,q}\bra{\gamma,p}\\\label{eq:Omega_2_LehmannBasis}
&\times\bigg\{\mathrm{e}^{\mathrm{i}(E_{\alpha,q}-E_{\beta,l})t_1}\mathrm{e}^{\mathrm{i}(E_{\beta,l}-E_{\gamma,p})t_2}-\mathrm{e}^{\mathrm{i}(E_{\alpha,q}-E_{\beta,l})t_2}\mathrm{e}^{\mathrm{i}(E_{\beta,l}-E_{\gamma,p})t_1}\bigg\}\,.
\end{align}
Terms where all energy gaps are mutually \emph{nonresonant} yield the contribution 
%
\begin{align}\nonumber
\Omega_2^\text{nonres}(t)=&\,\frac{\lambda^2}{2}\sum_{\alpha,\beta,\gamma}\sum_{q,l,p}\frac{\big(E_{\alpha,q}-2E_{\beta,l}+E_{\gamma,p}\big)\big[\mathrm{e}^{\mathrm{i}(E_{\alpha,q}-E_{\gamma,p})t}-1\big]+\big(E_{\alpha,q}-E_{\gamma,p}\big)\big[\mathrm{e}^{\mathrm{i}(E_{\beta,l}-E_{\gamma,p})t}-\mathrm{e}^{\mathrm{i}(E_{\alpha,q}-E_{\beta,l})t}\big]}{(E_{\alpha,q}-E_{\beta,l})(E_{\beta,l}-E_{\gamma,p})(E_{\alpha,q}-E_{\gamma,p})}\\
&\times\bra{\alpha,q}H_1\ket{\beta,l}\bra{\beta,l}H_1\ket{\gamma,p}\ket{\alpha,q}\bra{\gamma,p}.
\end{align}
As is the case for $\Omega_1^\text{nonres}(t)$, the oscillating terms average away for $t\gg 1/J$, leading to a constant contribution that adds to the pre-onset plateau [{\bf{(2)}} in Fig.~\ref{fig:Schematics}].

There are, however, two \emph{resonant} contributions that generate further time evolution through $\Omega_2^\text{res}(t)=\Omega_2^\text{res,A}(t)+\Omega_2^\text{res,B}(t)$. 
First, the single resonances $E_{\alpha,q}=E_{\beta,l}$ or $E_{\beta,l}=E_{\gamma,p}$ give
\begin{align}\nonumber
\Omega_2^\text{res,A}(t)
=&\,-\frac{\lambda^2}{2}\sum_{\alpha,\beta,\gamma}\sum_{q,l,p}\frac{2+\mathrm{i}(E_{\beta,l}-E_{\gamma,p})t-[2-\mathrm{i}(E_{\beta,l}-E_{\gamma,p})t]\mathrm{e}^{\mathrm{i}(E_{\beta,l}-E_{\gamma,p})t}}{(E_{\beta,l}-E_{\gamma,p})^2}\\
&\times\bra{\alpha,q}H_1\ket{\beta,l}\bra{\beta,l}H_1\ket{\gamma,p}\ket{\alpha,q}\bra{\gamma,p}-\mathrm{H.c.},
\end{align}
and, second, the single resonance $E_{\alpha,q}=E_{\gamma,p}$ (with $E_{\alpha,q}\neq E_{\beta,l}$ and $E_{\beta,l}\neq E_{\gamma,p}$) leads to the term
\begin{align}
\Omega_2^\text{res,B}(t)
=&\,\mathrm{i}\lambda^2\sum_{\alpha,\beta,\gamma}\sum_{q,l,p}\bra{\alpha,q}H_1\ket{\beta,l}\bra{\beta,l}H_1\ket{\gamma,p}\ket{\alpha,q}\bra{\gamma,p}\bigg\{\frac{\sin[(E_{\alpha,q}-E_{\beta,l})t]}{(E_{\alpha,q}-E_{\beta,l})^2}-\frac{t}{E_{\alpha,q}-E_{\beta,l}}\bigg\}.
\end{align}

We can again invoke our observations about the energy differences from the previous Section. At times $t \gg 1/J$, fast oscillating terms will have averaged out. Furthermore, we numerically find that the effect of $\Omega_2^\text{res,A}(t)$ is insignificant, leading us to the long-time limit
\begin{align}
\label{eq:Omega_2_eff}
\lim_{t\to\infty}\Omega_2^\text{res}(t)=-\mathrm{i} t \lambda^2  \sum_{\alpha,\beta}\sum_{q,l}\frac{1}{E_{\alpha,q}-E_{\beta,l}}\bra{\alpha,q}H_1\ket{\beta,l}\bra{\beta,l}H_1\ket{\gamma,p}\ket{\alpha,q}\bra{\gamma,p}\,.
\end{align}
Again, $\lim_{t\to\infty}\Omega_2^\text{res}(t)$ assumes the role of a time-independent effective Hamiltonian $H_\mathrm{eff}^{(2)}=\lim_{t\to\infty}\mathrm{i}\Omega_2^\text{res}(t)/t\propto\lambda^2$. 
In our numerics, we find improved convergence---in particular for the extended Bose--Hubbard model (eBHM); see Sec.~\ref{sec:GlobSym}---if we absorb the diagonal terms that do not change the gauge supersector (equivalently, particle-number sector in the case of the eBHM) into $H_0$. In principle, we should then recompute the Magnus expansion in a self-consistent manner. We find, however, that we obtain already excellent agreement with ED without this self-consistent adjustment.

The effective Hamiltonian $H_\mathrm{eff}^{(2)}$ becomes relevant only at times $t\propto \lambda^{-2}$, much after $\Omega_1(t)$ has averaged to a constant.
Thus, we have a separation of scales, and a time window opens between timescales $\tau_\mathrm{o}=\lambda^{-1}$ and $\tau_\mathrm{i}= \lambda^{-2}$ during which the first-order Magnus term in $\tilde{U}(t)$ does not invoke any dynamics anymore while the second-order Magnus term is not yet relevant---the gauge violation halts and reaches the stable onset plateau [see {\bf{(4)}} in Fig.~\ref{fig:Schematics}]. 
At a timescale $\lambda^{-2}$, $\Omega_2^\mathrm{res}(t)$ as per Eq.~\eqref{eq:Omega_2_eff} becomes important and admixes further gauge invariance-breaking states. Once $t\gg\lambda^{-2}$, also this dynamics halts and a second plateau is reached [see {\bf{(5)}} in Fig.~\ref{fig:Schematics}].

\end{widetext}

\subsubsection{Higher-order Magnus terms, further plateaus, and note about the thermodynamic limit}\label{sec:MEC}\label{sec:TL}
	
These arguments can be repeated to lead to plateaus for all orders of $\lambda^{s}$, $1\leq s\leq L/2$. The final plateau is reached at a timescale $\tau_\mathrm{f}=\lambda^{-L/2}$, at which point an equal probability of both gauge eigenvalues has been reached locally, corresponding to full gauge violation.

The above analytic observations are based on several assumptions, in particular the convergence of the Magnus expansion and the absence of small relevant gaps. 
Indeed, Tables \ref{tab:energyGapsZ2L4} and \ref{tab:energyGapsZ2L6} show that nonzero energy gaps $\Delta E<\lambda_\text{thresh}=0.01$ are quite rare, even for $L=6$ matter sites---i.e., when the system is $12$ sites in total. 
The predominance of gaps larger than $\lambda_\text{thresh}$ is also reflected in our numerical data for the time evolution. On the one hand, the pre-onset plateau typically occurs at times $t\approx 1/\lambda_\text{thresh}$. On the other hand, the 
value of $\lambda_\text{thresh}=0.01$ is the gauge invariance-breaking strength above which we observe in the ED results that some plateaus in the prethermal staircase begin to get compromised, since the separation of frequencies dominant in $H_0$ from those generated by the gauge breaking is no longer warranted. 

In addition, Tables \ref{tab:energyGapsZ2L4} and \ref{tab:energyGapsZ2L6}, which go up to second-order processes in $H_1$, show the presence of many exact resonances that lead to the resonant Magnus-expansion terms responsible for the successive destruction of the prethermal plateaus. The energy gap $E_{02}$ is due to a first-order process in $H_1$ where the initial state is being taken out of the gauge supersector $\{\alpha_{\{0\}}\}=\alpha_{\{0\}}$ into the gauge supersector $\{\alpha_{\{2\}}\}$. 
The gap $E_{020}$ describes a second-order process that first takes a gauge-invariant state in the supersector $\{\alpha_{\{0\}}\}=\alpha_{\{0\}}$ into the gauge supersector $\{\alpha_{\{2\}}\}$ where two local constraints are broken, and then back to the original supersector $\{\alpha_{\{0\}}\}$ where Gauss's law is restored to its initial value on all matter sites (see Glossary in Appendix~\ref{sec:glossary}). The energy gap $E_{024}$ is another second-order process, which first takes the initial state out of the gauge-invariant supersector $\{\alpha_{\{0\}}\}=\alpha_{\{0\}}$ into the supersector $\{\alpha_{\{2\}}\}$, and finally into the supersector $\{\alpha_{\{4\}}\}$ where four local constraints are broken with respect to the initial state.
(It is worth noting here that the spectrum of the $\mathrm{Z}_2$ LGT is symmetric between supersectors $\{\alpha_{\{s\}}\}$ and $\{\alpha_{\{L-s\}}\}$. 
Due to this symmetry, for $L=6$ matter sites $E_{64}$ provides as many zero-energy gaps as $E_{02}$, $E_{022}$ as many as $E_{644}$, $E_{024}$ as many as $E_{642}$, and so on. Similarly, in the case of $L=4$ matter sites, $E_{42}$ has as many zero-energy gaps as $E_{02}$, $E_{022}$ as many as $E_{422}$, $E_{024}$ as many as $E_{420}$, and so on.)

In a gauge theory, we naturally obtain a large number of exact degeneracies in these processes. Configurations where Gauss's law with respect to the initial value is violated at a set of sites $\{j_1,\ldots,j_m\}$ are exactly degenerate to a state where the violations are all shifted by a distance $\delta$, $\{j_1+\delta,\ldots,j_m+\delta\}$. The number of such distinct degenerate states increases in system size. Since $[H_0,G_j]=0$, there is no process in $H_0$ that could couple these different configurations and lift the degeneracy.
The abundance of such zero-energy gaps and the scarcity of nonzero energy gaps below $\lambda_\text{thresh}$ leads to the separation of timescales that makes staircase prethermalization possible.
In Sec.~\ref{sec:GlobSym}, we discuss why the phenomenon of staircase prethermalization has not been observed in a similar scenario where a global symmetry is slightly broken---the reason is the different behavior in the energy gaps. 

\begin{figure}[htp]
	\centering
	\hspace{-.25 cm}
	\includegraphics[width=.49\textwidth]{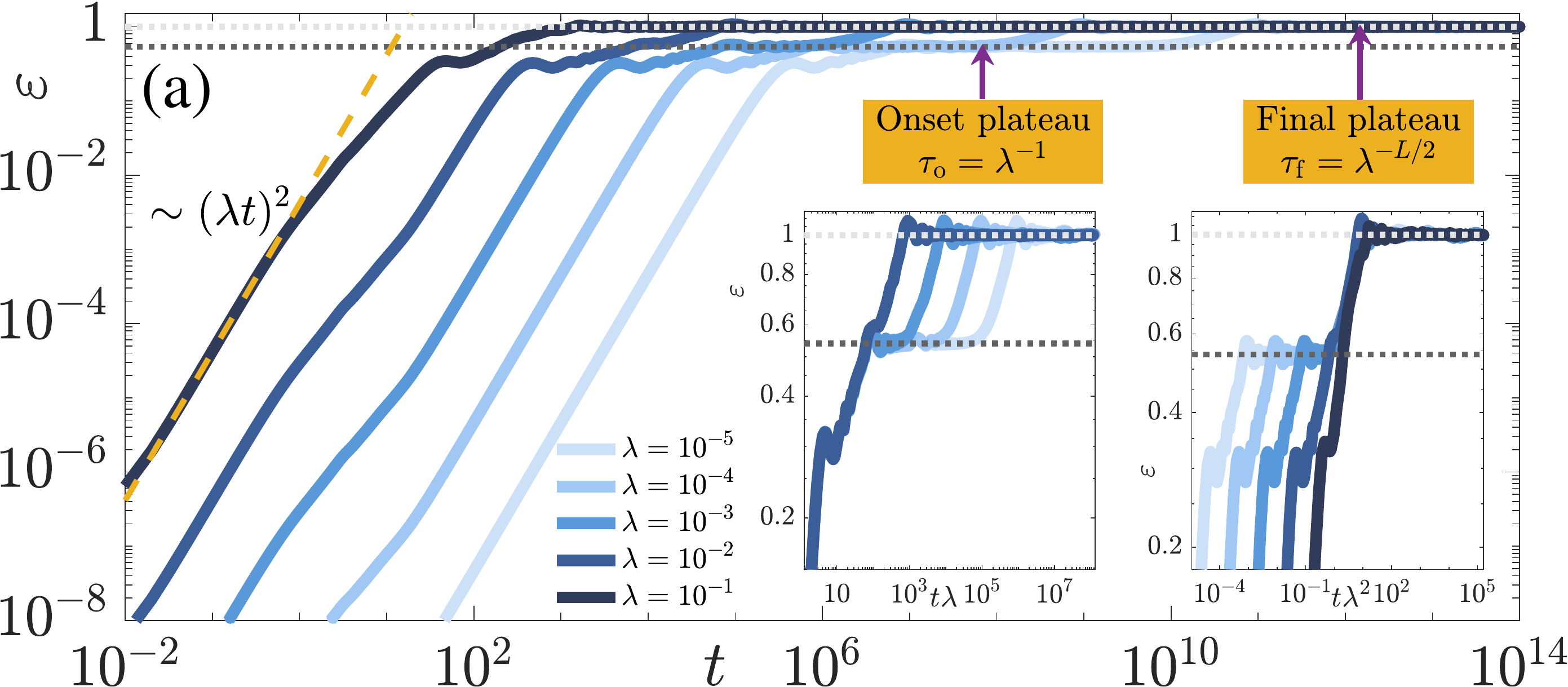}\quad\\
	\hspace{-.25 cm}
	\includegraphics[width=.49\textwidth]{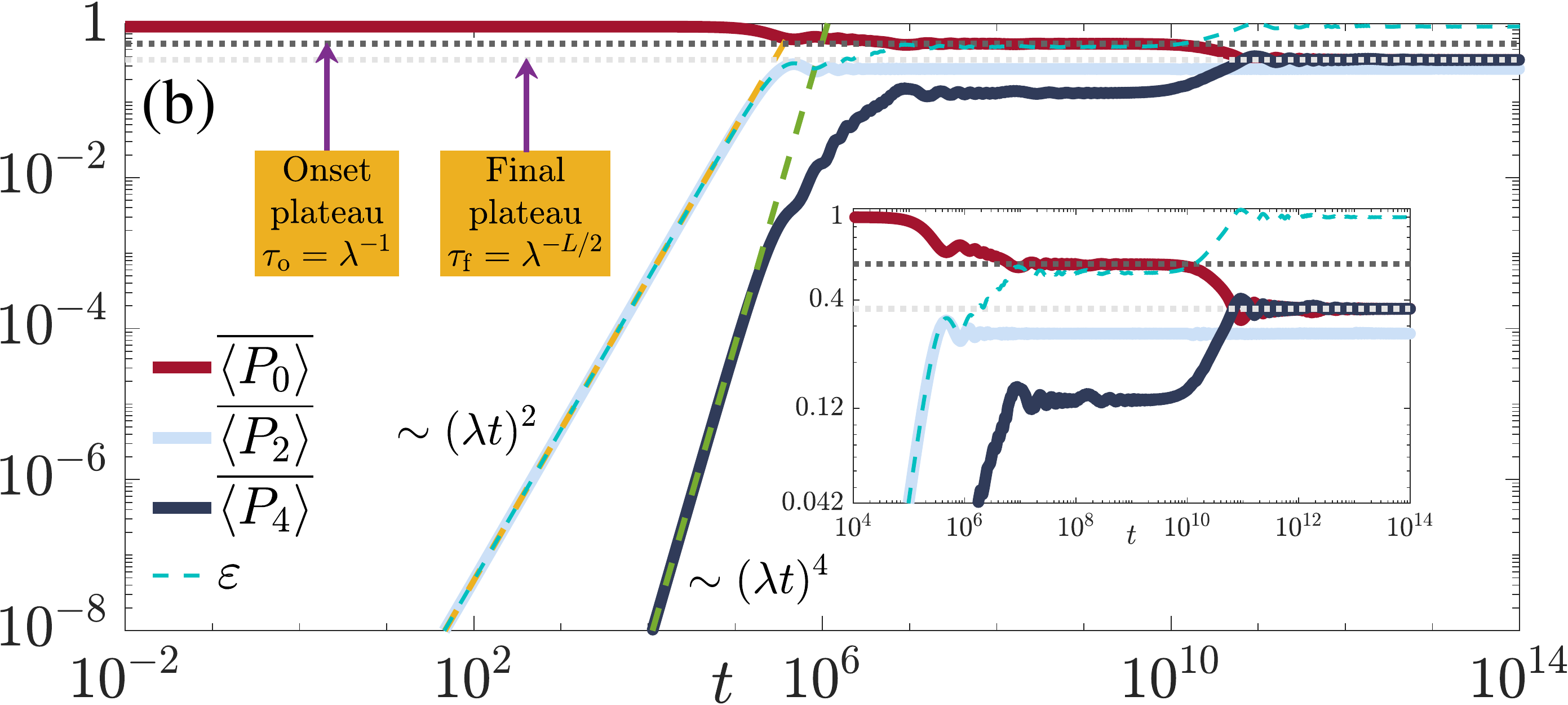}\quad
	\hspace{-.25 cm}
	\caption{(Color online). Quench dynamics of the $\mathrm{Z}_2$ gauge theory with $L=4$ matter sites, initial product state with staggered boson occupation and electric-link orientation lying in the sector $g_j=0$, $\forall j$, and with the same microscopic parameters as in the joint submission~\cite{Halimeh2020b} (also see text). Dynamics of the spatiotemporal averages of (a) the gauge-invariance violation of Eq.~\eqref{eq:error} for various values of breaking strength $\lambda$ (compare Fig.~\ref{fig:Schematics}), and (b) the expectation values of the projectors defined in Eq.~\eqref{eq:P} for $\lambda=10^{-5}$ (the behavior is qualitatively the same for other values of $\lambda$). The violation shows two ($=L/2$) plateaus. The first prethermal plateau is dominated by intermediate gauge invariance-violating processes quantified in $\langle P_2\rangle$. The final plateau is dominated by processes due to $\langle P_4\rangle$, which settles to the same value as $\langle P_0\rangle$ due to the spectral symmetry between gauge-invariant supersectors $\{\alpha_{\{s\}}\}$ and $\{\alpha_{\{L-s\}}\}$ in the $\mathrm{Z}_2$ LGT. Note that the final plateau does not appear to be thermal despite ushering in maximal gauge-invariance violation, i.e., it is equally likely to have locally either generator eigenvalue $g_j=0$ or $2$ (see discussion in Sec.~\ref{sec:MBL}). Worth noting here is that $\langle P_s\rangle=0$ identically for all odd $s$ since terms in $H_1$ can only break an even number of local constraints.
	}
	\label{fig:FigL4} 
\end{figure}

While we cannot prove that nonzero energy gaps $\Delta E<\lambda_\text{thresh}$ remain rare for general many-body gauge theories, the agreement between calculations based on the leading terms of the Magnus expansion and full exact numerics is striking (see Fig.~\ref{fig:Schematics}; also compare Fig.~\ref{fig:FigL4} to Fig.~\ref{fig:FigL4_Magnus}). One open question is in how far the observed phenomena extend to the thermodynamic limit---on the one hand, in the thermodynamic limit energy spectra become dense, so we cannot exclude the aforementioned separation of energy scales to break down; on the other hand, the number of exact degeneracies increases with system size, which may provide a balancing mechanism that retains staircase prethermalization even in the thermodynamic limit. 

Moreover, in Ref.~\onlinecite{Gorin2006} an unusually slow fidelity decay as a result of slightly breaking a \textit{global anti-unitary symmetry} has been analytically discussed and numerically corroborated in collective spin systems of as many as $400$ spin-$1/2$ particles. In that work, it has been argued that the effect is due to correlations between different subspectra of $H_0$, which are absent in a system with a global unitary symmetry (such as the particle number used in Sec.~\ref{sec:GlobSym}). Since it is precisely the resonances between different gauge-invariant supersectors that drive staircase prethermalization, the results of Ref.~\onlinecite{Gorin2006} give hope that our findings can persist to large systems, notwithstanding the conceptual and phenomenological orthogonality of their study to ours. 

It is worth adding that as the difficulty of computing unbiased long-time dynamics of quantum many-body systems restricts our numerical studies to rather small system sizes, it becomes an exciting prospect to use quantum-simulator experiments in order to investigate the persistence of the prethermal plateaus as system size increases.

\section{Staircase prethermalization in the $\mathrm{Z}_2$ gauge theory}\label{sec:Z2LGT}

It is illustrative to compare the analytic arguments derived in the preceding Section to exact numerical data, which is the purpose of this and the following Sec.~\ref{sec:U1LGT}. 	
In the present Section, we consider a $\mathrm{Z}_2$ gauge theory given by the Hamiltonian\cite{Zohar2017,Barbiero2019,Borla2019,Schweizer2019}
\begin{align}\label{eq:H0}
H_0=&\,\sum_{j=1}^L\big[J_a\big(a^\dagger_j\tau^z_{j,j+1}a_{j+1}+\mathrm{H.c.}\big)-J_f\tau^x_{j,j+1}\big],
\end{align}
with $L$ matter sites and periodic boundary conditions, i.e., also with $L$ links, each connecting two adjacent matter sites. The matter field on site $j$ is represented by a hard-core boson with creation and annihilation operators $a_j^\dagger$ and $a_j$, respectively, which satisfy the canonical commutation relations $[a_j,a_l]=0$ and $[a_j,a_l^\dagger]=\delta_{j,l}(1-2a_j^\dagger a_j)$. The gauge (electric) field linking matters sites $j$ and $j+1$ is represented by the Pauli matrix $\tau^{z(x)}_{j,j+1}$. We define the local symmetry generators of the $\mathrm{Z}_2$ gauge group as
\begin{align}\label{eq:Gauss}
G_j=1-(-1)^j\tau^x_{j-1,j}\mathrm{e}^{\mathrm{i}\pi a_j^\dagger a_j}\tau^x_{j,j+1},
\end{align}
where each $G_j$, living at matter site $j$, takes on two eigenvalues, $g_j=0,2$. The Hamiltonian $H_0$ is gauge-invariant, i.e., $[H_0,G_j]=0$, $\forall j$, and as such $G_j$ are local conserved quantities at each matter site $j$. We set throughout this paper $J_a=1$ and $J_f=0.54$---values inspired from Ref.~\onlinecite{Schweizer2019}---though we have checked that other generic values of these parameters lead to the same qualitative picture.

We initialize the system with a single boson on every even matter site, while the odd matter sites are empty. The electric fields are initialized in a similar staggered fashion such that the initial state lies in the sector $g_j=0$, $\forall j$. 
We refer to all deviations from the initial configuration of $g_j$ as `gauge violations' or `violations of Gauss's law with respect to the initial state' (see Glossary in Appendix~\ref{sec:glossary}).
At time $t=0$, this state is subjected to a sudden quench by the Hamiltonian $H=H_0+\lambda H_1$, where the gauge invariance-breaking term
\begin{align}\nonumber
H_1=\,\sum_{j=1}^L\Big[&\big(c_1a_j^\dagger\tau^+_{j,j+1} a_{j+1}+c_2a_j^\dagger \tau^-_{j,j+1} a_{j+1}+\mathrm{H.c.}\big)\\\label{eq:H1}
&\,+a_j^\dagger a_j\big(c_3\tau^z_{j,j+1}-c_4\tau^z_{j-1,j}\big)\Big],
\end{align}
models inherent errors in a recent ultracold-atom experiment.\cite{Schweizer2019} The overall strength is given by $\lambda>0$, and the coefficients $c_n$ governing the relative weight of the various terms depend on experimental parameters (see Ref.~\onlinecite{Schweizer2019} and Sec.~\ref{sec:Z2LGT_B} for detailed information). 
It is worth noting that $H_1$ does not break any global symmetries in the system that are found in $H_0$. For example, both $H_0$ and $H_1$ conserve particle number. For a discussion about systems with global-symmetry breaking, see Sec.~\ref{sec:GlobSym}.

Using ED,\cite{Weinberg2017,Weinberg2019,Johansson2012,Johansson2013} we compute the time evolution under $H_0+H_1$ of the gauge-invariance violation, whose spatiotemporal average is given by
\begin{align}\nonumber
\varepsilon(t)=&\,\frac{1}{L}\sum_{j=1}^L\,\Big[\bra{\psi_0} G_j\ket{\psi_0}\\\label{eq:error}
&+\frac{\mathrm{e}^{\mathrm{i}\frac{\pi}{2}\bra{\psi_0} G_j\ket{\psi_0}}}{t}\int_0^t\d \tau\,\bra{\psi(\tau)} G_j\ket{\psi(\tau)}\Big],
\end{align}
and which quantifies the deviation of the expectation value of the gauge generator from its initial value. Here, $\ket{\psi(\tau)}=\exp[-\mathrm{i}(H_0+\lambda H_1)\tau]\ket{\psi_0}$.
In addition, we compute the spatiotemporally averaged expectation values of the projectors onto the various gauge-invariant supersectors
\begin{align}\label{eq:P}
P_s=\sum_{\alpha_{\{s\}}}\sum_q|\alpha_{\{s\}},q\rangle\langle\alpha_{\{s\}},q|.
\end{align}
Here, $|\alpha_{\{s\}},q\rangle$ are eigenstates of $H_0$ living in the gauge-invariant supersector $\{\alpha_{\{s\}}\}$, which is the set of all gauge-invariant sectors $\alpha_{\{s\}}=(\alpha_1,\alpha_2,\ldots,\alpha_L)$ such that $\sum_j\alpha_j=2s$, indicating that Gauss's law is violated at $s$ matter sites with respect to an initial state where $g_j=0$, $\forall j$. 
In the main text, and in the joint submission,\cite{Halimeh2020b} we present temporal averages of these quantities. However, as we show in Appendix \ref{sec:Numerics}, the time-resolved as well as running maximal violations show the same qualitative behavior.  

In the following Sections, we provide in-depth comparisons of the influence of different system sizes (Sec.~\ref{sec:Z2LGT_A}), different microscopic parameters (Sec.~\ref{sec:Z2LGT_B}), different initial conditions (Secs.~\ref{sec:Z2LGT_C}-\ref{sec:Z2LGT_E}), and of relaxing the hard-core constraint on the matter fields (Sec.~\ref{sec:Z2LGT_F}). 
As these Sections show, the qualitative picture sketched in Fig.~\ref{fig:Schematics} is generic, although some details can depend on the specific realization (e.g., the pre-onset plateau is not always resolved).

\subsection{Effect of number of local constraints}\label{sec:Z2LGT_A}

\begin{figure}[htp]
	\centering
	\hspace{-.25 cm}
	\includegraphics[width=.49\textwidth]{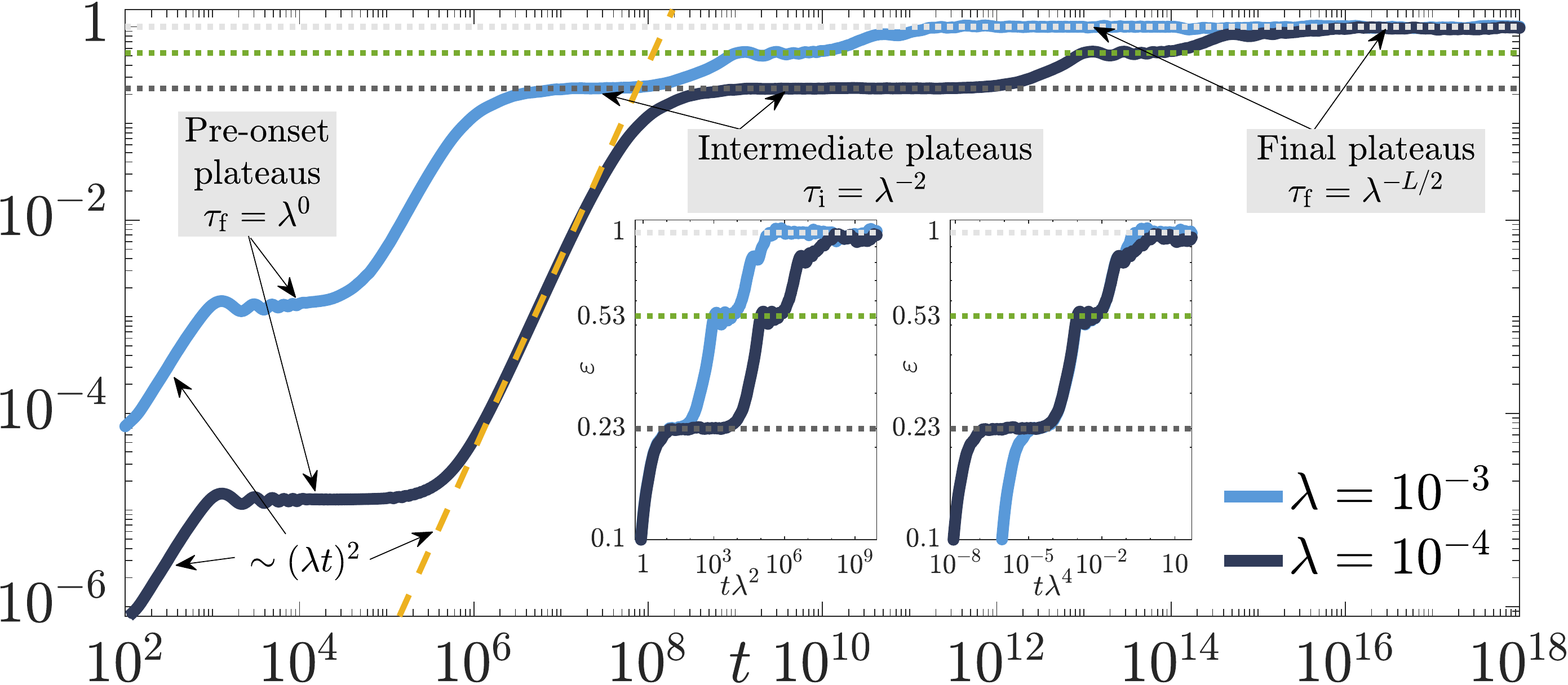}\quad
	\hspace{-.25 cm}
	\caption{(Color online). Gauge-violation dynamics as in Fig.~\ref{fig:FigL4}(a), but for $L=8$ matter sites. Four plateaus appear, with the pre-onset plateau exhibiting a timescale $\lambda^0$ (i.e., independent of $\lambda$), while the second (intermediate) plateau exhibits a timescale $\lambda^{-2}$, and the third and fourth (final) plateaus have the same timescale $\lambda^{-4}=\lambda^{-L/2}$ due to a separation of scales within $\Omega_{L/2}^\text{res}$. 
	} 
	\label{fig:FigL8} 
\end{figure}

We start by analyzing the effect of the number of local constraints on staircase prethermalization. Whereas in the joint submission\cite{Halimeh2020b} we have mostly focused on a $\mathrm{Z}_2$ LGT with $L=6$ matter sites, here we calculate in ED the time evolution of the gauge-invariance violation in Eq.~\eqref{eq:error} for $L=4$ and $L=8$ matter sites; see Figs.~\ref{fig:FigL4} and~\ref{fig:FigL8}, respectively. 

Following the early-time regime, systems of all considered sizes enter the prethermalization staircase. The gauge violation in Fig.~\ref{fig:FigL4}(a) displays a feature at $t\approx10/J_a$. It is related to the `pre-onset' plateau that comes about due to nonresonant terms in the Magnus expansion, as discussed in Sec.~\ref{sec:ME} and illustrated in Fig.~\ref{fig:Schematics}. For $L=8$, the pre-onset plateau becomes fully prominent (see Fig.~\ref{fig:FigL8}). 
For $L=4$ matter sites, two plateaus appear: the prethermal onset plateau at timescale $\lambda^{-1}$ and the final steady-state plateau at timescale $\lambda^{-2}=\lambda^{-L/2}$. 
For $L=8$, the pre-onset and intermediate plateaus at timescales $\lambda^0$ and $\lambda^{-2}$, respectively, are prominent. 
In contrast, no plateaus are found at the timescales $\lambda^{-1}$ and $\lambda^{-3}$, indicating the dependence of the precise structure of the prethermal staircase on microscopic conditions, as also seen, e.g., in Sec.~\ref{sec:Z2LGT_C}. 
Again, the gauge violation reaches the final steady state at the timescale $\lambda^{-4}=\lambda^{-L/2}$, exhibiting the same exponential-in-system-size delay we see for the cases of $L=4$ and $6$ matter sites. 
In addition, however, there is a penultimate plateau that has the same timescale $\lambda^{-4}=\lambda^{-L/2}$ as the final plateau, suggestive of a separation of scales within $\Omega_{L/2}^\mathrm{res}$. This can happen when there are several contributions to the same order of the Magnus expansion with strongly varying constant prefactors. 

To obtain further insights into the microscopic processes governing the prethermalization staircase, it is instructive to resolve $P_s$, the projectors onto different gauge supersectors; see Eq.~\eqref{eq:P}.
Similarly to the case for $L=6$ matter sites shown in Fig.~2 of the joint submission,\cite{Halimeh2020b} for the $\mathrm{Z}_2$ LGT with $L=4$ matter sites we see in Fig.~\ref{fig:FigL4}(b) that at short times $\langle P_s\rangle\sim(\lambda t)^s$ for even $s$, whereas $\langle P_s\rangle=0$ for odd $s$, which is not suprising since $H_1$ only breaks an even number of local constraints, and thus supersectors $\{\alpha_{\{s\}}\}$ with odd $s$ cannot be accessed. The scalings $\langle P_s\rangle\sim(\lambda t)^s$ for even $s$ are derived explicitly in TDPT in Appendix~\ref{sec:TDPT}. 

Each plateau in the gauge violation $\varepsilon$ is reflected in $P_0$ and at least one of the other supersectors, but not necessarily in all of them. For example, comparing Fig.~\ref{fig:FigL4}(b) of the present article and Fig.~2 of the joint submission,\cite{Halimeh2020b} one can see that the projector $P_2$ remains, up to the longest computed evolution times, at the plateau preceding the final one, both for $L=4$ matter sites (where it exhibits a single plateau at timescale $\lambda^{-1}$) as well as for $L=6$ matter sites (where it exhibits two plateaus at timescales $\lambda^{-1}$ and $\lambda^{-2}$). While at timescale $\lambda^{-L/2}$ the population in $P_2$ does get perturbed, it does not settle to a distinct new plateau; see inset of Fig.~\ref{fig:FigL4}(b) for $L=4$ matter sites, and the more strongly visible case in the inset of, e.g., Fig.~\ref{fig:FigOther}(b) for $L=6$ matter sites. In contrast, the associated final plateau is clearly reflected in $P_L$. Thus, we conclude that here the processes that couple gauge supersectors of increasing strength, such as $0\to 2\to 4$, play a much more dominant role than processes such as $0\to 2\to 2$. For $L=4$ matter sites, e.g., at the final timescale the former shift population from $P_0$ to $P_{L=4}$ while the latter only slightly perturb the plateau of $P_2$.

\subsection{Effect of microscopic experimental parameters}\label{sec:Z2LGT_B}
The error term and its coefficients $c_n$ in Eq.~\eqref{eq:H1} are inspired by the Floquet platform employed in Ref.~\onlinecite{Schweizer2019}. Pair tunneling with both bosonic species used in the two-component ultracold-atom experiment dominate the error terms in Eq.~\eqref{eq:H1}. The corresponding dimensionless driving parameter $\chi$ that determines $c_n$ can thus be used to tune the microscopic parameters of the error Hamiltonian.\cite{Schweizer2019} Specifically, in the double-well setup of Ref.~\onlinecite{Schweizer2019}, $\chi=A/\omega$ derives from the atomic species-independent driving $A\cos(\omega t+\phi)$ of the on-site potential, where $A$ is the modulation amplitude, $\omega$ the frequency, and $\phi$ a phase shift. We additionally ensure that $c_1+c_2+c_3+c_4=1$ in order to quantify the strength of the gauge invariance-breaking error by the parameter $\lambda$ only. The explicit expressions for $c_n$ can be found in Appendix~\ref{sec:Numerics}.

For the results of the joint submission,\cite{Halimeh2020b} we have used $\chi=1.84$, which was also used in the experiment of Ref.~\onlinecite{Schweizer2019}. In order to confirm that our conclusions are not dependent on the specific choice of the coefficients $c_n$, we have also checked other values of $\chi$. As an example, we illustrate in Fig.~\ref{fig:FigOther} the gauge violation and projectors onto the different gauge-invariant supersectors for the same quench as in Figs.~1 and~2 of the joint submission,\cite{Halimeh2020b} but for $\chi=1.3$. The qualitative picture is exactly the same. Indeed, not only does the prethermal staircase of the gauge violation in Fig.~\ref{fig:FigOther}(a) persist with three distinct timescales (onset $\lambda^{-1}$, intermediate $\lambda^{-2}$, and final $\lambda^{-3}=\lambda^{-L/2}$), but the expectation values of the projectors in Fig.~\ref{fig:FigOther}(b) also capture these timescales, with those of $P_2$ and $P_4$ exhibiting the onset and intermediate plateaus, and those of $P_0$ and $P_6$ exhibiting all three timescales. In fact, $P_2$ and $P_4$ are perturbed at the final timescale, but since they are already at their final steady-state values by then, the latter do not change; cf.~associated discussion in Sec.~\ref{sec:Z2LGT_A}. 
	
Furthermore, our choice of $J_a=1$ and $J_f=0.54$ is inspired by the experimental values used in Ref.~\onlinecite{Schweizer2019}. However, we have checked other generic values of $J_f/J_a$, and we have found that the qualitative picture remains the same. Interestingly, the parameter pairs $J_f/J_a=\pm h$, with $h>0$, lead to identical results.

\subsection{Effect of initial condition within the sector $g_j=0$, $\forall j$}\label{sec:Z2LGT_C}

\begin{figure}[htp]
	\centering
	\hspace{-.25 cm}
	\includegraphics[width=.48\textwidth]{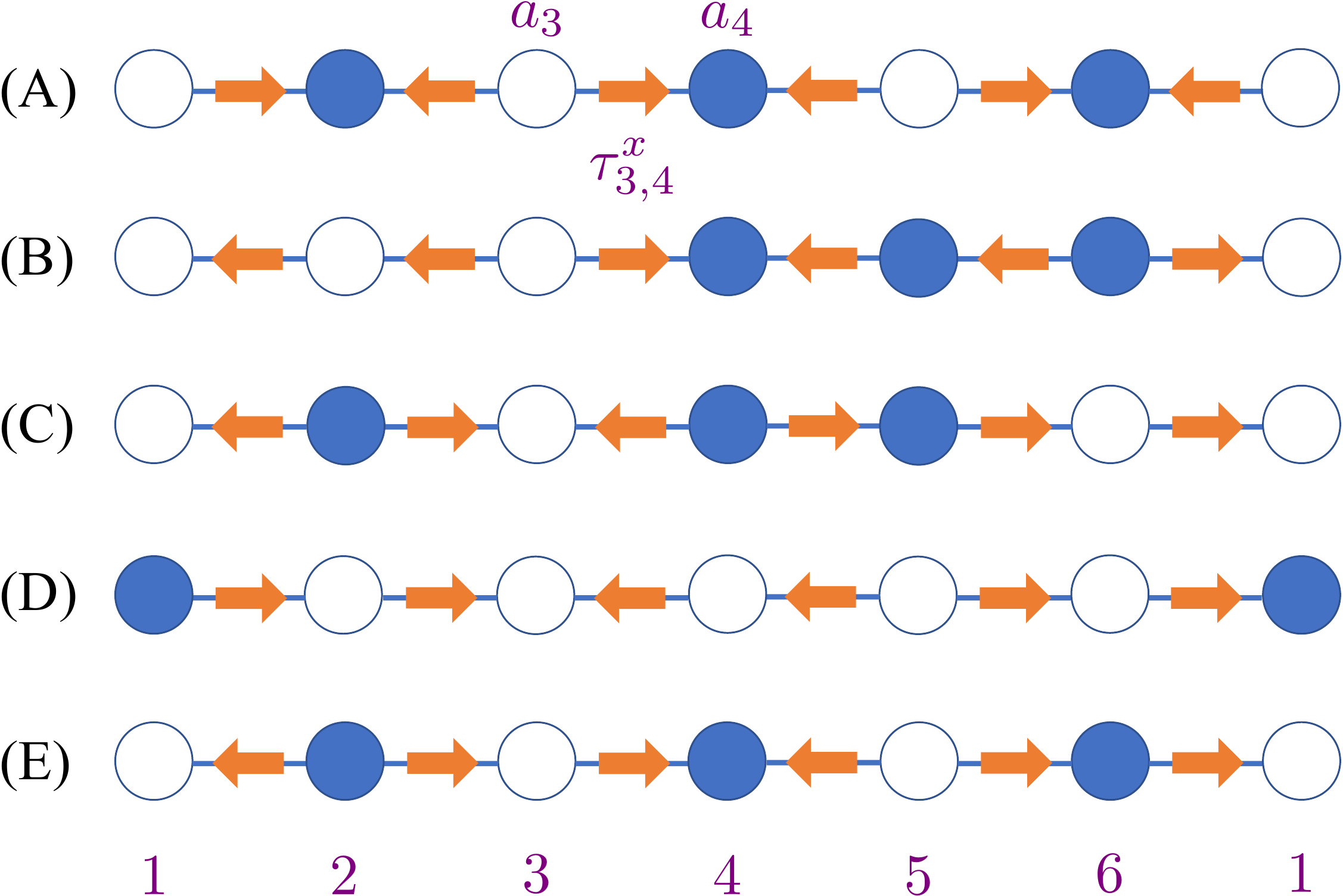}\quad
	\hspace{-.15 cm}
	\caption{(Color online). Initial states used for the quenches in this work with $L=6$ matter sites on a periodic lattice---there are $L=6$ links, each connecting two adjacent matter sites. Periodic boundary conditions are indicated by the site indexing at the bottom. In the joint submission of Ref.~\onlinecite{Halimeh2020b}, we have focused on the initial state (A). Circles with a solid fill denote a matter site occupied by a single hard-core boson. The orange arrows on links in between the matter sites $j$ and $j+1$ represent eigenstates of the Pauli operator $\tau^x_{j,j+1}$, which represents the electric field, with eigenvalues $\pm1$ when pointing right or left, respectively. Initial states (A), (B), (C), and (D) are in the gauge-invariant sector $G_j\ket{\psi}=0$, $\forall j$, while the initial state (E) lives in the sector $G_j\ket{\psi}=0$, for $j=1,2,4,5$ and $G_j\ket{\psi}=2\ket{\psi}$ for $j=3,6$; cf.~Eq.~\eqref{eq:Gauss}. All initial states are at half filling except for (D), which is at sixth filling.
	} 
	\label{fig:FigStates} 
\end{figure}

\begin{figure}[htp]
	\centering
	\hspace{-.25 cm}
	\includegraphics[width=.49\textwidth]{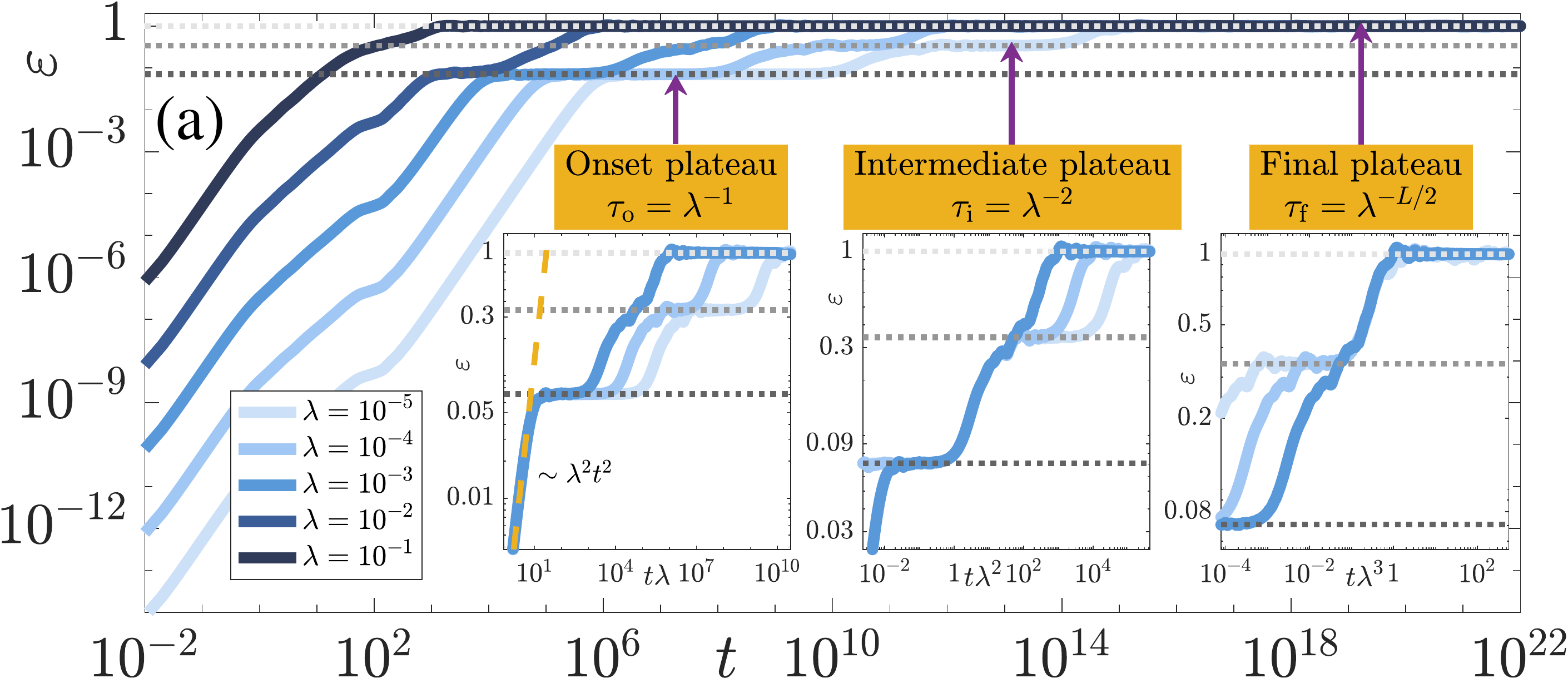}\quad\\
	\hspace{-.25 cm}
	\includegraphics[width=.49\textwidth]{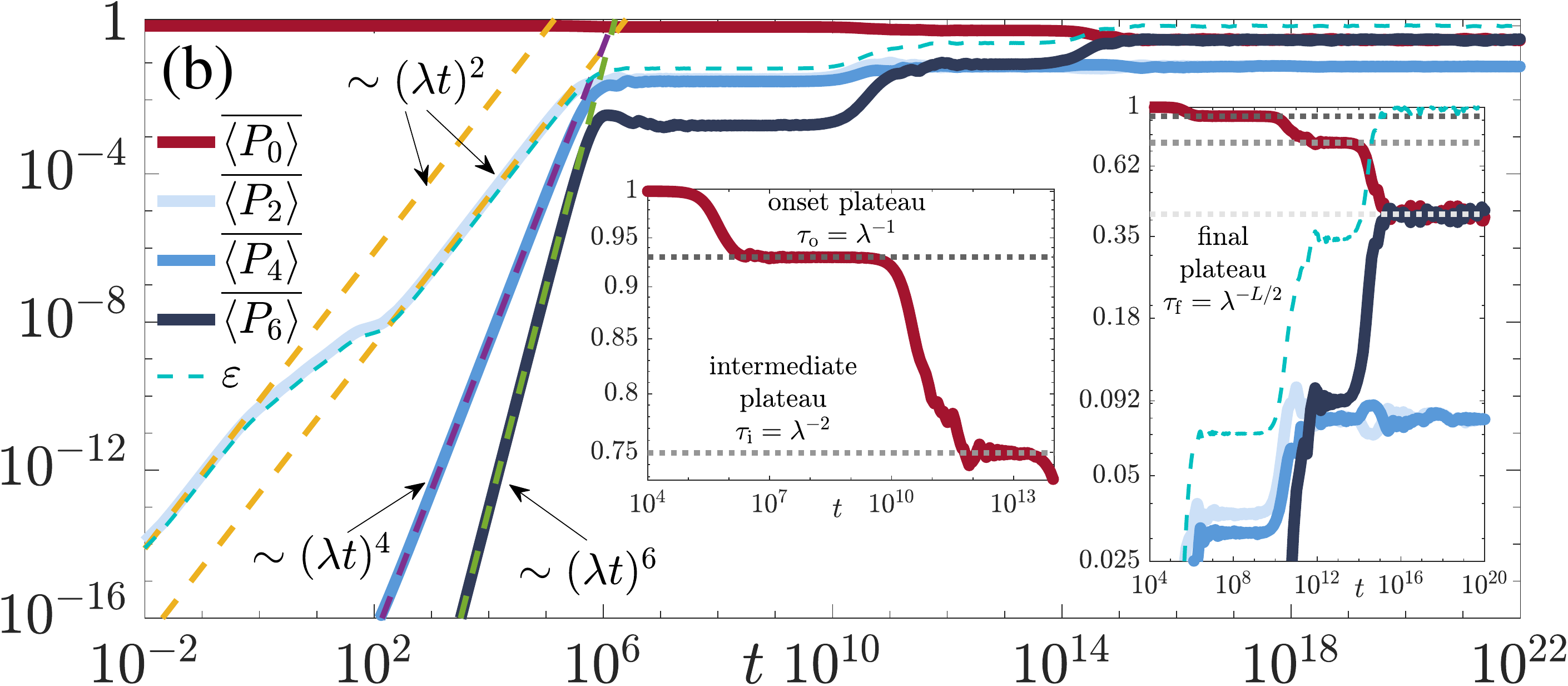}\quad
	\hspace{-.25 cm}
	\caption{(Color online). Same as Fig.~\ref{fig:FigL4}, with initial state (A) of Fig.~\ref{fig:FigStates} but for $L=6$ matter sites and different coefficients $c_n$ in the error Hamiltonian $H_1$ given in Eq.~\eqref{eq:H1} ($\chi=1.3$ instead of $1.84$). (a) The gauge-invariance violation shows plateaus at three distinct timescales $\tau_\text{o}=\lambda^{-1}$, $\tau_\text{i}=\lambda^{-2}$, and $\tau_\text{f}=\lambda^{-3}=\lambda^{-L/2}$. (b) The projectors exhibit the same plateau timescales as the violation. For short times, they scale perturbatively as $\langle P_s\rangle\sim(\lambda t)^s$ (for even $s$, whereas for odd $s$ they are identically zero).
	} 
	\label{fig:FigOther} 
\end{figure}

\begin{figure}[htp]
	\centering
	\hspace{-.25 cm}
	\includegraphics[width=.49\textwidth]{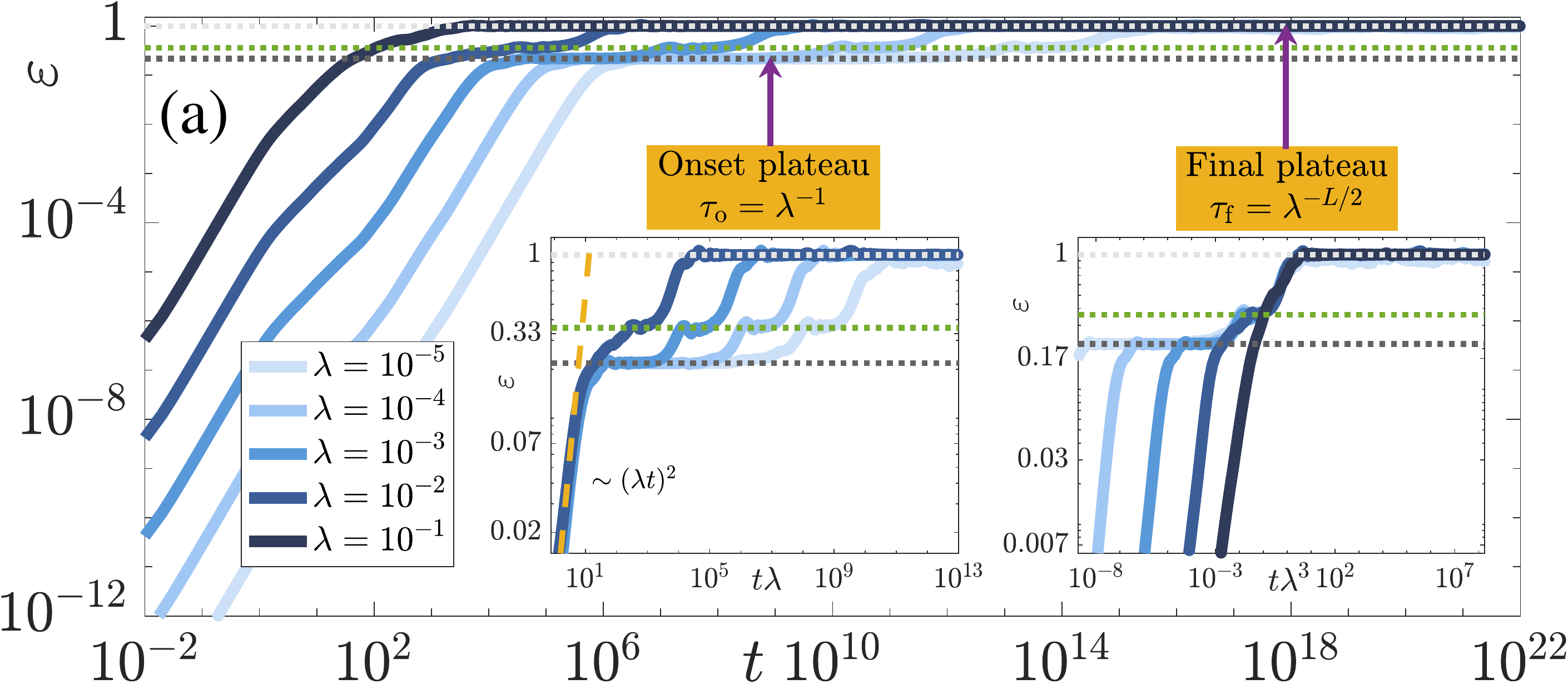}\quad\\
	\hspace{-.25 cm}
	\includegraphics[width=.49\textwidth]{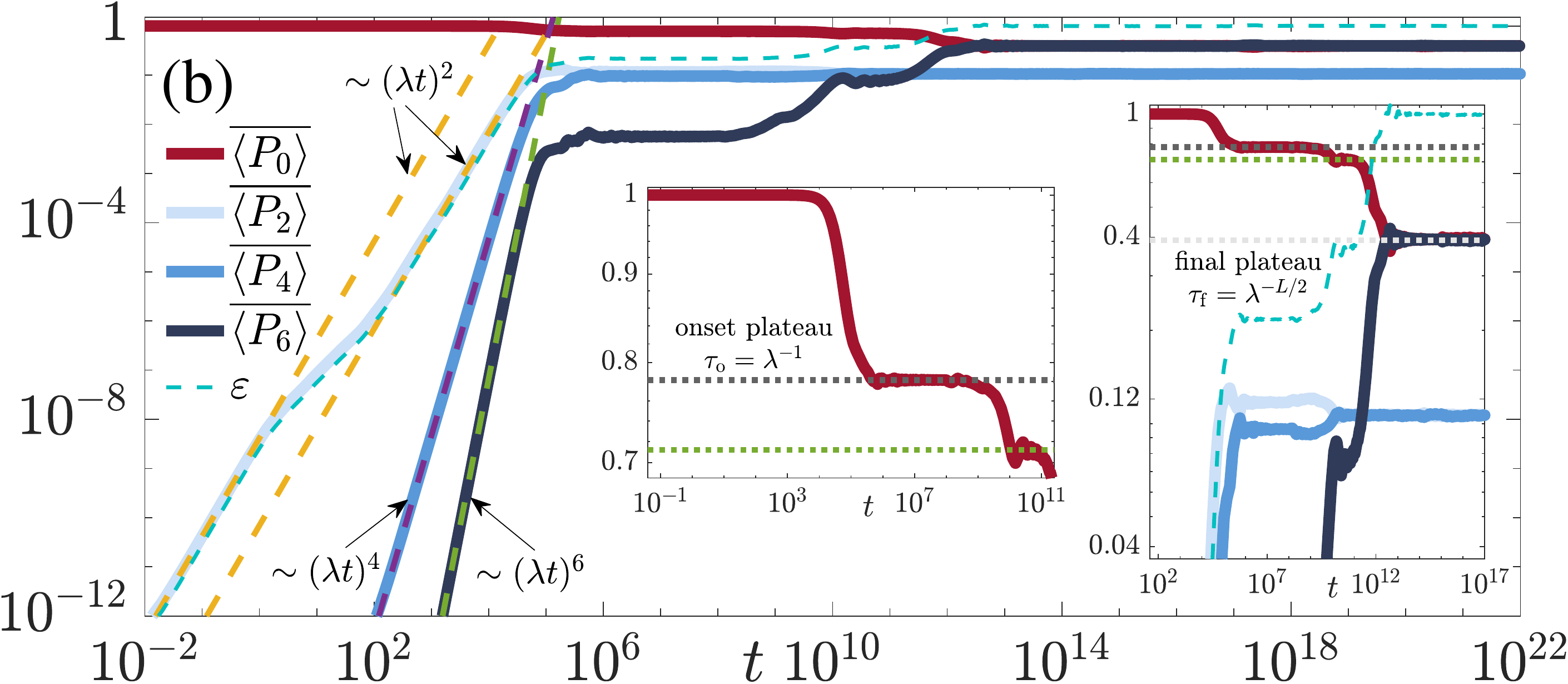}\quad
	\hspace{-.25 cm}
	\caption{(Color online). Same as Fig.~\ref{fig:FigOther}, but for the initial state (B). Even though three plateaus clearly emerge, only two timescales appear in this case, $\tau_\mathrm{o}=\lambda^{-1}$ and $\tau_\mathrm{f}=\lambda^{-3}=\lambda^{-L/2}$. Projectors in (b) computed for $\lambda=10^{-4}$.
	} 
	\label{fig:FigInitStateII} 
\end{figure}

\begin{figure}[htp]
	\centering
	\hspace{-.25 cm}
	\includegraphics[width=.49\textwidth]{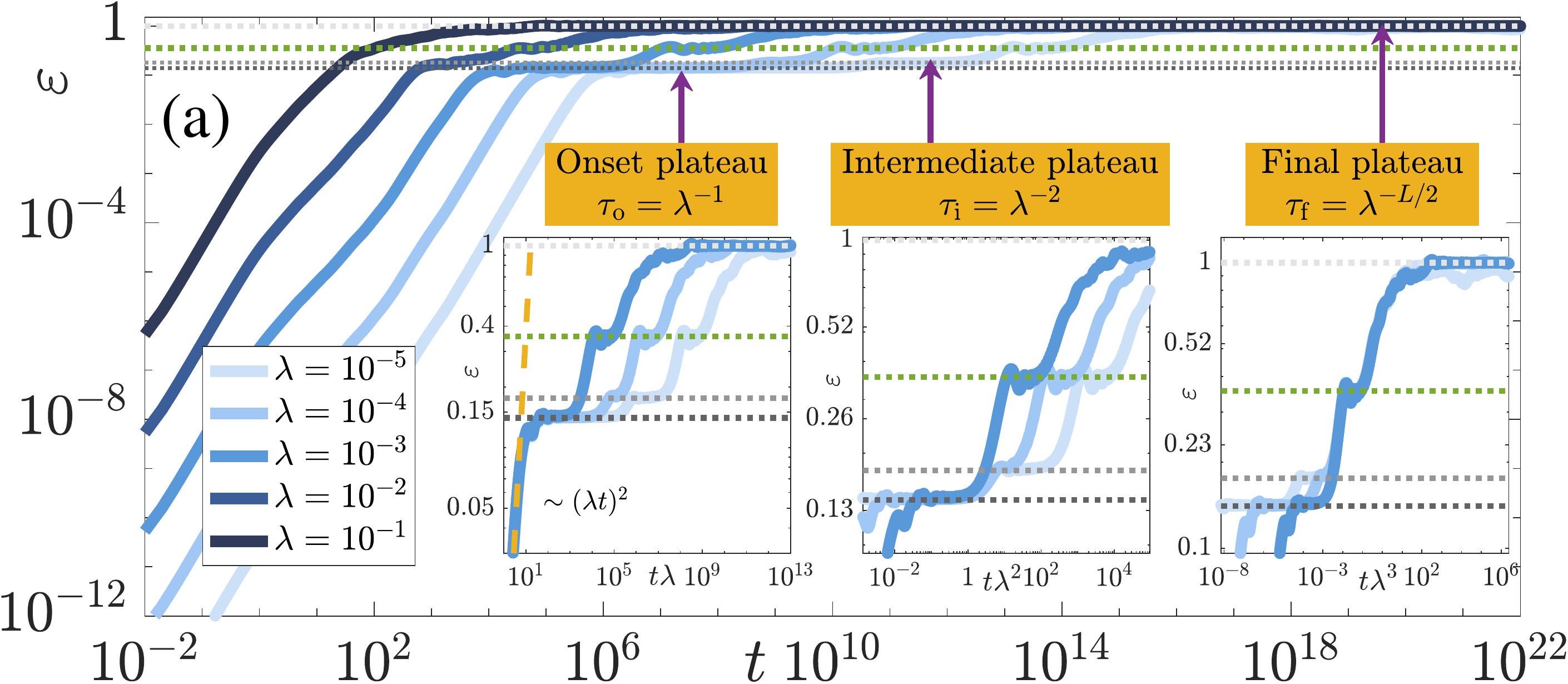}\quad\\
	\hspace{-.25 cm}
	\includegraphics[width=.49\textwidth]{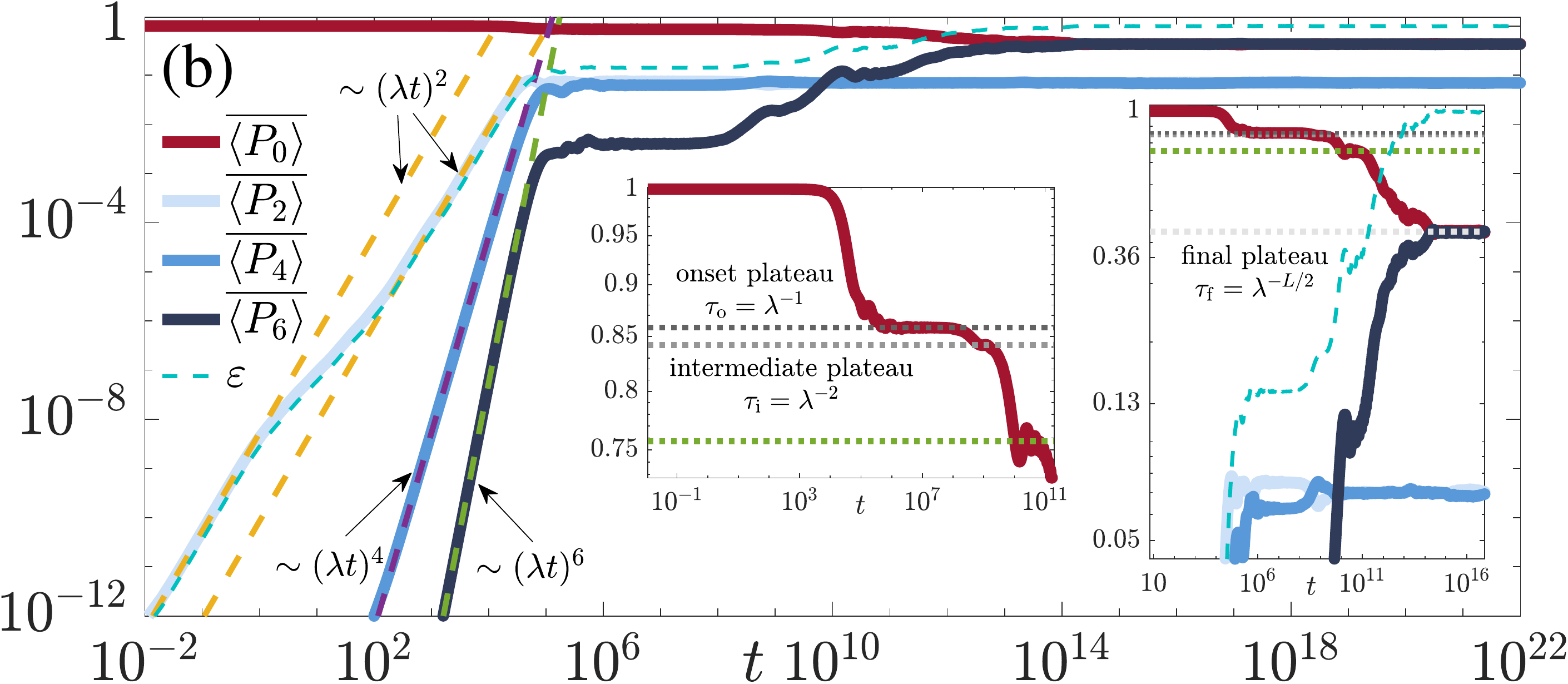}\quad
	\hspace{-.25 cm}
	\caption{(Color online). Same as Fig.~\ref{fig:FigInitStateII}, but for the initial state (C). Here, four plateaus emerge, and the same three timescales appear as in Fig.~\ref{fig:FigOther}.
	} 
	\label{fig:FigInitStateIII} 
\end{figure}

In this and the following two sections, we analyse the dependence of staircase prethermalization on different initial conditions; cf.~Fig.~\ref{fig:FigStates}. We start in this Section with states from the sector $g_j=0$, $\forall j$. 
In the joint submission,\cite{Halimeh2020b} our initial state is always set to the configuration (A) of Fig.~\ref{fig:FigStates}. 
When starting the quench from initial state (B) of Fig.~\ref{fig:FigStates}, again three plateaus emerge in the gauge violation, as shown Fig.~\ref{fig:FigInitStateII}(a). 
In this case, the onset and final plateaus clearly appear at the usual timescales $\tau_\text{o}=\lambda^{-1}$ and $\tau_\text{f}=\lambda^{-3}=\lambda^{-L/2}$, respectively, but the intermediate plateau (green dotted line) appears at timescale $\lambda^{-3}$ instead of $\tau_\text{i}=\lambda^{-2}$. 
Note that here the initial plateau appears at a larger value than that in Fig.~1 of Ref.~\onlinecite{Halimeh2020b}, and this may suppress the true intermediate plateau that would exhibit the timescale $\lambda^{-2}$. Lending credence to this explanation is the gauge violation for initial state (C) shown in Fig.~\ref{fig:FigInitStateIII}(a), where the initial plateau is at a value between those for initial states (A) and (B). In this case, an intermediate plateau appears at the intermediate timescale $\tau_\text{i}=\lambda^{-2}$, while another intermediate plateau (also marked with a green dotted line) also appears at the final timescale $\tau_\text{f}=\lambda^{-L/2}$. As such, the gauge violation for initial state (C) exhibits the features of the violations for both initial states (A) and (B).

The spatiotemporally averaged expectation values of the supersector projectors $P_s$ in the case of initial states (B) and (C) are shown in Figs.~\ref{fig:FigInitStateII}(b) and~\ref{fig:FigInitStateIII}(b), respectively. All timescales in the prethermal staircase of the gauge violation can be found in these projectors, similarly to what has been explained in Sec.~\ref{sec:Z2LGT_A}. 

Hence, we see that regardless of what initial state we start in within the (super)sector of $g_j=0$, $\forall j$, a staircase of prethermal plateaus emerges and the maximal violation is always delayed by a timescale $\lambda^{-L/2}$. In Sec.~\ref{sec:Z2LGT_E}, we will show that this also happens for initial states where $g_j=0$ is not satisfied at every matter site $j$. But first, we explore the effect of matter filling on staircase prethermalization.

\subsection{Effect of matter filling}\label{sec:Z2LGT_D}
\begin{figure}[htp]
	\centering
	\hspace{-.25 cm}
	\includegraphics[width=.49\textwidth]{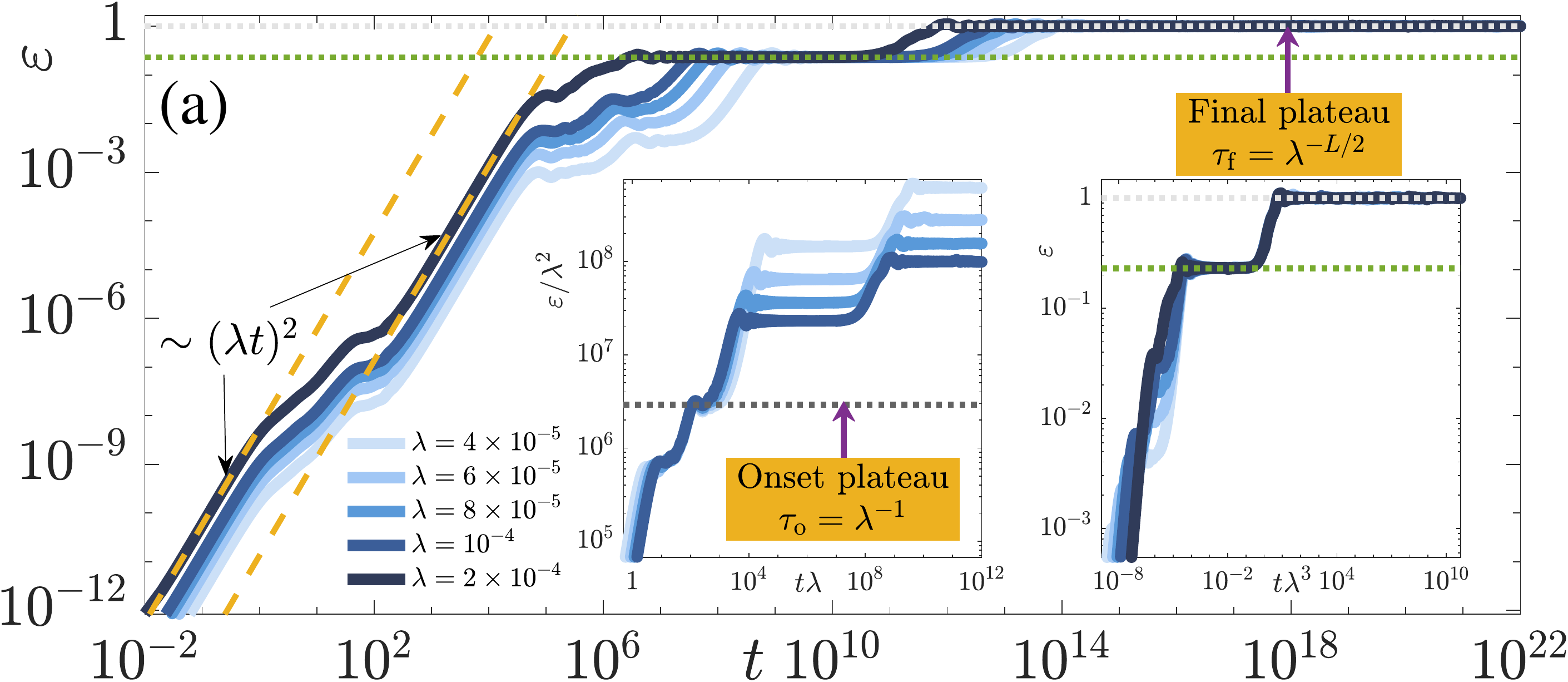}\quad\\
	\hspace{-.25 cm}
	\includegraphics[width=.49\textwidth]{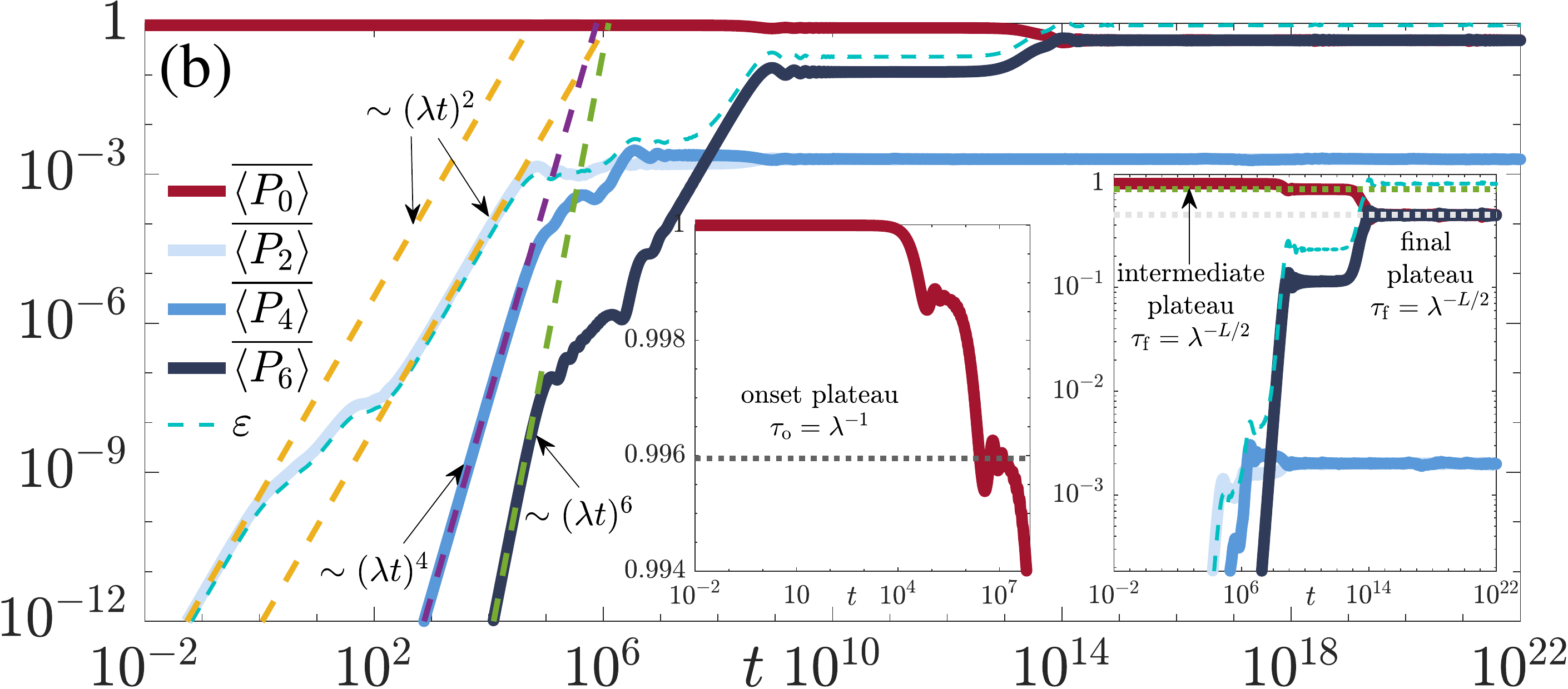}\quad
	\hspace{-.25 cm}
	\caption{
		(Color online). Same as Fig.~\ref{fig:FigOther}, but for initial state (D). In contrast to Figs.~\ref{fig:FigOther}-\ref{fig:FigInitStateIII}, this state lives at sixth filling of matter sites. The error behaves qualitatively similar to the cases of initial states (A-C) in that we have distinct prethermal plateaus with a final plateau at timescale $\tau_\text{f}=\lambda^{-L/2}$. Here, only two timescales appear, as in the case of initial state (B) in Fig.~\ref{fig:FigOther}, with the intermediate plateau (marked in green dotted line) exhibiting the \textit{final} timescale $\tau_\text{f}=\lambda^{-L/2}$. As a qualitative difference from the other states, the onset plateau occurring at timescale $\tau_\text{o}=\lambda^{-1}$ also scales in its value as $\lambda^2$. Projectors in (b) shown for $\lambda=4\times10^{-5}$.
	} 
	\label{fig:FigInitStateIV} 
\end{figure}

Initial states (A), (B), and (C) are at half filling, i.e., each carries three hard-core bosons on a periodic lattice of $L=6$ matter sites and $L=6$ links. 
In contrast, the initial state (D) of Fig.~\ref{fig:FigStates} hosts just a single boson in the $\mathrm{Z}_2$ LGT. 
As shown in Fig.~\ref{fig:FigInitStateIV}, again a prethermal staircase of stable plateaus appears. 
As with all initial states we have considered so far, the final plateau exhibits the exponentially large timescale $\tau_\text{f}=\lambda^{-L/2}$. 
We also see an onset plateau occurring at timescale $\tau_\text{o}=\lambda^{-1}$, but, in contrast to the initial states at half filling, here we find that its height scales as $\lambda^2$. Moreover, as in the case of initial state (B), the intermediate plateau exhibits the final timescale $\tau_\text{f}=\lambda^{-L/2}$.

\begin{figure}[htp]
	\centering
	\hspace{-.25 cm}
	\includegraphics[width=.45\textwidth]{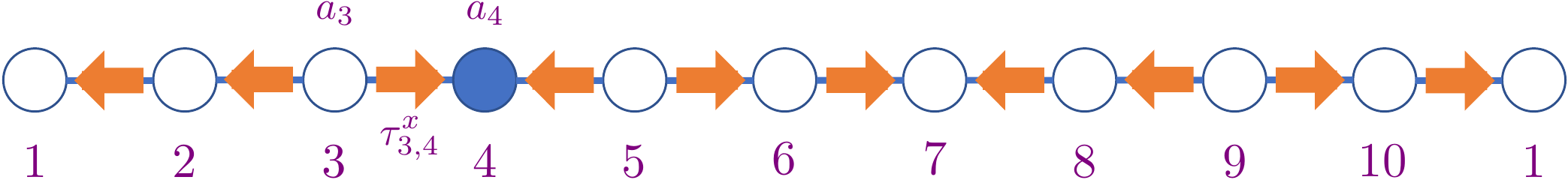}\quad\\
	\vspace{0.2 cm}
	\hspace{-.25 cm}
	\includegraphics[width=.49\textwidth]{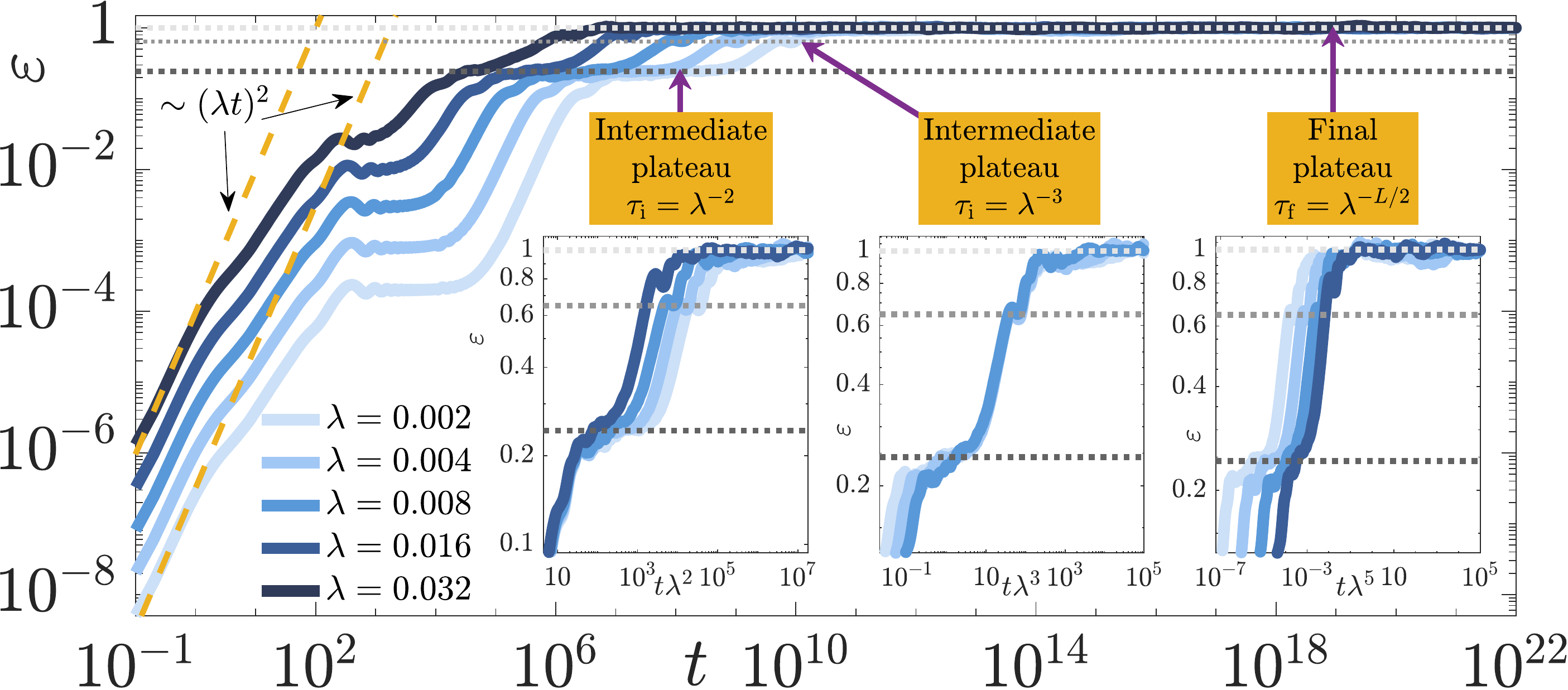}\quad
	\hspace{-.25 cm}
	\caption{(Color online). The spatiotemporal average of the gauge violation, given in Eq.~\eqref{eq:error}, for the $\mathrm{Z}_2$ LGT at tenth filling with $L=10$ matter sites (and $L=10$ links) for various values of the gauge-breaking strength $\lambda$ (see legend) starting in the initial state shown on top. Staircase prethermalization is prominent with plateaus at the pre-onset timescale $\lambda^0$, the intermediate timescales $\lambda^{-2}$ and $\lambda^{-3}$, and the final timescale of $\lambda^{-5}=\lambda^{-L/2}$ (see insets).}
	\label{fig:L10} 
\end{figure}

Here, we take advantage of the smaller number of bosons, which eases computational demands, in order to compute the dynamics of the gauge violation in the case of a $\mathrm{Z}_2$ LGT with $L=10$ matter sites (and $L=10$ links), with only a single boson on the lattice; cf.~Fig.~\ref{fig:L10}. Staircase prethermalization manifests itself in four plateaus at the respective timescales $\tau_\text{po}=\lambda^0$ (pre-onset), $\tau_{\text{i},1}=\lambda^{-2}$ (intermediate), $\tau_{\text{i},2}=\lambda^{-3}$ (intermediate), and $\tau_\text{f}=\lambda^{-5}=\lambda^{-L/2}$ (final). 
As these results show, even for the largest lattice sizes we can access in our ED code, staircase prethermalization is prominent and, importantly, the final plateau at timescale $\lambda^{-L/2}$ is always present even when a few of the earlier plateaus may vanish. Again, this indicates that at least NISQ-era quantum simulators with small inherent gauge invariance-breaking errors will offer reliable gauge-invariant dynamics up to times that are exponentially long in system size. 

Thus, we have shown in this Section that matter filling does not change the principal picture of staircase prethermalization.

\begin{figure}[htp]
	\centering
	\hspace{-.25 cm}
	\includegraphics[width=.49\textwidth]{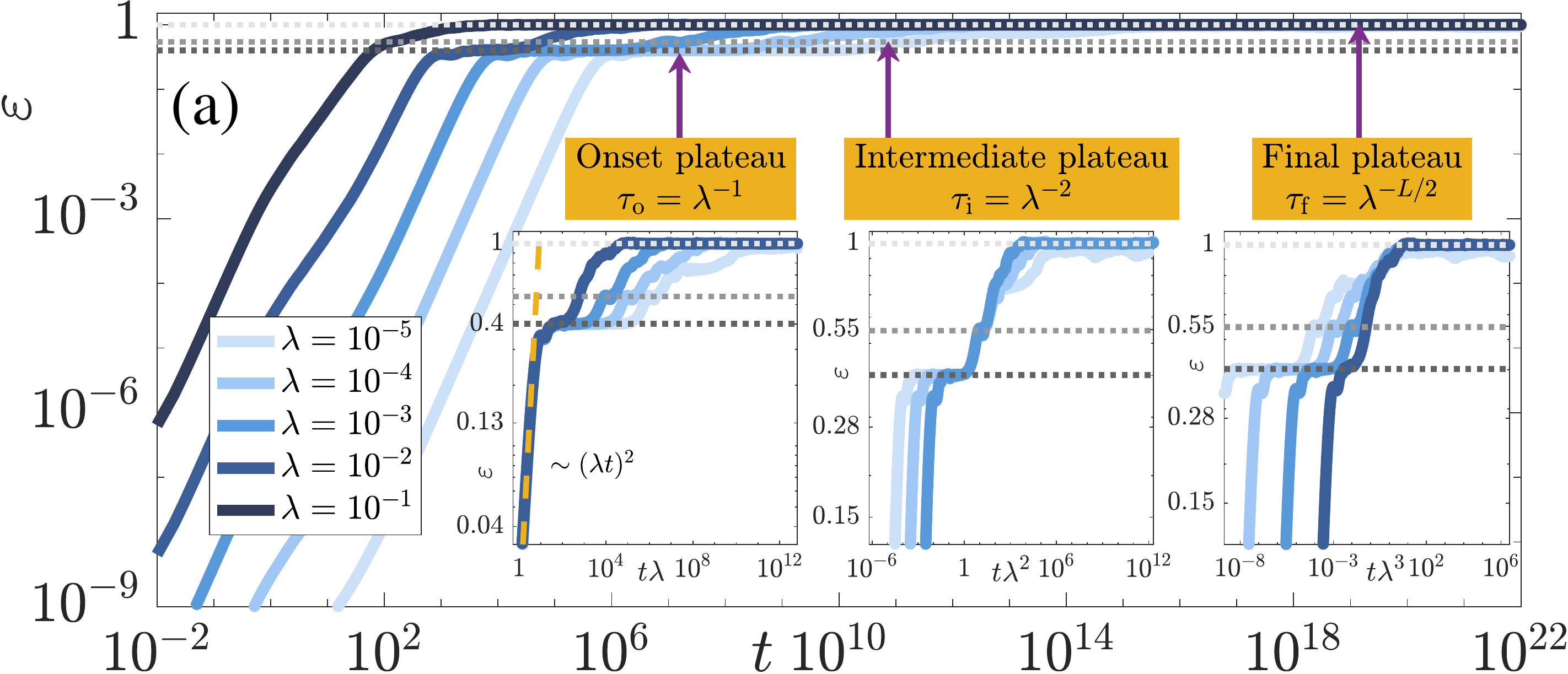}\quad\\
	\hspace{-.25 cm}
	\includegraphics[width=.49\textwidth]{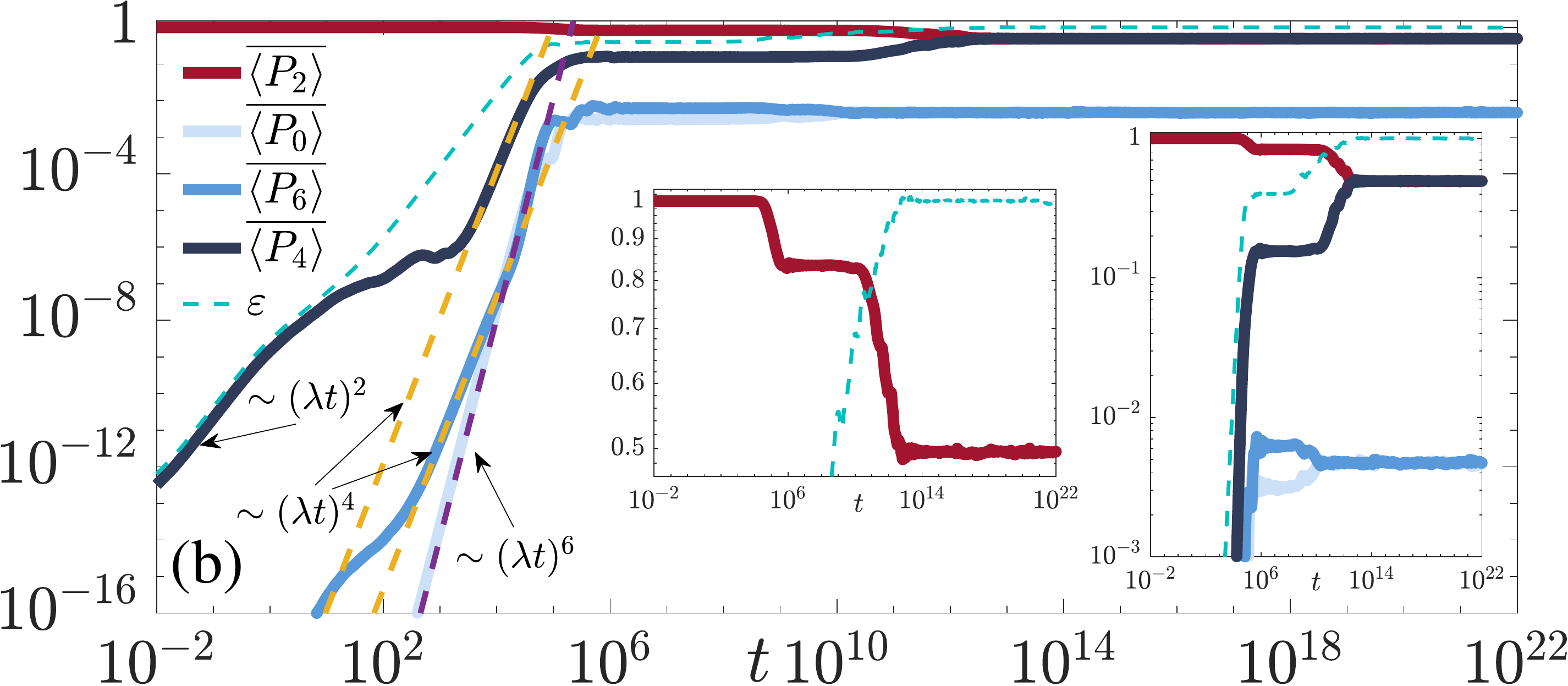}\quad
	\hspace{-.25 cm}
	\caption{
		(Color online). Same as Fig.~\ref{fig:FigInitStateII}, but for initial state (E) from Fig.~\ref{fig:FigStates}. In contrast to initial states (A-D) in Figs.~\ref{fig:FigOther}-\ref{fig:FigInitStateIV}, this state lies in the gauge sector $g_3=g_6=2$, with all other $g_j=0$. Nevertheless, the qualitative behavior of staircase prethermalization in the gauge violation and projectors is unchanged, with all timescales $\tau_\text{o}=\lambda^{-1}$, $\tau_\text{i}=\lambda^{-2}$, and $\tau_\text{f}=\lambda^{-3}=\lambda^{-L/2}$ of the prethermal plateau emerging in the spatiotemporal averages of (a) the gauge violation and (b) the expectation values of the projectors $P_s$ onto the gauge-invariant supersectors.
	} 
	\label{fig:FigNonzeroGj} 
\end{figure}

\subsection{Effect of starting in a different gauge-invariant sector}\label{sec:Z2LGT_E} 
Up to now, all discussed initial states were chosen from the sector $g_j=0$, $\forall j$, which is its own supersector (see Glossary in Appendix~\ref{sec:glossary}). To further corroborate the generality of our findings, we now consider the initial state (E) from Fig.~\ref{fig:FigStates}, in which $g_j\neq 0$ at matter sites $j=3,6$, i.e., state (E) is in the gauge sector $(0,0,2,0,0,2)$ within the gauge supersector $\{\alpha_{\{2\}}\}$.
The gauge-invariance violation, defined in Eq.~\eqref{eq:error} to measure the deviation from the initial gauge-invariant sector regardless of what supersector it may be in, is displayed in Fig.~\ref{fig:FigNonzeroGj}(a). 
Again, it shows three clear plateaus at the three timescales $\tau_\text{o}=\lambda^{-1}$, $\tau_\text{i}=\lambda^{-2}$, and $\tau_\text{f}=\lambda^{-3}=\lambda^{-L/2}$, respectively. 
As seen in Fig.~\ref{fig:FigNonzeroGj}(b), the timescales are resolved again in the projectors from Eq.~\eqref{eq:P}; see discussion in Sec.~\ref{sec:Z2LGT_A}. The supersectors $P_0$ and $P_6$ do not display a final plateau, which instead is clearly visible in the interplay of $P_2$ and $P_4$, in contrast to what we have seen for initial states (A-D), in panels (b) of Figs.~\ref{fig:FigOther}-\ref{fig:FigInitStateIV}, respectively. Indeed, $P_0$ and $P_6$ exhibit only two plateaus, while $P_2$ and $P_4$ exhibit all three timescales present in the associated gauge violation of Fig.~\ref{fig:FigNonzeroGj}(a). The switching of roles between $P_2$ and $P_4$ on the one hand and $P_0$ and $P_6$ on the other is not surprising. This is because the initial state now lives in the supersector of $P_2$ rather than $P_0$, while supersector $P_4$ ($P_6$) is still always symmetric to the former (latter) for the case of $L=6$ matter sites.

As such, we see that the phenomenology of staircase prethermalization is independent of the initial gauge-invariant supersector.

\subsection{Effect of relaxing the hard-core constraint}\label{sec:Z2LGT_F}
\begin{figure}[htp]
	\centering
	\hspace{-.25 cm}
	\includegraphics[width=.49\textwidth]{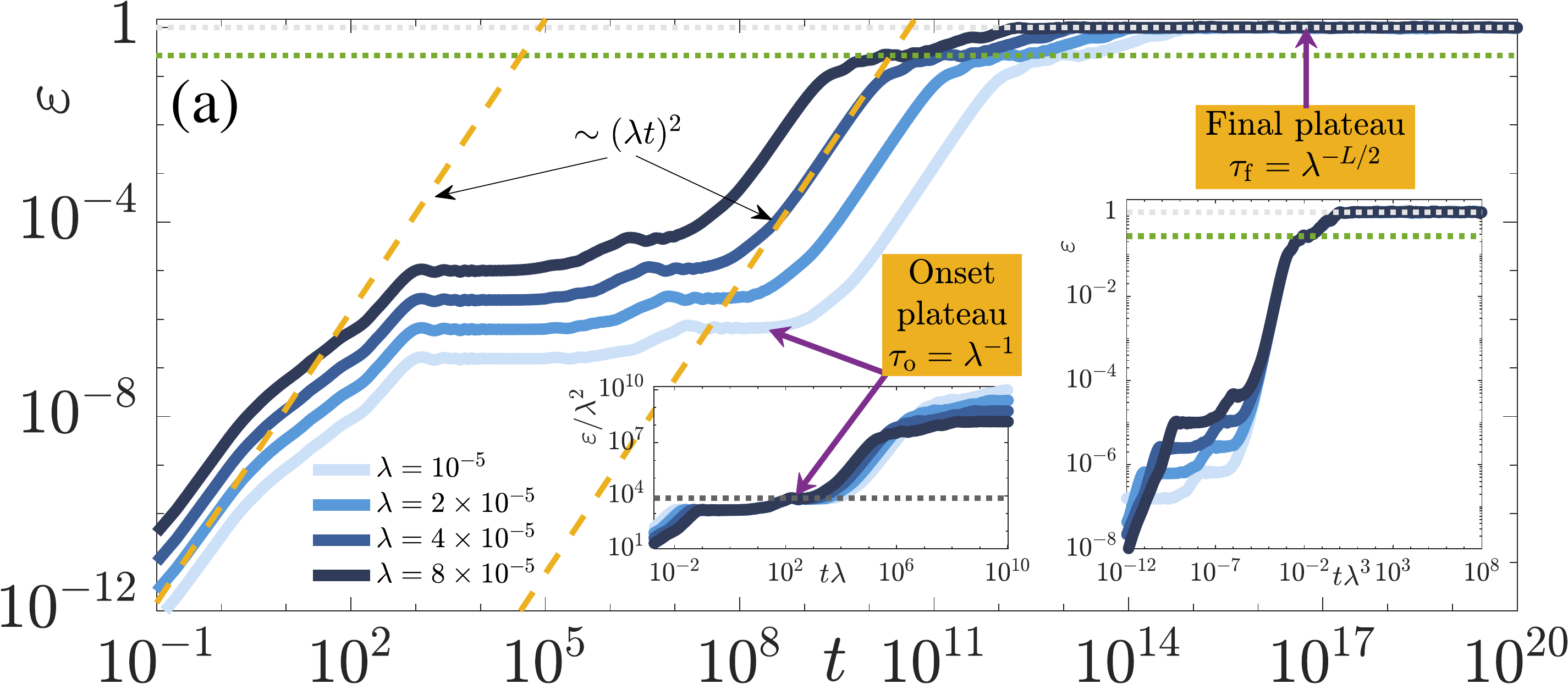}\quad\\
	\hspace{-.25 cm}
	\includegraphics[width=.49\textwidth]{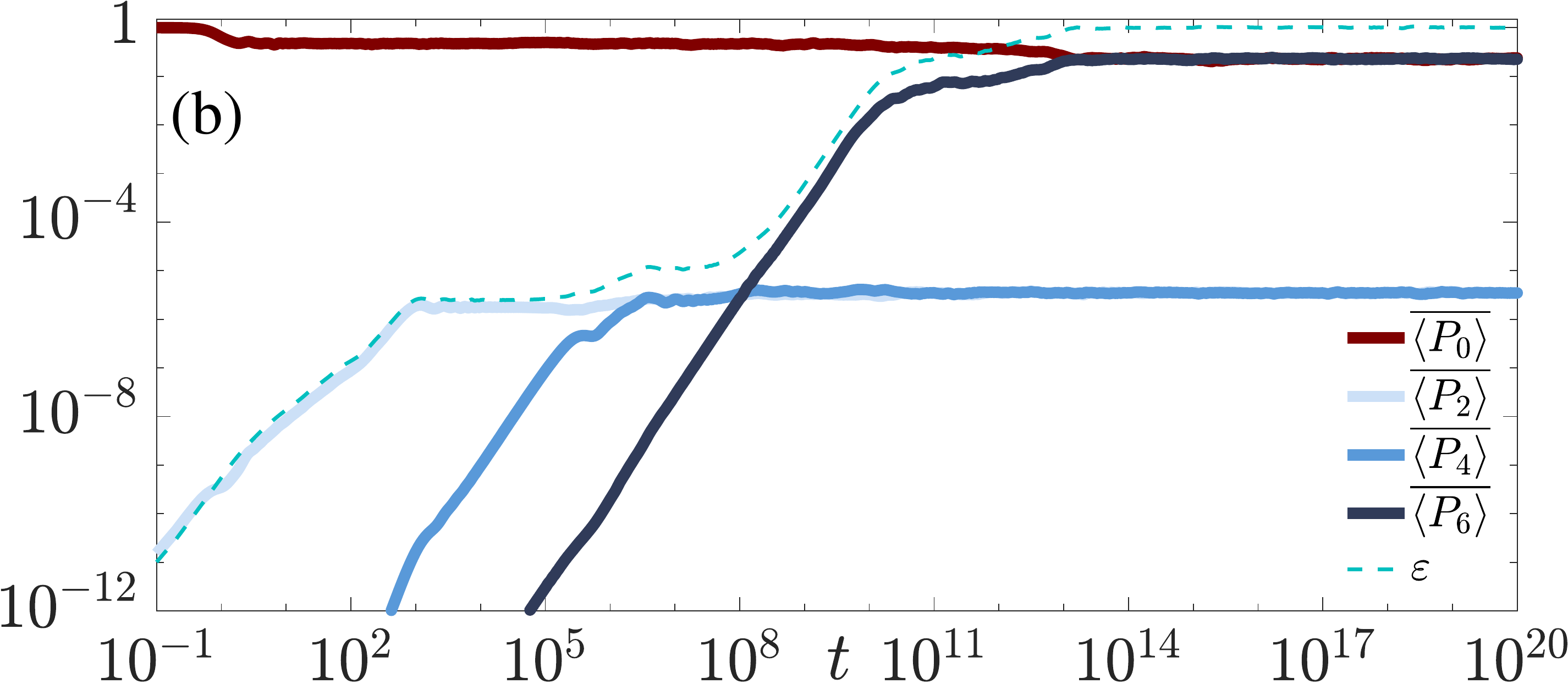}\quad
	\hspace{-.25 cm}
	\caption{(Color online). 
		Same as Fig.~1 of the joint submission\cite{Halimeh2020b} for initial state (A) but with relaxing the hardcore constraint to allow for a maximal occupation of two bosons per site. (a) In the gauge violation, the pre-onset, onset, and final plateaus at timescales $\tau_\text{po}=\lambda^0$, $\tau_\text{o}=\lambda^{-1}$, and $\tau_\text{f}=\lambda^{-L/2}$, respectively, emerge while the intermediate plateau of timescale $\tau_\text{i}=\lambda^{-2}$ is missing. Projectors in (b) are shown for $\lambda=4\times10^{-5}$.
	}	 
	\label{fig:MaxOcc2} 
\end{figure}

The hard-core constraint is not a necessary requirement for the model in Eq.~\eqref{eq:H0} to be a $\mathrm{Z}_2$ LGT, though it is rather useful for experimental (and numerical) feasibility.\cite{Schweizer2019} 
In this Section, we study the generality of staircase prethermalization when relaxing the hard-core constraint to allow a maximal occupation of two bosons per matter site. We again start in initial state (A) of Fig.~\ref{fig:FigStates}. The numerical results are shown in Fig.~\ref{fig:MaxOcc2} for the spatiotemporal average of the gauge violation in panel (a) and the expectation values of the supersector projectors $P_s$ in panel (b).

In this case, we find a clear pre-onset plateau at timescale $\tau_\text{po}=\lambda^0$, followed by an onset plateau at timescale $\tau_\text{o}=\lambda^{-1}$. As was the case for sixth filling (Sec.~\ref{sec:Z2LGT_D}), the height of the onset plateau scales as $\lambda^2$. 
No intermediate plateau at timescale $\tau_\text{i}=\lambda^{-2}$ can be discerned. Nevertheless, the final plateau at timescale $\tau_\text{f}=\lambda^{-3}=\lambda^{-L/2}$ is clearly visible, as in all cases considered above. 

We have also repeated this quench with a maximum on-site occupation of three bosons per matter site, with qualitatively identical results. Therefore, we conclude that the maximal on-site occupation does not change the qualitative nature of staircase prethermalization.

\section{Staircase prethermalization in the $\mathrm{U}(1)$ gauge theory}\label{sec:U1LGT}

\begin{figure}[htp]
	\centering
	\hspace{-.25 cm}
	\includegraphics[width=.48\textwidth]{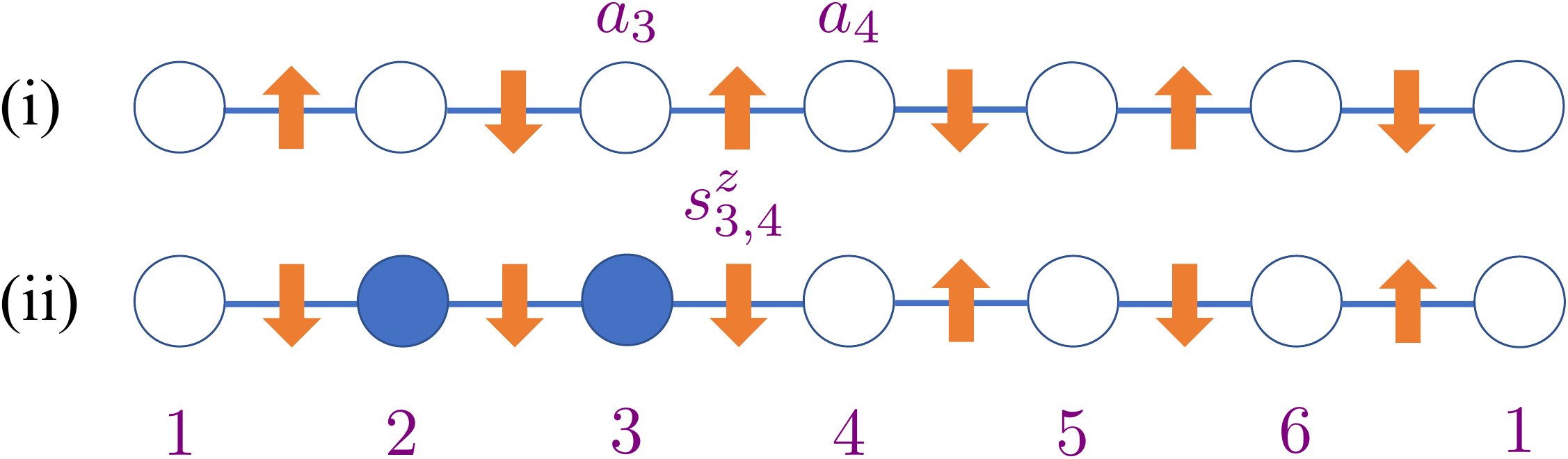}\quad
	\hspace{-.15 cm}
	\caption{(Color online). Initial states used for the quenches in the $\mathrm{U}(1)$ quantum link model, here shown for $L=6$ matter sites. Just as in Fig.~\ref{fig:FigStates}, periodic boundary conditions are assumed, as indicated by the site indexing at the bottom.} 
	\label{fig:U1LGTInitialState} 
\end{figure}

\begin{figure}[htp]
\centering
\hspace{-.25 cm}
\includegraphics[width=.49\textwidth]{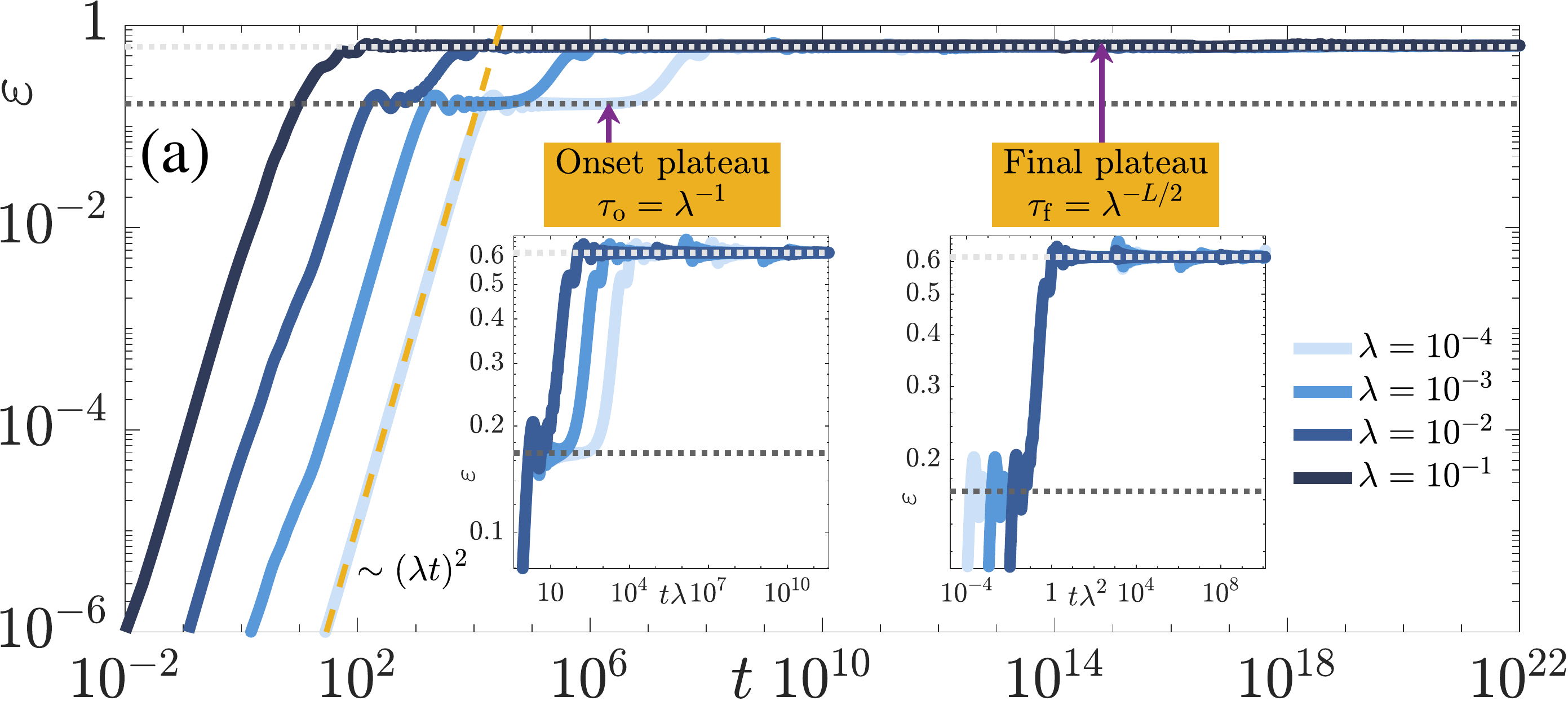}\quad\\
\hspace{-.25 cm}
\includegraphics[width=.49\textwidth]{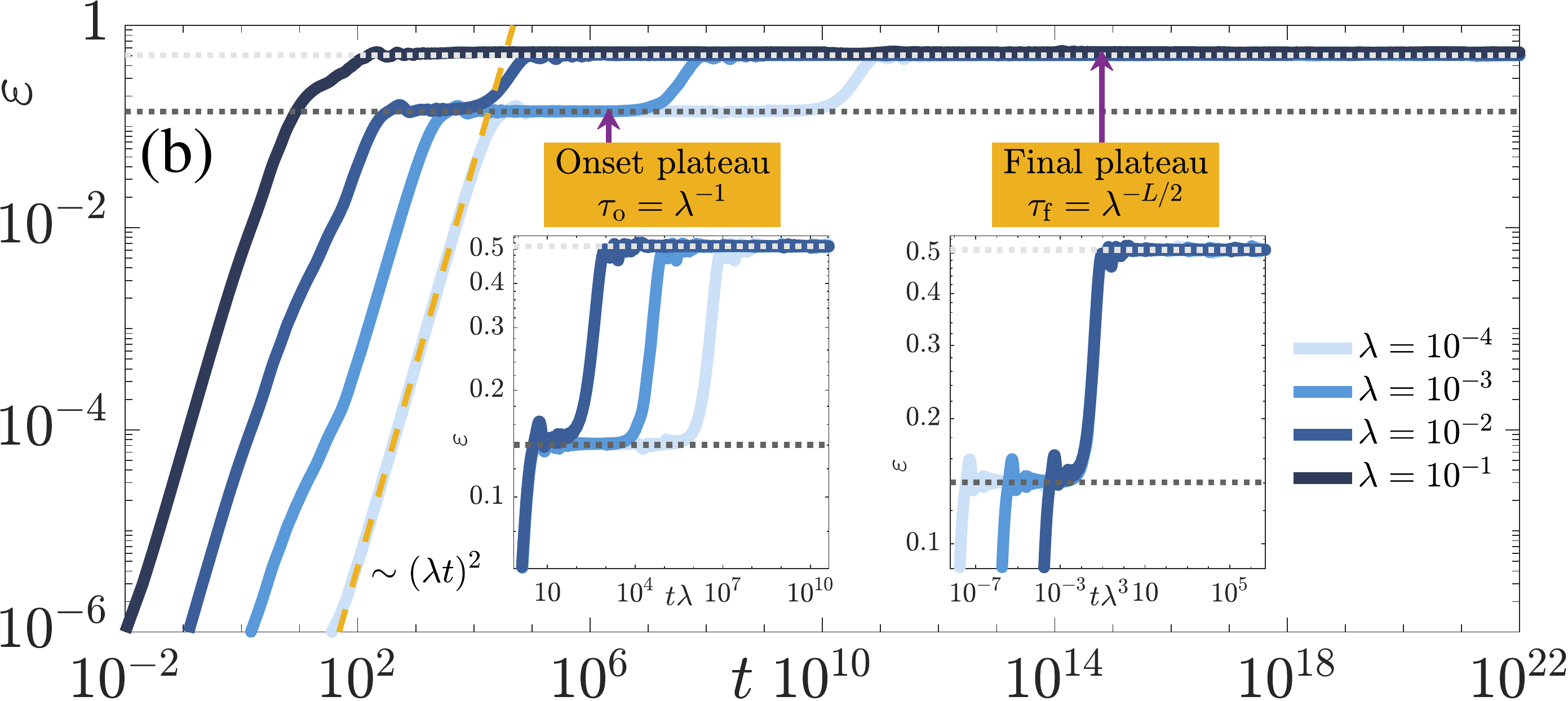}\quad\\
\hspace{-.25 cm}
\includegraphics[width=.49\textwidth]{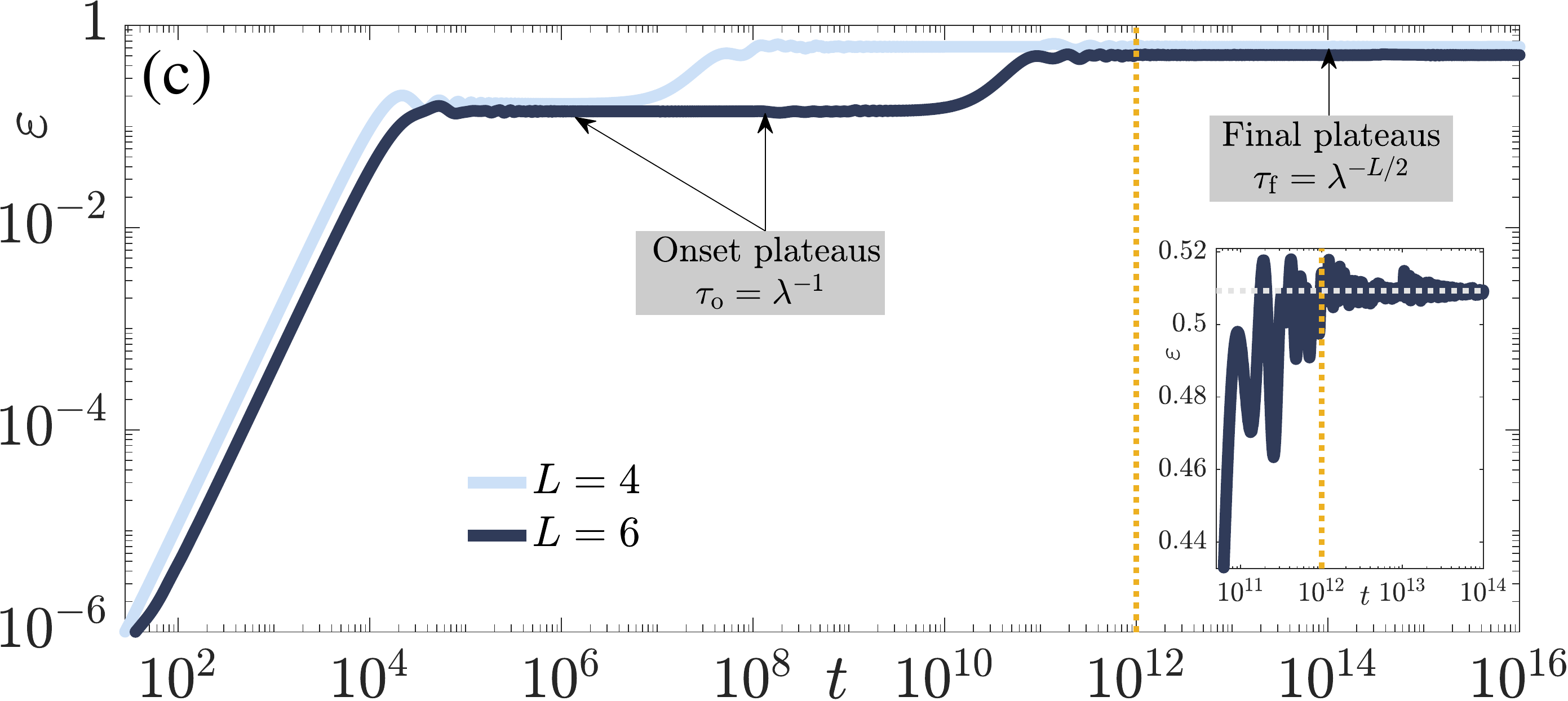}\quad
\hspace{-.25 cm}
\caption{(Color online). $\mathrm{U}(1)$ LGT: Time evolution of the spatiotemporally averaged gauge-invariance violation, with initial state (i) from Fig.~\ref{fig:U1LGTInitialState}. (a,b) Gauge violation for $L=4$ and $L=6$ matter sites, respectively, comparing various values of $\lambda$ (see legend). In both cases, two timescales appear as expected, an onset timescale $\tau_\text{o}=\lambda^{-1}$ and a final timescale $\tau_\text{f}=\lambda^{-L/2}$. (c) Direct comparison of gauge violation as a function of the number of local gauge constraints ($\lambda=10^{-4}$). Inset: the final steady state plateau for $L=6$ matter sites indeed begins at $\tau_\mathrm{f}=\lambda^{-3}=\lambda^{-L/2}$.
}
\label{fig:FigU1} 
\end{figure}

\begin{figure}[htp]
\centering
\hspace{-.25 cm}
\includegraphics[width=.48\textwidth]{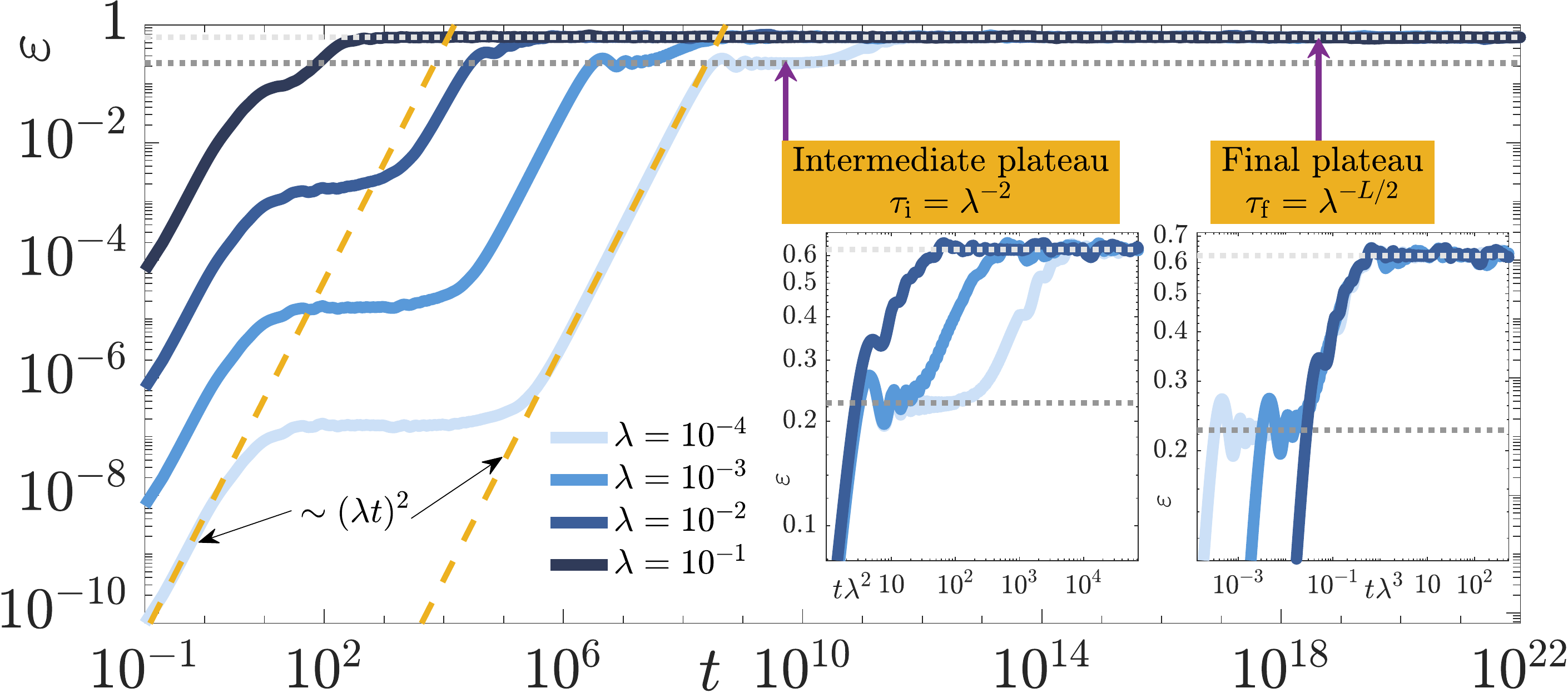}\quad
\hspace{-.15 cm}
\caption{(Color online). Same as Fig.~\ref{fig:FigU1}(b) but with initial state (ii) from Fig.~\ref{fig:U1LGTInitialState}. Here, a pre-onset plateau at timescale $\tau_\text{po}=\lambda^{0}$ as well as plateaus at timescales $\tau_\text{i}=\lambda^{-2}$ and $\tau_\text{f}=\lambda^{-3}=\lambda^{-L/2}$ emerge. As this shows, which intermediate timescales are realized does depend on the initial condition (just as in the case of the $\mathrm{Z}_2$ LGT), though the general phenomenon of staircase prethermalization does not.}
\label{fig:FigU1other} 
\end{figure}

We now show evidence for the same behavior in a $\mathrm{U}(1)$ gauge theory as has been realized in a recent experiment.\cite{Yang2016} The model is described by the Hamiltonian 
\begin{align}\label{eq:U1H0}
H_0=-J\sum_{j=1}^L\big(a_j s_{j,j+1}^+ a_{j+1}+\text{H.c.}\big)+\mu\sum_{j=1}^La_j^\dagger a_j,
\end{align}
with $a_j,a_j^\dagger$ again representing ladder operators for hard-core bosons on site $j$, but where now, employing the quantum link model (QLM) formalism,\cite{Chandrasekharan1997,Wiese_review} the gauge (electric) field is represented by the spin-$1/2$ matrix $s_{j,j+1}^+$ ($s_{j,j+1}^z$). The local gauge operator is
\begin{align}
G_j=s^z_{j-1,j}+a_j^\dagger a_j+s^z_{j,j+1},
\end{align}
and $[H_0,G_j]=0$, $\forall j$. Unlike the $\mathrm{Z}_2$ LGT of Eq.~\eqref{eq:H0}, which has two local gauge sectors, the $\mathrm{U}(1)$ LGT of Eq.~\eqref{eq:U1H0} hosts four local gauge-invariant sectors. We choose again the initial gauge-invariant sector $G_j\ket{\psi}=0$, $\forall j$, which is its own supersector. 

We first prepare our system in the initial state (i) of Fig.~\ref{fig:U1LGTInitialState}. This state has zero bosons in the chain, while links are initialized in an alternating fashion between spin-up and spin-down (in the $z$ basis), such that the state satisfies Gauss's law with $G_j\ket{\psi}=0$ at each matter site. We now quench the system with the Hamiltonian $H_0+\lambda H_1$ with $J=1$ and $\mu=0$, where the gauge-invariance breaking is now
\begin{align}\label{eq:H1_U1QLM}
H_1=\sum_{j=1}^L\big(a_ja_{j+1}+\text{H.c.}\big).
\end{align}
We note that we choose $H_1$ intentionally such that it breaks only the local symmetry of $H_0$ and not its global symmetry corresponding to pairing conservation. We also remark that we have tried different forms of $H_1$, including one where $\sum_js^x_{j,j+1}$ is added to Eq.~\eqref{eq:H1_U1QLM}, but this did not alter our qualitative picture.

The corresponding time evolution of the gauge-invariance violation, now given by
\begin{align}\label{eq:error_U1LGT}
\epsilon(t)=\frac{1}{Lt}\int_0^t\d \tau\,\sum_{j=1}^L\bra{\psi(\tau)} G_j^2\ket{\psi(\tau)},
\end{align}
is shown in Fig.~\ref{fig:FigU1} for $L=4$ and $6$ matter sites in panels (a) and (b), respectively. Just as in the case of the $\mathrm{Z}_2$ LGT of Fig.~\ref{fig:FigL4}(a), when $L=4$ matter sites the gauge violation exhibits two timescales $\tau_\text{o}=\lambda^{-1}$ and $\tau_\text{f}=\lambda^{-2}=\lambda^{-L/2}$. On the other hand, the case of $L=6$ matter sites in the case of initial state (i) is more reminiscent of that of the $\mathrm{Z}_2$ LGT with initial state (B) shown in Fig.~\ref{fig:FigInitStateII}(a), in that only the timescales $\tau_\text{o}=\lambda^{-1}$ and $\tau_\text{f}=\lambda^{-3}=\lambda^{-L/2}$ appear, while the intermediate timescale $\tau_\text{i}=\lambda^{-2}$ is absent. We further illustrate the system size-dependence in Fig.~\ref{fig:FigU1}(c) by plotting the gauge violation for $L=4$ and $6$ matter sites at error strength $\lambda=10^{-4}$. Clearly, the maximal violation is delayed exponentially in system size.

For comparison, Fig.~\ref{fig:FigU1other} displays the gauge violation when initializing the system in state (ii) of Fig.~\ref{fig:U1LGTInitialState}. 
While the intermediate timescale $\tau_\text{i}=\lambda^{-2}$ as well as the final timescale $\tau_\text{f}=\lambda^{-3}=\lambda^{-L/2}$ are clearly present, the onset timescale $\tau_\text{o}=\lambda^{-1}$ is absent.
Instead, a pre-onset plateau appears with a $\lambda$-independent timescale $\tau_\text{po}=\lambda^0$ and height $\propto \lambda^2$, similar to the case of the $\mathrm{Z}_2$ LGT with $L=8$ matter sites (see Fig.~\ref{fig:FigL8}). As these results show, the phenomenon of staircase prethermalization is not specific to discrete gauge theories but also appears in gauge theories with continuous symmetries.

\section{Contrast to global-symmetry breaking}\label{sec:GlobSym}

\begin{table*}\footnotesize
	\caption{Energy gaps and their properties in the extended Bose--Hubbard model of Eq.~\eqref{eq:BHM} with $L=8$ sites up to second-order processes in $H_1$, which is defined in Eq.~\eqref{eq:H1eBHM}. For the sake of notational consistency with Tables~\ref{tab:energyGapsZ2L4} and~\ref{tab:energyGapsZ2L6} for the $\mathrm{Z}_2$ LGT, we denote here by a subscript $0$ a filling of $L/2$, by a subscript $2$ a filling of $L/2\pm2$, and by a subscript $4$ a filling of $L/2\pm4$. Note how only global symmetry-preserving transitions lead to any resonances. The latter processes can all be absorbed into $H_0$.}
	\label{tab:energyGapsBHML8}
	\begin{tabular}{| l | l | l | l | l |}\hline
		Gap & Nonzero minimum & Total number of accessible states & Number of states with $0<E<\lambda_\text{thresh}$ & Number of states with $E=0$ \\ \hline\hline
		$E_{02}$ & $0.044371$ & $780$ & $0$ & $0$ \\ \hline
		$E_{020}$ & $0.074607$ & $12140$ & $0$ & $1468$ ($12.09\%$ of total)\\ \hline
		$E_{022}$ & N/A & $0$ & $0$ & $0$ \\ \hline
		$E_{024}$ & $0.145393$ & $72$ & $0$ & $0$ \\ \hline
	\end{tabular}
\end{table*}

\begin{table*}\footnotesize
	\caption{Same as Table~\ref{tab:energyGapsBHML8} but for $L=12$ sites. Not only are the global symmetry-breaking processes without resonances, but they also exhibit nonzero energy gaps below $\lambda_\text{thresh}$, which further renders impossible any possibility of nonpreturbative prethermal plateaus emerging.}
	\label{tab:energyGapsBHML12}
	\begin{tabular}{| l | l | l | l | l |}\hline
		Gap & Nonzero minimum & Total number of accessible states & Number of states with $0<E<\lambda_\text{thresh}$ & Number of states with $E=0$ \\ \hline\hline
		$E_{02}$ & $0.000083$ & $132384$ & $294$ ($0.22\%$ of total) & $0$ \\ \hline
		$E_{020}$ & $0.000044$ & $19457752$ & $17312$ ($0.09\%$ of total) & $257644$ ($1.32\%$ of total)\\ \hline
		$E_{022}$ & N/A & $0$ & $0$ & $0$ \\ \hline
		$E_{024}$ & $0.000168$ & $1386558$ & $3428$ ($0.25\%$ of total) & $0$ \\ \hline
	\end{tabular}
\end{table*}

From the arguments brought forward in Sec.~\ref{sec:ME}, we can also understand why we do not observe similar plateaus when breaking a global symmetry. When $H_1$ breaks a \emph{local} symmetry, the gauge violations $g_j\neq 0$ are localized at site $j$. This results in an  abundance of degeneracies, which can be understood by the fact that gauge violations at different sites are equivalent due to translational invariance. Alternatively, the localized gauge violations can be seen as giving rise to flat degenerate bands. These degeneracies are what generate the series of effective Hamiltonians $H_\mathrm{eff}^{(s)}$ that drive the transitions between the different plateaus. 

When, instead, $H_1$ breaks a \emph{global} symmetry, $H_0$ only conserves the total charge $\sum_j g_j$, so locally generated violations can move through the entire system. In this way, they acquire kinetic energy which spreads their energy into an extended band structure. 
Thus, in the case of global-symmetry breaking there are much less exact degeneracies than for local-symmetry breaking, as can be seen in Tables~\ref{tab:energyGapsBHML8} and~\ref{tab:energyGapsBHML12}. 
Indeed, exact degeneracies appear only for processes where $H_1$ is applied an even number of times to leave a given symmetry sector and return to it. Such diagonal processes commute with the symmetry generator, and can thus be absorbed in a renormalized $H_0$. As a consequence, in the considered scenarios for global-symmetry breaking, we find the effective Hamiltonians $H_\mathrm{eff}^{(s)}$ to be completely absent. As a result, the system stabilizes on a generic pre-onset plateau, and the prethermalization staircase does not appear. 

These discussions are consistent with results such as those of Ref.~\onlinecite{Gorin2006}. There, it has been demonstrated that breaking an anti-unitary global symmetry---even in large many-body systems---can lead to a slow fidelity decay, potentially characterized by a long-lived plateau (the pre-onset plateau in our language), which is compromised at the timescale $\lambda^{-1}$. In Ref.~\onlinecite{Gorin2006}, it has been further argued that this behavior is due to correlations in the subspectra of $H_0$, which are not present for a unitary global symmetry such as the particle number conservation discussed here. Analogously, we find resonances between the gauge-symmetry supersectors of our theories to be responsible for the staircase prethermalization. Even though Ref.~\onlinecite{Gorin2006} comprises a study orthogonal to our own, it nevertheless lends credence to the spectral analysis and associated explanation carried out in our work.

\begin{figure}[htp]
	\centering
	\hspace{-.25 cm}
	\includegraphics[width=.49\textwidth]{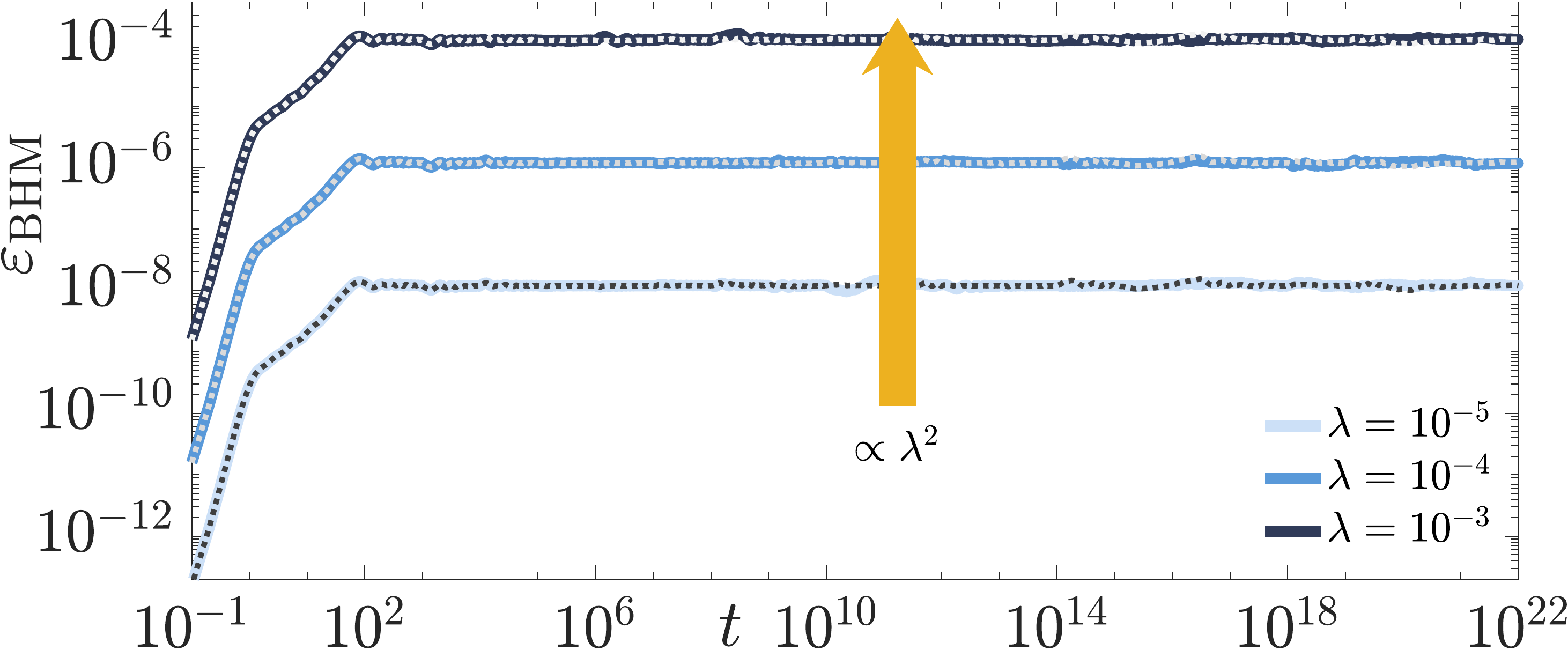}\quad
	\hspace{-.25 cm}
	\caption{(Color online). Absence of (staircase) prethermalization under global-symmetry breaking, illustrated for the extended Bose--Hubbard model with $L=8$ sites, subjected to a small breaking of particle-number conservation. Shown is the time evolution of the spatiotemporally averaged total connected fluctuations in the particle number from half filling. The solid lines are ED results, while the dotted lines are the corresponding results from a second-order Magnus expansion. There is no trace of prethermalization or $\lambda$-dependent timescales. Instead, we only observe the pre-onset plateau that can be described by nonresonant contributions in TDPT [see also {\bf{(1)}} in Fig.~\ref{fig:Schematics}]. We get qualitatively the same behavior for larger and smaller lattice sizes (not shown).}
	\label{fig:BHMdynamics} 
\end{figure}

To bring our above statements on a concrete footing, let us consider the paradigmatic extended Bose--Hubbard model\cite{Kuehner1998,Kuehner2000,Dutta2015} (eBHM)
\begin{align}\nonumber
H_0=&\,-\sum_{j=1}^L\big(J_1a_j^\dagger a_{j+1}+J_2 a_j^\dagger a_{j+2}+\text{H.c.}\big)\\\label{eq:BHM}
&+V\sum_{j=1}^Ln_jn_{j+1}\,.
\end{align} 
This model has a global $\mathrm{U}(1)$ symmetry that embodies particle-number conservation. 
To improve comparability with the above results, we employ a hard-core constraint (thus, there is no on-site interaction term in Eq.~\eqref{eq:BHM}). 
For a finite $J_1$, $J_2$, and $V$, this model is nonintegrable, while it is integrable when $J_1=J_2=0$ (atomic limit) as well as when $V=0$ (free bosons). Under the constraint of hard-core bosons, the model is also integrable at $J_2=0$ as then it is equivalent to the XXZ model.\cite{Kollath2010} 
To avoid prethermalization due to small integrability breaking\cite{Santos2010} and thus to ensure that the resulting dynamics is solely due to global-symmetry breaking, we set $J_1=1$, $J_2=0.83$, and $V=0.11$ (other generic values of these parameters, exluding integrable points, yield the same qualitative conclusions). 

We initialize the system at half-filling with staggered occupation (odd sites contain zero bosons, while every even site has a single boson). 
We quench this initial state with $H=H_0+\lambda H_1$, where
\begin{align}\label{eq:H1eBHM}
\lambda H_1&=\lambda \sum_{j=1}^L(a_ja_{j+1}+\text{H.c.}),
\end{align}
is a small global-symmetry breaking. The time evolution of the total connected particle-number fluctuations
\begin{align}
\varepsilon_\text{BHM}(t)=\frac{1}{Lt}\int_0^t\d s\,\langle\bigg[\sum_{j=1}^Ln_j(s)-\frac{L}{2}\bigg]^2\rangle,
\end{align}
which mimics in form the gauge violation in Eq.~\eqref{eq:error}, is presented in Fig.~\ref{fig:BHMdynamics} for a chain of $L=8$ sites. Here, $n_j=a_j^\dagger a_j$ is the particle-number operator. There is no signature of staircase prethermalization. Regardless of $\lambda$, the deviation grows at small times until it saturates at a timescale $\lambda^0$. 
The behavior is fully captured in a leading-order Magnus expansion with $\tilde{U}(t)=\exp[\Omega_1^\mathrm{nonres}(t)]$ (for consistency with the above figures, the dotted lines in Fig.~\ref{fig:BHMdynamics} also contain the second-order terms in the Magnus expansion, which contribute subleading corrections to the plateau height). 
We have also simulated the same dynamics in larger and smaller system sizes, but the picture remains qualitatively the same. 

\section{(Absence of) thermalization}\label{sec:MBL}
The $\mathrm{Z}_2$ LGT in Eq.~\eqref{eq:H0} with finite nonzero $J_a$ and $J_f$ is, in a broad sense, not integrable. Indeed, it maps onto an interacting spin model for finite $J_a,J_f\neq0$.\cite{Borla2019} It has two integrable points. At $J_f=0$, the gauge field decouples and the model is that of free fermions, while at $J_a=0$, the model maps onto an effective integrable Hamiltonian for bound states.\cite{Borla2019} Note that for $J_f>0$ the bare fermions confine into dimers. Similarly, the $\mathrm{U}(1)$ QLM in Eq.~\eqref{eq:U1H0} is not integrable for generic $J$ and $\mu$. As such, the prethermalization behavior we observe for both models in this work seems to be distinct from that due to small integrability breaking, though this cannot be finally confirmed until further investigations such as those based on a Bethe-Ansatz rule out integrability for generic values of $J_a$ and $J_f$ for the $\mathrm{Z}_2$ LGT and of $J$ and $\mu$ for the $\mathrm{U}(1)$ QLM. At any rate, the staircase of prethermal plateaus that we show in this work and its joint submission\cite{Halimeh2020b} is not observed in the usual scenario of prethermalization due to small integrability breaking,\cite{Moeckel2008,Mazets2008,Rigol2009a,Rigol2009b,Kollar2011,Bertini2015,Marcuzzi2016,Halimeh2017} nor is the timescale of the final (second in this case) plateau expected to be delayed exponentially in system size.\cite{Mori_review} Also, in this traditional setting of weak-integrability breaking, only the first plateau is prethermal as the integrable part of the system tries to settle into a generalized Gibbs ensemble,\cite{Barthel2008} while the second is thermal and characterized by a Gibbs ensemble. Indeed, it is known from ED studies that in integrable systems subjected to an integrability-breaking term, the breaking strength required to observe signatures of nonintegrability (such as equilibration to a thermal steady state) is inversely proportional to system size.\cite{Santos2010} 
Thus, the larger the system size the more prominent are signatures of nonintegrability for a fixed integrability-breaking strength $\lambda$, and hence a larger system is expected to reach the onset of the final (second) plateau at times not later than those for smaller system sizes.\cite{Mallayya2019} This is in stark contrast to our findings in gauge theories with $L$ local constraints subjected to small gauge-invariance breaking. In the latter case, the final---in general $(L/2+1)^\text{th}$---plateau is delayed exponentially with system size as $\lambda^{-L/2}$. Such an exponential delay with system size appears in the case of weak integrability breaking only in the case when $\lambda$ is itself exponentially \textit{small} in system size.\cite{Pandey2020} For such a $\lambda$, the final plateau in the case of small gauge-invariance breaking would lead to a final timescale that is doubly exponential with system size.

Even more, our numerical results indicate that the final steady-state plateau of timescale $\lambda^{-L/2}$ may not be thermal. In particular, the expectation values of a given set of two or more local observables with respect to a thermal density matrix do not agree with those obtained from the time evolution. Of course, this may be due to the small system sizes we can access in ED, or it can be due to an unforeseen integrability, as mentioned in the previous paragraph. Unfortunately, our attempts at level statistics with the system sizes we are able to achieve in ED could not provide a definitive answer on whether or not there is level repulsion. Nevertheless, it has been shown in a recent study,\cite{Yao2020} in which the extended $\mathrm{Z}_2$ LGT is mapped to a transverse-field Ising model with a random longitudinal field, that this model hosts both ergodic and many-body localized phases, depending on $J_f/J_a$. According to those results, it seems we see staircase prethermalization in both of these phases. A further avenue for future work would be to simulate our dynamics using advanced numerical methods such as the Lanczos algorithm\cite{Lanczos1950} or matrix product states,\cite{Uli2010} and thus to try and push the achievable system sizes. Another promising route is drawing a concrete connection between our work on one hand and disorder-free localization and many-body-localized dynamics on the other. The latter have been observed in gauge theories without terms that explicitly break gauge invariance.\cite{Smith2017a,Smith2017b,Brenes2018,Turner2018,Magnifico2019,Karpov2020,Papaefstathiou2020} Indeed, our results indicate that constrained dynamics are a direct consequence of an abundance of local constraints. The latter are also the mechanism through which many-body localization dynamics arises in lattice gauge theories.\cite{Brenes2018}

\section{Conclusion and outlook}\label{sec:conclusion}
In the joint submission of Ref.~\onlinecite{Halimeh2020b}, we present analytic and numerical evidence of staircase prethermalization. 
In this paper, we have provided a thorough numerical analysis of the robustness of staircase prethermalization and laid out a detailed Magnus-expansion derivation of the nonperturbative timescales associated with the prethermal plateaus. In particular, we have shown that our conclusions hold for various initial conditions, including those in other gauge-invariance sectors, with different matter filling, with larger maximal on-site occupation of matter fields, and for continuous gauge groups. 

The generic picture our conclusions draw is that in lattice gauge theories with small gauge-invariance breaking, local constraints enforce a sequence of local-symmetry breaking that gives rise to staircase prethermalization. In particular, the local constraints lead to exact degeneracies in the spectrum of the gauge theory. In turn, these give rise to effective Hamiltonians that we have derived in a Magnus expansion, with excellent quantitative agreement to exact numerics. The associated exponents of these timescales are inversely proportional to powers of the strength of gauge-invariance breaking, generating a clear separation of timescales. Even though certain initial conditions can lead to the vanishing of a plateau at an intermediate timescale, the maximal violation seems to always occur at the final timescale, which is exponentially delayed in system size.

An open question is whether our observations in terms of prethermalization are somehow connected to a hidden integrability at small symmetry breaking, as discussed in Sec.~\ref{sec:MBL}. We stress that the significance of our conclusions does not hinge on this possibility. Rather, it lies in the fact that full gauge violation is exponentially delayed in time as a function of system size, at least for finite-size LGTs. In traditional prethermalization subjected to weak integrability-breaking perturbations, an onset plateau occurs at a timescale independent of the integrability-breaking strength, which is usually a generalized Gibbs ensemble steady state, before rapid thermalization after a timescale of the inverse square of the breaking strength. In our case, instead, the system size enters in a nontrivial way, by increasing the number of plateaus and by delaying the full violation of gauge invariance to exponentially large times. Furthermore, a larger system size is not expected to lead to slower relaxation to the second (and last) plateau in the case of weak integrability breaking at a fixed strength, which is in stark contrast to what we observe in our case where the timescale of the final plateau is exponential in system size. Also, a very recent study\cite{Yao2020} indicates that the $\mathrm{Z}_2$ LGT hosts an ergodic phase for the parameters we have used in most of our results. Nevertheless, it is to be noted that we still see staircase prethermalization even for parameter values within the many-body localized phase.

The conclusions of our work promise the possibility of engineering special initial states that can altogether avoid maximal gauge violation. As seen in Secs.~\ref{sec:Z2LGT} and~\ref{sec:U1LGT}, depending on what initial state the system is prepared in, the intermediate prethermal plateaus can be made of shorter or longer duration, or can be eliminated altogether. One attractive prospect is to design the initial state such that an intermediate plateau can last indefinitely, allowing the system to circumvent maximal violation. Another intriguing avenue is to test our conclusions in the thermodynamic limit. As discussed in Sec.~\ref{sec:TL}, even though we can ascertain from our analytics and numerics that our conclusions hold for lattice gauge theories of finite sizes relevant to current state-of-the-art quantum simulators of the NISQ era, this is not guaranteed in the thermodynamic limit. A promising direction to proceed further would be to simulate our quench dynamics in the thermodynamic limit using uniform matrix product states.\cite{Vanderstraeten2019,Haegeman2011} Although with such methods the longest-accessible evolution times can be limited, it will be sufficient to reach the onset plateau, which suffices to confirm that the delay of full gauge violation is still valid in the thermodynamic limit.

\section*{Acknowledgments}
The authors are grateful to Michael Hartmann, Alessio Recati, Pablo Sala, and Torsten V.~Zache for stimulating discussions. This work is part of and supported by the DFG Collaborative Research Centre SFB 1225 (ISOQUANT), the Provincia Autonoma di Trento, and the ERC Starting Grant StrEnQTh (Project-ID 804305). 

\appendix

\section{Glossary}\label{sec:glossary}
With the aim of making our article as accessible as possible, we provide here a few definitions of frequently used terms.  

\bigskip

\noindent\textbf{\textit{Pre-onset plateau.}} One of the main findings of our work is staircase prethermalization, which involves several timescales. The very first of these is referred to as the \textit{pre-onset} timescale. It is \textit{perturbative} and shows no dependence on the gauge-breaking strength $\lambda$. Its associated plateau can be fully described in TDPT. For this reason, we prefix it with `\textit{pre}' in order to separate it from the later $\lambda$-dependent timescales whose associated plateaus cannot be described by TDPT, but rather require a Magnus expansion. As the Magnus expansion encompasses TDPT, the pre-onset plateau can be captured by the nonresonant terms in the Magnus expansion (see Sec.~\ref{sec:MEA}). 

\medskip

\noindent\textbf{\textit{Onset plateau.}} The onset prethermal plateau is that which occurs at the onset timescale $\lambda^{-1}$. Even though TDPT can predict its own breakdown at this timescale, it cannot describe the subsequent dynamics. Instead, the onset plateau can be captured by a first-order Magnus expansion (see Sec.~\ref{sec:MEA}).

\medskip

\noindent\textbf{\textit{Intermediate plateau.}} An intermediate plateau in staircase prethermalization occurs at any of the intermediate timescales $\lambda^{-s}$ with $1<s<L/2$ and $L$ the number of local constraints. These timescales and their associated plateaus require second- and higher-order Magnus expansions to capture their dynamics (see Secs.~\ref{sec:MEB} and~\ref{sec:MEC}).

\medskip

\noindent\textbf{\textit{Final plateau.}} The final plateau in staircase prethermalization occurs at the final timescale $\lambda^{-L/2}$ with $L$ the number of local constraints. At this plateau, the error generally reaches its maximal value, indicating equal probability of no violation and maximal violation. This timescale and its associated plateau can be captured by a $(L/2)^\mathrm{th}$-order Magnus expansion (see Sec.~\ref{sec:MEC}). 

\medskip

\noindent\textbf{\textit{Staircase prethermalization.}} The pre-onset, onset, intermediate, and final plateaus at the distinct timescales $\lambda^{-s}$---with $s$ even and in the range $0\leq s\leq L/2$ in a lattice gauge theory with $L$ matter sites---lie at different values of the gauge violation. These values increase over evolution time until they reach maximal violation at the final timescale $\lambda^{-L/2}$. This leads to a \textit{staircase prethermalization} structure, also referred to as a \textit{prethermal staircase}, whose steps are these prethermal plateaus. This is the central phenomenon of this work and its joint submission Ref.~\onlinecite{Halimeh2020b}.

\medskip

\noindent\textbf{\textit{Gauge-invariant sector.}} A state $\ket{\psi}$ is said to be gauge-invariant when there exist local-symmetry generators $G_j$ of the gauge group at matter sites $j$ such that $G_j\ket{\psi}=g_j\ket{\psi}$, $\forall j$, where the eigenvalues $g_j$ can take on any of a number of values, depending on the gauge symmetry. A gauge-invariant sector is one where the $g_j$ take on a fixed set of values at each $j$. Since the generators $G_j$ commute with the gauge-theory Hamiltonian $H_0$, for an initial state within a given gauge-invariant sector, $H_0$ drives dynamics only within that sector.

\medskip

\noindent\textbf{\textit{Gauge-invariant supersector.}} A gauge-invariant supersector $M$ is the set of all gauge-invariant sectors where $\sum_j G_j^p\ket{\psi}=M\ket{\psi}$, with $p=1$ for the $\mathrm{Z}_2$ LGT and $p=2$ for the $\mathrm{U}(1)$ LGT. For the former, we can denote the supersector as $\{\alpha_{\{s\}}\}$, which is now the set of all $\binom{L}{s}$ unique gauge-invariant sectors $\alpha_{\{s\}}=(\alpha_1,\ldots,\alpha_L)$ satisfying $\sum_j G_j\ket{\psi}=2s\ket{\psi}$, i.e., $\sum_j\alpha_j=2s$, according to the definition of the gauge-group generator $G_j$ in Eq.~\eqref{eq:Gauss}. The projector onto the gauge-invariant supersector $\{\alpha_{\{s\}}\}$ is $P_s$ given in Eq.~\eqref{eq:P}. 

\medskip

\noindent\textbf{\textit{Gauge-invariance breaking.}} The gauge-invariance breaking terms $H_1$---Eq.~\eqref{eq:H1} for the $\mathrm{Z}_2$ LGT and Eq.~\eqref{eq:H1_U1QLM} for the $\mathrm{U}(1)$ LGT---drive the dynamics out of the gauge-invariant sector within which the initial state lives. This is sometimes also referred to simply as gauge breaking.

\medskip

\noindent\textbf{\textit{Gauge-invariance violation.}} The gauge-invariance violation, not to be confused with gauge-invariance breaking (defined in the Hamiltonian $H_1$), is the measure of how much the system has deviated from its initial gauge-invariant sector over evolution time, and is given by Eq.~\eqref{eq:error} for the $\mathrm{Z}_2$ LGT and Eq.~\eqref{eq:error_U1LGT} for the $\mathrm{U}(1)$ LGT. This is sometimes also referred to simply as gauge violation.

\section{Further results from the Magnus expansion}\label{sec:furtherME}
\begin{figure}[htp]
	\centering
	\hspace{-.25 cm}
	\includegraphics[width=.49\textwidth]{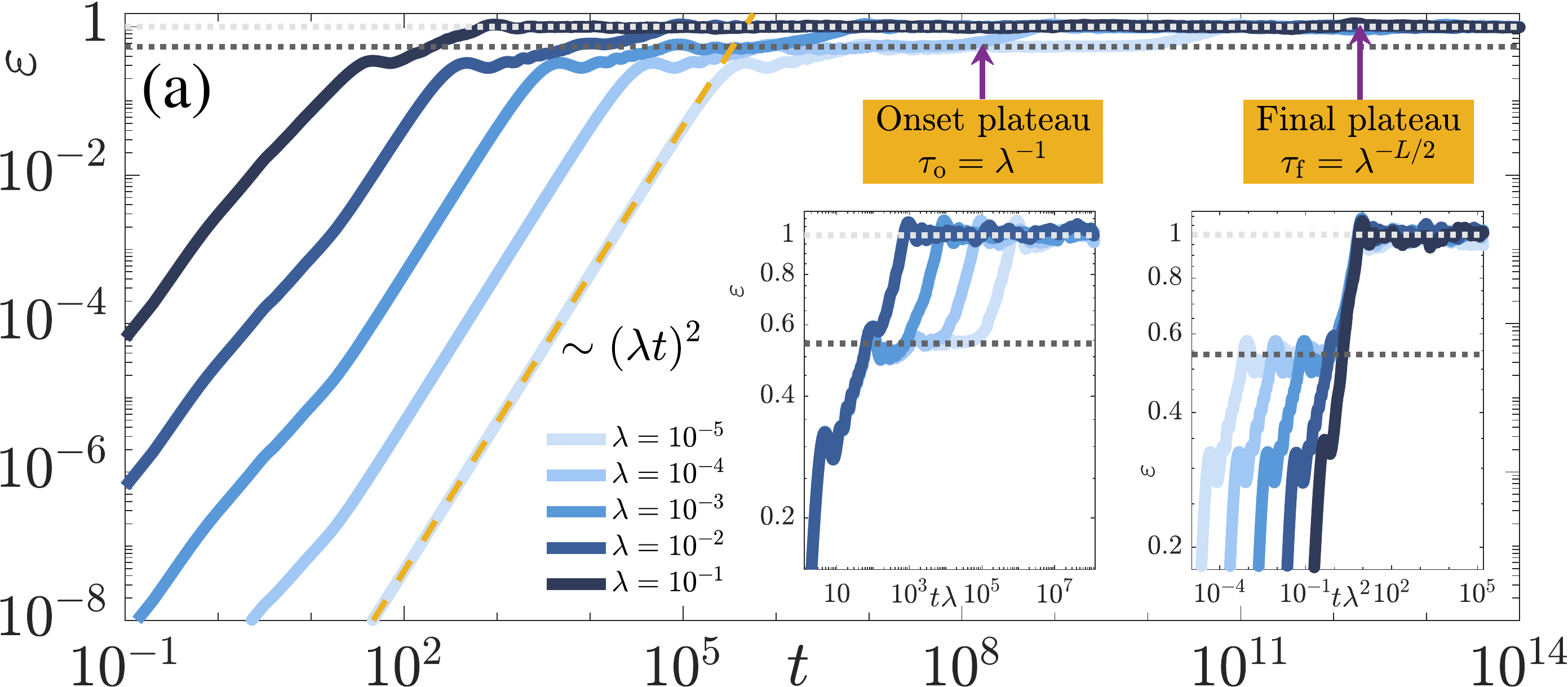}\quad\\
	\hspace{-.25 cm}
	\includegraphics[width=.49\textwidth]{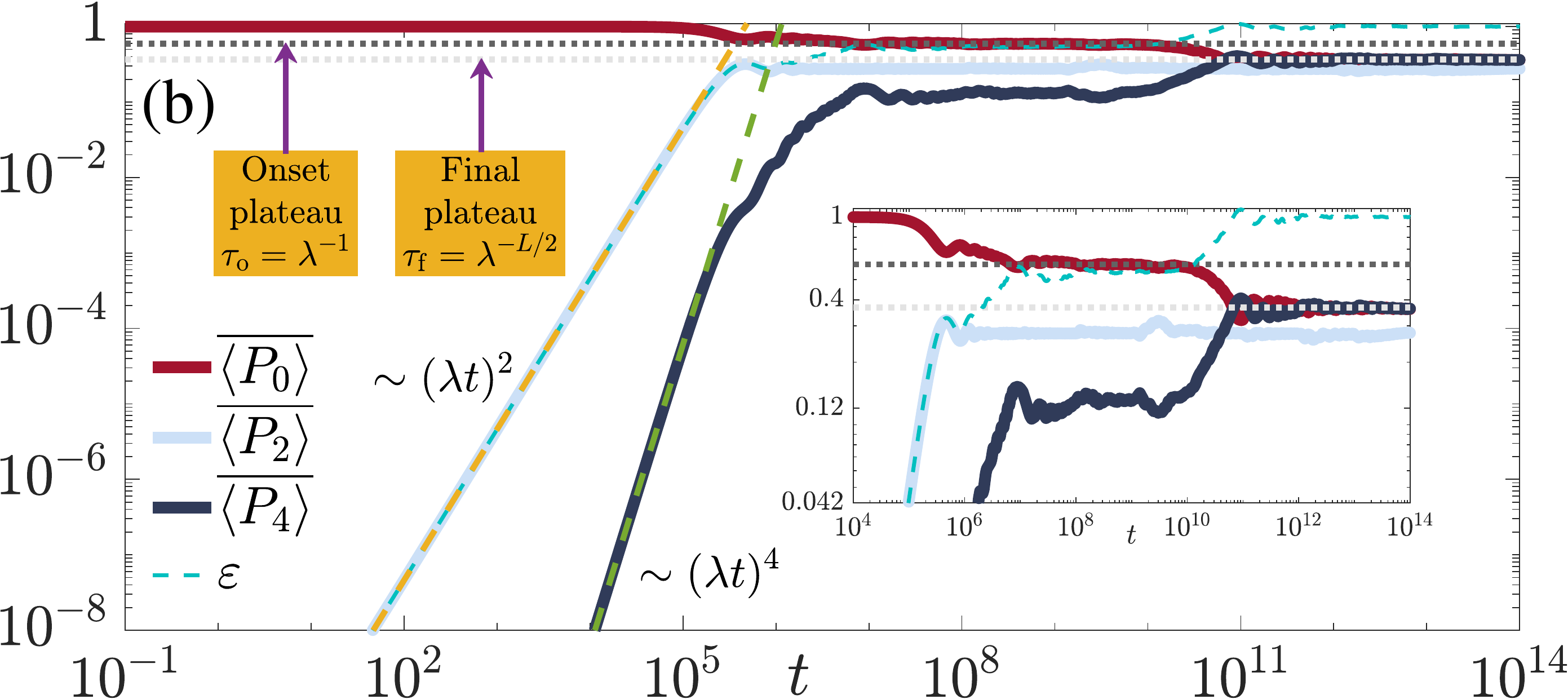}\quad
	\hspace{-.25 cm}
	\caption{(Color online). Same as Fig.~\ref{fig:FigL4} but with the time evolution carried out in terms of the second-order Magnus expansion in Eq.~\eqref{eq:MagnusAppendix} instead of ED. The agreement between both methods is excellent.}
	\label{fig:FigL4_Magnus} 
\end{figure}

In order to further demonstrate the accuracy and power of the Magnus expansion employed to analytically derive the timescales associated with staircase prethermalization, we repeat Fig.~\ref{fig:FigL4} using a second-order Magnus expansion for the time evolved state, 
\begin{align}\label{eq:MagnusAppendix}
\ket{\psi(t)}=\mathrm{e}^{-\mathrm{i}H_0t}\mathrm{e}^{[\Omega_1(t)+\Omega_2(t)]}|\psi_0\rangle,
\end{align}
(see Sec.~\ref{sec:ME} for details). The corresponding results are shown in Fig.~\ref{fig:FigL4_Magnus} for the spatiotemporal averages of the gauge violation in Eq.~\eqref{eq:error} and the projectors onto the gauge-invariant sectors, given in Eq.~\eqref{eq:P}. The similarity to the corresponding ED results of Fig.~\ref{fig:FigL4} is excellent.

\section{Numerics specifics}\label{sec:Numerics}
For our ED simulations, we have used the toolkits QuSpin\cite{Weinberg2017,Weinberg2019} and QuTiP,\cite{Johansson2012,Johansson2013} in addition to our in-house code in order to cross-check and verify our results. Specifically, we have used our own time-evolution routine. It is based on exact exponentiation and is thus able to reach very large evolution times that would require a very small time-step in methods employed by traditional toolkits, as they usually employ a solution of an ordinary differential equation for time evolution. The code for the Magnus expansion is entirely built by us and is based on the formalism presented in Sec.~\ref{sec:ME}.

Here, we note that we are able to achieve larger maximum system sizes in the $\mathrm{Z}_2$ LGT than in its $\mathrm{U}(1)$ counterpart due to the former conserving particle number (even when subjected to $H_1$, which also conserves particle number). This further restricts the effective Hilbert space in which our system resides, thereby facilitating larger chains. 

For the $\mathrm{Z}_2$ gauge theory, we use the following coefficients in the error term $H_1$ given in Eq.~\eqref{eq:H1}, inspired from the experiment of Ref.~\onlinecite{Schweizer2019}:  
\begin{subequations}
	\begin{align}\nonumber
	c_1=&\,\sum_{k>0}\frac{\mathcal{N}(\chi)}{k}\big[\mathcal{J}_{-k-1}(\chi)\mathcal{J}_{-k-2}(\chi)+\mathcal{J}_k(\chi)\mathcal{J}_{k+1}(\chi)\\
	&-\mathcal{J}_{k-1}(\chi)\mathcal{J}_{k-2}(\chi)-\mathcal{J}_{-k}(\chi)\mathcal{J}_{-k+1}(\chi)\big],\\\nonumber
	c_2=&\,\sum_{k>0}\frac{\mathcal{N}(\chi)}{k}\big[\mathcal{J}_{-k+1}(\chi)\mathcal{J}_{k-2}(\chi)+\mathcal{J}_{-k}(\chi)\mathcal{J}_{k-1}(\chi)\\
	&-\mathcal{J}_{k+1}(\chi)\mathcal{J}_{-k-2}(\chi)-\mathcal{J}_{k}(\chi)\mathcal{J}_{-k-1}(\chi)\big],\\\nonumber
	c_3=&\,\sum_{k>0}\frac{\mathcal{N}(\chi)}{k}\big[\mathcal{J}_{k-1}^2(\chi)+\mathcal{J}_{k-2}^2(\chi)\\
	&-\mathcal{J}_{-k-1}^2(\chi)-\mathcal{J}_{-k-2}^2(\chi)\big],\\\nonumber
	c_4=&\,\sum_{k>0}\frac{\mathcal{N}(\chi)}{k}\big[\mathcal{J}_{-k+1}^2(\chi)+\mathcal{J}_{-k}^2(\chi)\\
	&-\mathcal{J}_{k+1}^2(\chi)-\mathcal{J}_{k}^2(\chi)\big],
	\end{align}
\end{subequations}
where $\chi$ is a dimensionless driving parameter (see Sec.~\ref{sec:Z2LGT_B}), $\mathcal{J}_q(\chi)$ is the $q^\text{th}$-order Bessel function of the first kind and $\mathcal{N}(\chi)$ is a nonzero factor ensuring that $\sum_{n=1}^4c_n=1$. In this work, we display results for $\chi=1.84$ and $1.3$. 

\begin{figure}[htp]
	\centering
	\hspace{-.25 cm}
	\includegraphics[width=.48\textwidth]{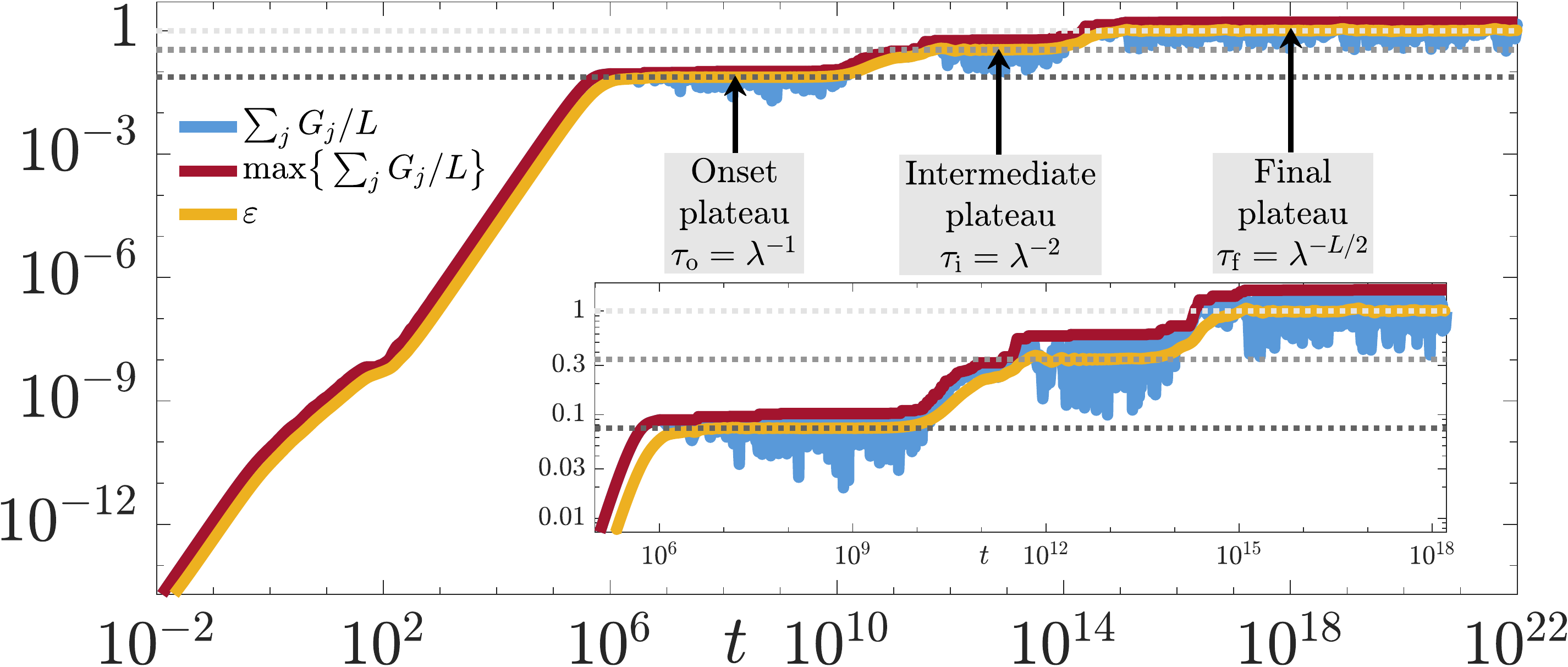}\quad
	\hspace{-.15 cm}
	\caption{(Color online). ED results for the case of the $\mathrm{Z}_2$ LGT for $L=6$ matter sites, initial state (A) from Fig.~\ref{fig:FigStates}, and $\lambda=10^{-5}$. In the main text we only display time average of the gauge violation $\varepsilon$, but exactly the same features are shared by the raw time evolution of the gauge violation as well as its running maximum.  } 
	\label{fig:NumericsSpecifics} 
\end{figure}

In the results of our paper and the joint submission,\cite{Halimeh2020b} we show the spatiotemporal averages of observables. However, the staircase prethermalization is also just as evident in the raw gauge violation (no temporal averaging) and in the running maximum thereof, as shown in Fig.~\ref{fig:NumericsSpecifics}. In the raw gauge violation, the prethermal plateau manifests itself in persistent oscillations around a mean value.

\section{Time-dependent perturbation theory}\label{sec:TDPT}
Even though our discussion is general for any Abelian gauge symmetry, here we focus for concreteness on the $\mathrm{Z}_2$ gauge theory. Our numerical results in panels (b) of Figs.~\ref{fig:FigL4},~\ref{fig:FigOther}--\ref{fig:FigInitStateIV},~\ref{fig:FigNonzeroGj}, and~\ref{fig:MaxOcc2} show that for nonzero even $s\leq L$ the expectation value of the projector onto supersector $\{\alpha_{\{s\}}\}$, denoted by $P_{s}$ in Eq.~\eqref{eq:P}, grows $\sim(\lambda t)^s$ for times $t\lesssim\lambda^{-1}$. We recall that our Hamiltonian is $H=H_0+\lambda H_1$, where $[H_0,G_j]=0$ and $[H_1,G_j]\neq0$, $\forall j$, with $G_j$ being a local gauge generator at position $j$, and that the gauge-invariant supersector $\{\alpha_{\{s\}}\}$ is the set of all gauge-invariant sectors $\alpha_{\{s\}}=(\alpha_1,\ldots,\alpha_L)$ where $\sum_jG_j\ket{\psi}=2s\ket{\psi}$, i.e., $\sum_j\alpha_j=2s$, in the case of the $\mathrm{Z}_2$ LGT (see Glossary in Appendix~\ref{sec:glossary}). Consequently, the gauge invariance encapsulated in $H_0$ is violated up to a strength $\lambda$ by the gauge-noninvariant term $H_1$. The gauge-invariance violation grows as $\sim(\lambda t)^2$ at short times $t\lesssim\lambda^{-1}$, as can also be shown in time-dependent pertubration theory.\cite{Halimeh2020a} Here, we calculate the scaling of the expectation values of projectors $P_s$ at short times, employing TDPT.

The gauge symmetry takes on a value of either $0$ or $2$ locally. If our initial state is in a given sector, $H_0$ induces dynamics solely within the associated sector containing it (see Appendix~\ref{sec:glossary} for a glossary defining gauge-invariant sectors and supersectors), whereas $H_1$ drives the dynamics into other sectors within the total Hilbert space of the system that are not necessarily within the initial supersector.

Now since $[H_0,G_j]=[H_0,P_s]=[G_j,P_s]=[G_j,G_l]=[P_s,P_r]=0$, $\forall$ spatial indices $j,x$ and supersector indices $s,r$, we can find a common eigenbasis $\{\ket{\alpha,q}\}$ for $H_0$ and all $G_j$ and $P_s$, where $\alpha=(\alpha_1,\alpha_2,\ldots,\alpha_L)$ denotes the gauge-invariant sector defined by the unique set of local values $\alpha_j$, and $q$ stands for all remaining good quantum numbers, i.e., we have $H_0\ket{\alpha,q}=E_{\alpha,q}\ket{\alpha,q}$, $G_j\ket{\alpha,q}=\alpha_j\ket{\alpha,q}$, and $P_s\ket{\alpha,q}=\delta_{\sum_j\alpha_j,2s}\ket{\alpha,q}$. Our initial state $\ket{\psi_0}$ at $t=0$ is such that $G_j\ket{\psi_0}=0$, $\forall j$, i.e., it is in the gauge-invariant sector $0$---this sector is its own supersector, same as for the sector where $G_j\ket{\psi}=2\ket{\psi}$, $\forall j$. It is to be noted that the scalings obtained in the following analysis hold also when starting in different sectors, albeit not for the same projectors (see Sec.~\ref{sec:Z2LGT_E} and Fig.~\ref{fig:FigNonzeroGj} therein). As such, our initial state can be written as
\begin{align}
\ket{\psi_0}=\sum_q\ket{0,q}\bra{0,q}\ket{\psi_0}.
\end{align}
The projector $P_s$ of Eq.~\eqref{eq:P} is gauge-invariant, i.e., it satisfies $[P_s,G_j]=0$, $\forall j,s$, and it also commutes with $H_0$, $[P_s,H_0]=0$, $\forall s$. The time evolution of $H_1$ under $H_0$ is
\begin{align}\label{eq:H11}
H_1(t)=\sum_{\alpha,\beta}\sum_{q,l}\mathrm{e}^{\mathrm{i}(E_{\alpha,q}-E_{\beta,l})t}\bra{\alpha,q}H_1\ket{\beta,l}\ket{\alpha,q}\bra{\beta,l}.
\end{align}
The expression for the commutator of Eqs.~\eqref{eq:P} and~\eqref{eq:H11} will be useful later on:  
\begin{align}\nonumber
P_sH_1(\tau)=&\,\sum_{\alpha_{\{s\}},\beta}\sum_{q,l}\mathrm{e}^{\mathrm{i}\big(E_{\alpha_{\{s\}},q}-E_{\beta,l}\big)\tau}\\\label{eq:OH}
&\times\langle\alpha_{\{s\}},q| H_1|\beta,l\rangle|\alpha_{\{s\}},q\rangle\langle\beta,l|,\\\nonumber
H_1(t)P_s=&\,\sum_{\alpha_{\{s\}},\beta}\sum_{q,l}\mathrm{e}^{-\mathrm{i}\big(E_{\alpha_{\{s\}},q}-E_{\beta,l}\big)t}\\\label{eq:HO}
&\times\langle\beta,l| H_1|\alpha_{\{s\}},q\rangle|\beta,l\rangle\langle\alpha_{\{s\}},q|.
\end{align}

Within TDPT, the time-evolution operator can be written as
\begin{widetext}
	\begin{align}\nonumber
	U(t)=&\,\mathrm{e}^{-\mathrm{i}(H_0+\lambda H_1)t}=\mathrm{e}^{-\mathrm{i}H_0t}\mathcal{T}\big\{\mathrm{e}^{-\mathrm{i}\lambda\int_0^t \d\tau H_1(\tau)}\big\}\\\nonumber
	=&\,\mathrm{e}^{-\mathrm{i}H_0t}\bigg\{1-\mathrm{i}\lambda\int_0^t\d t_1H_1(t_1)-\lambda^2\int_0^t\d t_2\int_0^{t_2}\d t_1H_1(t_2)H_1(t_1)+\mathrm{i}\lambda^3\int_0^t\d t_3\int_0^{t_3}\d t_2\int_0^{t_2}\d t_1H_1(t_3)H_1(t_2)H_1(t_1)\\\label{eq:U}
	&+\lambda^4\int_0^t\d t_4\int_0^{t_4}\d t_3\int_0^{t_3}\d t_2\int_0^{t_2}\d t_1H_1(t_4)H_1(t_3)H_1(t_2)H_1(t_1)+\mathcal{O}(\lambda^5)\bigg\}.
	\end{align}
\end{widetext}
Using this expansion, whose validity requires $t\ll\lambda^{-1}$, we can now derive $\bra{\psi_0}U^\dagger(t)P_sU(t)\ket{\psi_0}$ up to fourth order in TDPT.
As the gauge invariance-breaking terms of Eq.~\eqref{eq:H1} act on two adjacent $G_j$ simultaneously, a process of any given order in $H_1$ would allow breaking only an even number of local constraints. As a consequence, $\langle P_s\rangle=0$ when $s$ is odd. 
Only $\langle P_s\rangle$ with even $s$ give nonvanishing contributions.

The zeroth-order contribution is
\begin{align}\nonumber
&\bra{\psi_0}P_s\ket{\psi_0}=\sum_{p,q}\bra{\psi_0}\ket{0,p}\bra{0,q}\ket{\psi_0}\bra{0,p}P_s\ket{0,q}\\\label{eq:ZerothOrderPT}
&=\sum_{q}|\bra{0,q}\ket{\psi_0}|^2\bra{0,q}P_s\ket{0,q}=\delta_{s,0}.
\end{align}
This expression is nonzero only if $s=0$, since otherwise $\bra{0,q}P_{s\neq0}\ket{0,q}=0$. This is because $\ket{\psi_0}$ lies in the gauge-invariant sector $0$, and thus $\bra{\psi_0}P_0\ket{\psi_0}=1$. Therefore, the zeroth-order contribution from TDPT dominates $\langle P_0\rangle$ at short times.

The first-order contribution is
\begin{align}
&\mathrm{i}\lambda\int_0^t\d t_1\bra{\psi_0}[H_1(t_1),P_s]\ket{\psi_0}=0,
\end{align}
for all $P_s$. This makes sense because $P_s$ is not only gauge-invariant, but it also commutes with $H_0$. Gauge-invariant observables that do not commute with $H_0$ in the $\mathrm{Z}_2$ gauge theory, such as the staggered electric field, can have nonzero linear-in-$\lambda$ contributions from TDPT. \cite{Halimeh2020a}

The second-order contribution from TDPT is of two components, with the first taking the form $-\lambda^2\int_0^t\d t_2\int_0^{t_2}\d t_1\bra{\psi_0}H_1(t_1)H_1(t_2)P_s+P_sH_1(t_2)H_1(t_1)\ket{\psi_0}$, which vanishes for $s\neq0$ since $P_{s\neq0}\ket{\psi_0}=0$, while the second component is
\begin{align}\label{eq:PT2ndOrder}
&\lambda^2\int_0^t\d t_1\int_0^t\d \tau_1\bra{\psi_0}H_1(t_1)P_sH_1(\tau_1)\ket{\psi_0}\\\nonumber
=&\,-\lambda^2\sum_{\alpha_{\{s\}}}\sum_{p,q,l}\frac{\mathrm{e}^{\mathrm{i}\big(E_{\alpha_{\{s\}},q}-E_{0,l}\big)t}-1}{E_{\alpha_{\{s\}},q}-E_{0,l}}\frac{\mathrm{e}^{\mathrm{i}\big(E_{0,p}-E_{\alpha_{\{s\}},q}\big)t}-1}{E_{0,p}-E_{\alpha_{\{s\}},q}}\\\nonumber
&\times\langle 0,p|H_1|\alpha_{\{s\}},q\rangle\langle\alpha_{\{s\}},q| H_1|0,l\rangle\bra{\psi_0}\ket{0,p}\langle 0,l|\psi_0\rangle\,.
\end{align}
Here, we implied only non-resonant contributions in the summation. 
At times much shorter than the relevant gaps, $t\ll \left| E_{\alpha_{\{s\}},q}-E_{0,l} \right|^{-1},\left| E_{0,p}-E_{\alpha_{\{s\}},q} \right|^{-1}$, this term can be approximated as 
\begin{align}\nonumber
&\lambda^2t^2\sum_{\alpha_{\{s\}}}\sum_{p,q,l}\langle 0,p|H_1|\alpha_{\{s\}},q\rangle\langle\alpha_{\{s\}},q| H_1|0,l\rangle\\\label{eq:leadingOrderTDPTshortTimes}
&\times\bra{\psi_0}\ket{0,p}\langle 0,l|\psi_0\rangle,
\end{align}
giving a polynomial increase [see {\bf{(1)}} in Fig.~\ref{fig:Schematics}]. 
The error term $H_1$ in Eq.~\eqref{eq:H1} is composed of local terms each of which acts simultaneously on two adjacent local constraints. Moreover, $H_1$ includes processes that do not drive the dynamics \textit{completely} out of the gauge-invariant sector.\cite{Halimeh2020a,Schweizer2019} As such, $\langle 0,p|H_1|\alpha_{\{s\}},q\rangle$ and $\langle\alpha_{\{s\}},q| H_1|0,l\rangle$ in Eq.~\eqref{eq:leadingOrderTDPTshortTimes} can only be nonzero for $s\leq2$. This explains why at short times $\langle P_2\rangle\sim(\lambda t)^2$, but $\langle P_{s>2}\rangle\nsim(\lambda t)^2$; cf.~Fig.~\ref{fig:FigL4}(b), for example. This also shows that the subleading contribution from TDPT to $\langle P_0\rangle$ is $\propto(\lambda t)^2$.

At times $t\gg \left| E_{\alpha_{\{s\}},q}-E_{0,l} \right|^{-1},\left| E_{0,p}-E_{\alpha_{\{s\}},q} \right|^{-1}$ the oscillating exponentials in Eq.~\ref{eq:leadingOrderTDPTshortTimes} average away, yielding a constant contribution that can give rise to the pre-onset plateau [see {\bf{(2)}} in Fig.~\ref{fig:Schematics}].
In addition, similarly to $\Omega_1$ in Sec.~\ref{sec:ME}, there are resonant contributions with $E_{\alpha_{\{s\}},q}-E_{0,l}=0$ or $E_{0,p}-E_{\alpha_{\{s\}},q}=0$. These generate a dependence $\propto\lambda t$ that---in contrast to the nonresonant contributions---does not saturate at any time. Their increasing norm eventually drives the system out of the validity regime of TDPT, where it becomes necessary to resort to the Magnus expansion developed in Sec.~\ref{sec:ME}. 

The third-order contribution from TDPT involves two terms, with the first term vanishing as $-\mathrm{i}\lambda^3\int_0^t\d t_3\int_0^{t_3}\d t_2\int_0^{t_2}\d t_1\langle\psi_0|H_1(t_1)H_1(t_2)H_1(t_3)P_s|\psi_0\rangle+\text{c.c.}=0$ for $s\neq0$, because $P_{s\neq0}\ket{\psi_0}=0$, while the second term is given by
\begin{align}\nonumber
&-\mathrm{i}\lambda^3\int_0^t\d t_1\int_0^t\d \tau_2\int_0^{\tau_2}\d \tau_1\\\nonumber
&\times\langle\psi_0|\big(H_1(t_1)P_sH_1(\tau_2)H_1(\tau_1)-\text{H.c.}\big)|\psi_0\rangle\\\nonumber
\approx&\,-\frac{\mathrm{i}}{2}\lambda^3t^3\sum_{\alpha_{\{s\}},\beta}\sum_{q,l,p,k}\langle 0,l| H_1|\alpha_{\{s\}},q\rangle\langle\alpha_{\{s\}},q|H_1|\beta,p\rangle\\
&\times\bra{\beta,p}H_1\ket{0,k}\bra{0,k}\ket{\psi_0}\langle\psi_0|0,l\rangle+\text{c.c.},
\end{align}
where the approximation as in Eq.~\eqref{eq:leadingOrderTDPTshortTimes} holds for times much shorter than the relevant gaps.
This expression reduces to zero for $s>2$ because then $\langle 0,l| H_1|\alpha_{\{s\}},q\rangle\mbeq0$. Nevertheless, $\langle 0,l| H_1|\alpha_{\{s\}},q\rangle$ does not necessarily vanish for $s=2$, and thus $\langle P_2\rangle$ can have a subleading correction $\propto(\lambda t)^3$, with the leading correction at short times being $\propto(\lambda t)^2$ from Eq.~\eqref{eq:leadingOrderTDPTshortTimes}.

Finally, the fourth-order contribution from TDPT has three terms. The first one reads  
\begin{align}
&\lambda^4\int_0^t\d t_4\int_0^{t_4}\d t_3\int_0^{t_3}\d t_2\int_0^{t_2}\d t_1\\\nonumber
&\times\langle\psi_0|\big(H_1(t_1)H_1(t_2)H_1(t_3)H_1(t_4)P_s+\text{H.c.}\big)|\psi_0\rangle=0,
\end{align}
for $s\neq0$ since $P_{s\neq0}\ket{\psi_0}=0$. The second fourth-order term can be shown to reduce to
	\begin{align}\label{eq:FourthOrderPTterm2}
	&-\lambda^4\int_0^t\d t_1\int_0^t\d\tau_3\int_0^{\tau_3}\d\tau_2\int_0^{\tau_2}\d\tau_1\\\nonumber
	&\times\langle\psi_0|H_1(t_1)P_sH_1(\tau_3)H_1(\tau_2)H_1(\tau_1)|\psi_0\rangle+\text{c.c.}
	\\\nonumber
	\approx&\,\frac{-2}{3!}\lambda^4t^4\sum_{\alpha_{\{s\}},\beta,\gamma}\sum_{q,l,p,k,m}\real\big[\langle 0,l|H_1|\alpha_{\{s\}},q\rangle\langle\alpha_{\{s\}},q|H_1|\beta,p\rangle\\\nonumber
	&\times\bra{\beta,p}H_1\ket{\gamma,k}\bra{\gamma,k}H_1\ket{0,m}\bra{0,m}\ket{\psi_0}\langle\psi_0|0,l\rangle\big],
	\end{align}
with the approximation again holding for times much shorter than the relevant gaps.
This expression is always zero for $s>2$ since $\langle 0,l|H_1|\alpha_{\{s>2\}},q\rangle\mbeq 0$ as linear-order processes in the terms of $H_1$ can break only pairs of local constraints. Nevertheless, Eq.~\eqref{eq:FourthOrderPTterm2} can still contribute to $\langle P_0\rangle$ and $\langle P_2\rangle$, but such a contribution will be dominated by lower-order contributions from TDPT.

The third and final fourth-order term from TDPT can for short times be approximated as
	\begin{align}\nonumber
	&2\lambda^4\int_0^t\d t_2\int_0^{t_2}\d t_1\int_0^t\d \tau_2\int_0^{\tau_2}\d \tau_1\\\nonumber
	&\times\langle\psi_0|H_1(t_1)H_1(t_2)P_sH_1(\tau_2)H_1(\tau_1)|\psi_0\rangle\\\nonumber
	\approx&\,\frac{1}{2}\lambda^4t^4\sum_{\alpha_{\{s\}},\beta,\gamma}\sum_{p,l,k,q,m}
	\frac{2E_{\alpha_{\{s\}},q}-E_{\gamma,k}-E_{0,m}}{E_{\gamma,k}-E_{0,m}}\\\nonumber
	&\times\frac{2E_{\alpha_{\{s\}},q}-E_{\beta,l}-E_{0,p}}{E_{\beta,l}-E_{0,p}}\bra{0,m}\ket{\psi_0}\bra{\psi_0}\ket{0,p}\\\nonumber
	&\times\bra{0,p}H_1\ket{\beta,l}\langle\beta,l| H_1|\alpha_{\{s\}},q\rangle\langle\alpha_{\{s\}},q|H_1|\gamma,k\rangle\\\label{eq:FourthOrderPTterm3}
	&\times\bra{\gamma,k}H_1\ket{0,m},
	\end{align}
which indeed is in general nonzero for $s\leq4$ with $s$ even. Specifically, Eq.~\eqref{eq:FourthOrderPTterm3} contributes nondominantly to $\langle P_0\rangle$ and $\langle P_2\rangle$ and dominantly to $\langle P_4\rangle$, explaining why the latter scales as $\langle P_4\rangle\sim(\lambda t)^4$ at short times; cf.~Fig.~\ref{fig:FigL4}(b). However, Eq.~\eqref{eq:FourthOrderPTterm3} is zero for $s>4$.

Similarly, one can go to sixth order in TDPT to illustrate why $\langle P_6\rangle\sim(\lambda t)^6$, and more generally, to $s^\text{th}$ order in TDPT to show that $\langle P_s\rangle\sim(\lambda t)^s$ for even $s$. Inspired by the forms of Eqs.~\eqref{eq:leadingOrderTDPTshortTimes} and~\eqref{eq:FourthOrderPTterm3}, we can deduce the nonzero $s^\text{th}$-order contribution from TDPT to $\langle P_s\rangle$ with even $s$ to be
\begin{widetext}
	\begin{align}\nonumber
	&(\lambda t)^s\sum_{\alpha_{\{w\}},\beta_{\{v\}}}\sum_{p,m,l_{\alpha_{\{w\}}},q_{\beta_{\{v\}}}}\mathcal{E}(\alpha_{\{w\}},\beta_{\{v\}},0;p,m,l_{\alpha_{\{w\}}},q_{\beta_{\{v\}}})\langle 0,m|\psi_0\rangle\langle\psi_0|0,p\rangle\\\nonumber
	&\times\overbrace{\langle 0,p|H_1|\alpha_{\{2\}},l_{\alpha_{2}}\rangle\langle\alpha_{\{2\}},l_{\alpha_{2}}|H_1|\alpha_{\{4\}},l_{\alpha_{4}}\rangle\ldots\langle\alpha_{\{s-2\}},l_{\alpha_{\{s-2\}}}|H_1|\alpha_{\{s\}},l_{\alpha_{\{s\}}}\rangle}^{\frac{s}{2}\,\text{terms}}\\
	&\times\underbrace{\langle\alpha_{\{s\}},l_{\alpha_{\{s\}}}|H_1|\beta_{\{s-2\}},q_{\beta_{\{s-2\}}}\rangle\ldots\langle\beta_{\{4\}},q_{\beta_{\{4\}}}|H_1|\beta_{\{2\}},q_{\beta_{\{2\}}}\rangle\langle\beta_{\{2\}},q_{\beta_{\{2\}}}|H_1|0,m\rangle}_{\frac{s}{2}\,\text{terms}},
	\end{align}
\end{widetext}
where $\mathcal{E}(\alpha_{\{w\}},\beta_{\{v\}},0;p,m,l_{\alpha_{\{w\}}},q_{\beta_{\{v\}}})$, with even $0\leq w,v\leq s$, are terms consisting of eigenenergies of $H_0$; cf.~Eqs.~\eqref{eq:PT2ndOrder}-\eqref{eq:FourthOrderPTterm3}. Even though the derivation we have carried out in this Section has been tailored for the $\mathrm{Z}_2$ gauge theory, our analytic conclusions extend to the $\mathrm{U}(1)$ gauge theory.

\bibliography{PrethermalBiblio}
\end{document}